\documentclass[format=acmsmall, review=false, screen=true]{acmart}

\usepackage[T1]{fontenc} 
\usepackage[utf8]{inputenc}
\usepackage{amsmath}
\usepackage{bm} 
\usepackage{graphicx}
\usepackage{tikz}
\usepackage{tcolorbox}
\usepackage{pdflscape}
\usepackage{rotating}
\usepackage{xcolor}
\usepackage{fancyhdr} 
\usepackage{color, colortbl}
\usepackage{longtable}
\usepackage{multirow}
\usepackage{graphicx}
\usepackage{enumitem}
\usepackage{fontawesome}
\usepackage{multirow}
\usepackage{balance}
\usepackage{tablefootnote}
\usepackage{colortbl}
\tcbuselibrary{skins}
\usepackage{fontawesome}
\usepackage{hyperref}
\usepackage[export]{adjustbox}
\usepackage{framed}
\usepackage{tabularx}
\usepackage{booktabs} 

\usepackage{arydshln}
\newcolumntype{M}[1]{>{\centering\arraybackslash}m{#1}}
\newcolumntype{N}{@{}m{0pt}@{}}

\usepackage{tikz}
\newcommand{\blackcirclednum}[1]{%
  \tikz[baseline=(char.base)]{
    \node[
      circle,
      fill=black,
      text=white,
      minimum size=3.5mm,
      inner sep=0pt,
      font=\scriptsize
    ] (char) {#1};
  }%
}

\hypersetup
{
 colorlinks = true, 
 urlcolor = blue, 
 linkcolor = blue, 
 citecolor = blue 
}

\usepackage{graphicx}
\usepackage{svg}
\usepackage{subfig}
\usepackage{array}
\definecolor{mybeige}{HTML}{FFF7F3}
\definecolor{myoffwhite}{HTML}{F1F1F1}
\definecolor{mydarkpurple}{HTML}{49006A}
\definecolor{mypurple}{HTML}{99017B}
\definecolor{mydarkpink}{HTML}{E23E99}
\definecolor{mypink}{HTML}{F767A1}
\definecolor{mypink2}{HTML}{F769A1}
\definecolor{mylightpink}{HTML}{F994B1}
\definecolor{mysalmon}{HTML}{FCC8C3}
\definecolor{mylightsalmon}{HTML}{FBBABD}

\AtBeginDocument{%
  }

\setcopyright{acmlicensed}
\copyrightyear{2026}
\acmYear{2026}
\acmDOI{XXXXXXX.XXXXXXX}

\acmJournal{TOSEM}
\acmVolume{0}
\acmNumber{0}
\acmArticle{0}
\acmMonth{0}

\begin{document}

\title{Decision Models for Selecting Architecture Patterns and Strategies in Quantum Software Systems}

\author{Mst Shamima Aktar}
\orcid{0009-0003-6058-2970}
\affiliation{%
  \institution{School of Computer Science, Wuhan University}
  \city{Wuhan}
  \country{China}
}
\email{shamima@whu.edu.cn}

\author{Peng Liang}
\orcid{0000-0002-2056-5346}
\affiliation{%
  \institution{School of Computer Science, Wuhan University}
  \city{Wuhan}
  \country{China}
  }
\email{liangp@whu.edu.cn}

\author{Muhammad Waseem}
\orcid{0000-0001-7488-2577}
\affiliation{%
  \institution{Tampere University}
  \city{Tampere}
  \country{Finland}
  }
\email{muhammad.waseem@tuni.fi}

\author{Amjed Tahir}
\orcid{0000-0001-9454-1366}
\affiliation{%
  \institution{Massey University}
  \city{4442 Palmerston North}
  \country{New Zealand}
  }
\email{a.tahir@massey.ac.nz}

\author{Mojtaba Shahin}
\orcid{0000-0002-9081-1354}
\affiliation{%
  \institution{RMIT University}
  \city{Melbourne}
  \country{Australia}
  }
\email{mojtaba.shahin@rmit.edu.au}

\author{Muhammad Azeem Akbar}
\orcid{0000-0002-4906-6495}
\affiliation{%
  \institution{LUT University}
  \city{Lappeenranta}
  \country{Finland}
  }
\email{Azeem.Akbar@lut.fi}

\author{Arif Ali Khan}
\orcid{0000-0002-8479-1481}
\affiliation{%
  \institution{University of Oulu}
  \city{Oulu}
  \country{Finland}
  }
\email{arif.khan@oulu.fi}

\author{Aakash Ahmad}
\orcid{0000-0002-3198-9638}
\affiliation{%
  \institution{University of Derby}
  \city{Derby}
  \country{United Kingdom}
  }
\email{a.abbasi@derby.ac.uk}

\author{Musengamana Jean de Dieu}
\orcid{0009-0000-5956-997X}
\affiliation{%
  \institution{School of Computer Science, Wuhan University}
  \city{Wuhan}
  \country{China}
  }
\email{jeanmusenga@whu.edu.cn}

\author{Ruiyin Li}
\orcid{0000-0001-8536-4935}
\affiliation{%
  \institution{School of Computer Science, Wuhan University}
  \city{Wuhan}
  \country{China}
  }
\email{ryli_cs@whu.edu.cn}

\thanks{This work has been partially supported by the National Natural Science Foundation of China (NSFC) with Grant No. 92582203 and 62402348.}

\renewcommand{\shortauthors}{Aktar et al.}


\begin{abstract}
\textcolor{black}{Quantum software is an emerging class of software systems, services, and applications that leverage the principles of quantum mechanics through programmable quantum bits (qubits) and quantum gates to perform computations that offer advantages for certain classes of problems.} Quantum software architecture enables quantum software developers to abstract away implementation-specific details (i.e., mapping of qubits and quantum gates to high-level architectural components and connectors). \textcolor{black}{Architecture patterns provide proven solutions to recurring design problems, while architecture strategies are approaches that guide the overall design and evolution of a system to meet specific goals.} \textcolor{black}{Such patterns and strategies are needed because quantum software systems involve complex hybrid quantum-classical design concerns. However, quantum software practitioners face significant challenges in selecting and implementing appropriate patterns and strategies.} To address these challenges, this study proposes decision models for selecting patterns and strategies in six critical design areas in quantum software systems: Communication, Decomposition, Data Processing, Fault Tolerance, Integration and Optimization, and Algorithm Implementation. These decision models are constructed based on data collected from both a mining study (i.e., GitHub and Stack Exchange) and a Systematic Literature Review (SLR), which were used to identify relevant patterns and strategies with their involved Quality Attributes (QAs). We then conducted semi-structured interviews with 30 quantum software practitioners to evaluate the familiarity, understandability, completeness, and usefulness of the proposed decision models. The results show that the proposed decision models can aid practitioners in selecting suitable patterns and strategies to address the challenges related to the architecture design of quantum software systems. The dataset is available at \cite{dataset2}, allowing the community to reproduce and build upon our findings.
\end{abstract}





\maketitle

\section{Introduction} \label{introduction} 
Quantum computing represents a transformative shift in computational paradigms, leveraging principles such as superposition, entanglement, and quantum interference to achieve capabilities beyond classical computing \cite{nielsen2010quantum}. \textcolor{black}{Unlike classical bits, which exist exclusively in binary states (0 or 1), qubits can exist simultaneously in multiple states, offering exponential speedup for certain problems \cite{feynman2018simulating}. This capability allows quantum computers to address highly complex problems, such as factoring large numbers \cite{shor1994algorithms}, and to enhance optimization tasks through quantum machine learning, where quantum algorithms offer potential advantages over classical approaches \cite{salloum2024integration, biamonte2017quantum}}. \textcolor{black}{The concept of quantum supremacy, wherein quantum computers outperform classical supercomputers on specific tasks, has driven substantial research in Quantum Software Engineering (QSE) \cite{arute2019quantum}. QSE is an emerging discipline that integrates the principles of quantum mechanics with software engineering practices to support the design, development, and evolution of quantum software systems \cite{ahmad2022towards}.} QSE has been identified as a critical research direction in the broader software engineering landscape \cite{pezze20252030}. However, developers often face challenges in building robust and scalable quantum applications due to the difficulty of applying classical software engineering practices \cite{murillo2024quantum, de2024quantum} and the lack of established quantum-specific best practices \cite{aparicio2024overview}. \textcolor{black}{Despite these challenges, the increasing potential of quantum computing has driven significant global investments \cite{humble2019quantum} and led to the collaborative development of platforms and tools tailored for quantum programming, such as IBM’s Qiskit, Google’s Cirq, Microsoft’s Q\#, and Intel’s quantum C++ extensions \cite{piattini2021toward}, along with frameworks like $C2|Q\rangle$ that provide modular workflows for translating classical specifications into quantum-executable programs \cite{ye2026c2}}. These initiatives support the growing ecosystem for QSE and facilitate structured development workflows \cite{akbar2024genetic}. Amidst these advancements, quantum software architecture has emerged as a critical subdiscipline \cite{yue2023challenges}. It provides high-level abstractions and design principles \cite{khan2023software} for constructing both quantum and hybrid quantum-classical applications \cite{weder2021hybrid}. Software architects play a crucial role in designing systems that align with business requirements while integrating quantum capabilities \cite{weder2020integrating}. However, designing robust quantum solutions remains highly challenging, especially for practitioners with little background in quantum mechanics \cite{garcia2021quantum} or quantum programming paradigms \cite{ahmad2023engineering}. \textcolor{black}{In response, recent studies have introduced early-stage modeling approaches \cite{yue2023challenges} and architecture-centric frameworks, such as a catalog of patterns for quantum AI systems \cite{klymenko2024architectural}. These frameworks include extensions to well-established modeling tools, such as UML \cite{perez2021modelling} and BPMN \cite{weder2020integrating}, which aim to streamline the integration of quantum and classical components by providing clearer abstractions and promoting modular design principles.}


\textcolor{black}{Architecture patterns provide reusable solutions within a specific context, defining components, their responsibilities, relationships, and interactions based on proven best practices, enabling developers to effectively tackle complex design problems \cite{buschmann2007pattern}. In contrast, architecture strategies are high-level design approaches that guide the architectural synthesis process, promoting the reuse of proven design solutions to meet specific quality requirements of a software system \cite{valle2021architectural}. In the context of quantum software systems, both architecture patterns and strategies offer reusable solutions to common design challenges \cite{baczyk2024towards}. They provide a structured framework for developing scalable and maintainable systems in quantum computing \cite{khan2023software, buhler2023patterns}.}  As quantum systems increasingly integrate with traditional system infrastructure, high-level architecture design patterns are required to orchestrate services \cite{murillo2024quantum} and manage workflows across quantum and classical boundaries \cite{zhao2024towards}. In classical software systems, architecture patterns such as layered \cite{richards2015software}, service-oriented \cite{endrei2004patterns}, and microservices \cite{richardson2018microservices} architectures have been widely adopted to improve various QAs (e.g., maintainability, scalability, and performance). \textcolor{black}{The choice of patterns and their effectiveness is highly dependent on the specific system context, including factors such as the business, operational, and technological constraints \cite{bi2018architecture, harper2015exploring}.} However, these classical patterns can be extended to address the unique constraints of quantum software systems, including quantum resource management and error correction \cite{baczyk2024patterns}, the integration of classical and quantum components through hybrid execution models \cite{weder2021hybrid}, and the challenges of orchestration and execution in hybrid quantum-classical systems \cite{perez2023generation}. Various architecture patterns have been introduced for designing quantum applications. For example, Khan et al. \cite{khan2023software} identified Layer, Pipe and Filter, Composite, and Prototype patterns tailored for quantum systems. Our previous work \cite{aktar2025architecture} highlighted that selecting appropriate architecture patterns is a key decision in quantum software development. Discussions on open-source platforms such as GitHub and Stack Exchange \cite{khan2025mining}, along with a systematic mapping of research literature on quantum computing services \cite{ahmad2023engineering}, demonstrate ongoing efforts to identify, refine, and apply architecture patterns within the emerging domain of quantum software systems. 


One key limitation in quantum software development is the absence of systematic support to guide practitioners in evaluating and selecting appropriate architectural solutions \cite{murillo2024quantum}. Especially, it is challenging for practitioners to balance competing QAs such as performance, scalability, reliability, and interoperability \cite{weder2021hybrid, perez2023generation}. While architectural design in classical software engineering benefits from decision models that aid in selecting architecture patterns and strategies within well-defined design contexts \cite{waseem2022decision, xu2021decision, jacob2018software}, QSE has yet to employ comparable frameworks. Previous studies have explored quality trade-offs in architectural decisions for quantum-classical hybrid systems \cite{khan2024advancing}, and proposed architecture patterns specifically designed for quantum-enhanced artificial intelligence systems \cite{klymenko2024architectural}. However, the absence of structured, quality-driven decision models that map architecture patterns and strategies to design concerns and quality trade-offs remains a critical gap, limiting informed architectural decision-making in quantum software systems.


To address this gap, this study aims to introduce comprehensive decision models tailored to designing quantum software systems, focusing on six crucial design areas: (1) Communication, (2) Decomposition, (3) Data Processing, (4) Fault Tolerance, (5) Integration and Optimization, and (6) Algorithm Implementation.  \textcolor{black}{The necessity of these six areas is reinforced by a growing body of recent literature. First, quantum communication remains a critical architectural challenge in distributed and networked quantum systems, requiring efficient communication patterns to minimize inter-node overhead \cite{dutta2024quantum, wu2023qucomm}. Second, quantum software systems benefit from decomposition into modular components and layered architectures, thereby enhancing maintainability, scalability, and integration, following established architecture patterns \cite{serrano2022quantum, schonberger2022peel, svore2006layered}. Third, data processing affects circuit width, state preparation, and algorithm execution, while patterns such as angle encoding and basis encoding enable efficient data representation in qubit states while balancing resource requirements \cite{weigold2020data, weigold2021encoding, weigold2021expanding}. Fourth, fault tolerance is a critical concern in NISQ-era hybrid quantum-classical architectures \cite{bensoussan2025taxonomy}, where recurring fault categories necessitate dedicated architecture patterns, such as comparison, voting, and sparing \cite{scheerer2022fault, ding2017classification}. Fifth, classical-quantum integration is among the most structurally complex challenges in hybrid architectures, requiring careful orchestration of heterogeneous computational models \cite{baczyk2024towards, saurabh2023conceptual}. Sixth, algorithm implementation encompasses critical decisions around circuit design, gate selection, and optimization that determine both the correctness and efficiency of quantum programs \cite{buhler2023patterns}.} These decision models were developed based on insights collected from an SLR and a mining study of software repositories (namely GitHub issues, Stack Exchange sites) about architecture patterns and strategies in quantum software systems (see Section \ref{ResearchMethodology}). These complementary data sources consistently highlight that these six domains capture the most frequent and consequential challenges encountered by both researchers and practitioners in quantum software development. These areas were initially derived from recurring architectural challenges reported in the QSE literature \cite{khan2023software, baczyk2024patterns}.


These six decision models can help practitioners select appropriate architecture patterns and strategies for quantum software systems.
\textcolor{black}{While these six decision models are presented to address specific design areas, they can be used together to address complex, interrelated challenges encountered in real-world quantum software development (see Section \ref{InterdependenciesBetweenModels}).} We evaluated the proposed models through semi-structured interviews with \textcolor{black}{30 quantum software practitioners}, providing empirical validation of their applicability and effectiveness. \textcolor{black}{The \textbf{main contributions} of this study are threefold: \textbf{1)} we developed six decision models that map architecture patterns and strategies to QAs in quantum software systems; \textbf{2)} we evaluated the proposed decision models through semi-structured interviews with quantum software practitioners; and \textbf{3)} we provide all identified patterns and strategies, together with their corresponding references, in a replication package for use by the research and practitioner communities \cite{dataset2}.}

\textbf{Paper Organization}: Section \ref{ResearchRelatedWork} discusses the related work, Section \ref{ResearchMethodology} presents the research methodology, Section \ref{DecisionModelsSection} presents the six decision models, Section \ref{EvaluationOfDecisionModels} details the evaluation process of the proposed decision models, Section \ref{Discussion} discusses the implications and insights of the study results, Section \ref{ResearchThreatsValidity} outlines the threats to validity, and Section \ref{ResearchConclusion} concludes this work with future directions.

\section{Related Work} \label{ResearchRelatedWork}
\textcolor{black}{This section reviews related work on patterns and decision models in software engineering and QSE. Section \ref{pattern_QSE} discusses patterns for QSE, Section \ref{DMS_SE} reviews decision models for selecting patterns in software development, Section \ref{DM_QSA} presents decision models for architecting quantum software systems, and Section \ref{Conclusive_Summary} summarizes the identified research gap.}

\subsection{Patterns for Quantum Software Engineering}\label{pattern_QSE}
QSE relies heavily on design and architecture patterns to address the unique challenges posed by quantum systems, especially in integrating quantum and classical components. Klymenko et al. \cite{klymenko2024architectural} identified several architecture patterns that facilitate the integration of quantum components into classical inference engines, emphasizing key QAs (e.g., efficiency, scalability, and portability). These patterns enable developers to balance performance with hardware limitations when designing complex quantum AI systems, a class of quantum software systems integrating quantum computing to enhance AI tasks such as training and inference. P\'erez-Castillo et al. \cite{perez2024preliminary} explored the application of design patterns in quantum circuits through an analysis of 80 Qiskit and OpenQASM source code files. Their study identified three recurring patterns: initialization, which involves preparing qubits in known basis states using gates; uniform superposition, a quantum-specific pattern that uses Hadamard gates to place all qubits into an equal probability superposed state; and oracle, which encodes problem-specific functions or conditions used in quantum algorithms such as Grover's search. These patterns provide insights into common circuit structures and support more efficient quantum algorithm design. \textcolor{black}{Fern\'andez-Osuna et al. \cite{fernandez2025exploring} investigated the prevalence and relationships between quantum design patterns in over 2,610 Qiskit programs from GitHub. Their work reveals how these patterns frequently occur together, offering practical guidance for quantum developers aiming to optimize scalability and performance.} Baczyk et al. \cite{baczyk2024patterns} argued for the development of higher-level architecture patterns for hybrid quantum-classical systems. While low-level patterns for quantum circuits are well-established, there remains a notable gap in architecture patterns that address critical quality concerns, such as efficiency (e.g., optimizing quantum-classical interactions and resource usage), maintainability (e.g., enabling modular, adaptable architectures as quantum technologies evolve), and security (e.g., safeguarding classical-quantum communication and integration). They emphasized the need for architectural-level guidance to improve the robustness and extensibility of hybrid quantum-classical system design. \textcolor{black}{Leymann \cite{leymann2019towards} introduced a pattern language for quantum algorithms, highlighting that many quantum solutions are built from recurring structures. His work identifies core patterns such as Initialization, Uniform Superposition, Entanglement, Oracle, and Amplitude Amplification, and emphasizes their interconnections in constructing complex algorithms. By organizing these patterns into a pattern language, this study laid a foundational basis for applying SE principles to quantum algorithm design and reuse. Baczyk and P\'erez-Castillo \cite{baczyk2025guidelines} reviewed hybrid quantum-classical design patterns, identifying recurring solutions across quantum algorithms and workflows. They organized these patterns into a unified framework and highlighted their role in improving modularity, reusability, and interoperability in NISQ-era quantum software systems. Bechtold et al. \cite{bechtold2023patterns} introduced pattern-based guidance for quantum circuit cutting to address hardware limitations in NISQ devices, while Jim\'enez-Fern\'andez et al. \cite{jimenez2023systematic} provided a systematic mapping study that categorizes existing quantum circuit design patterns, highlighting their potential for improving the quality and structure of quantum software.}

Recent studies have also highlighted emerging architectural solutions in hybrid quantum-classical systems. For example, Carneiro et al. \cite{carneiro2016microservices} demonstrated how quantum microservices enable modular design by organizing applications into loosely coupled services that communicate through messages. Rojo et al. \cite{rojo2021trials} further explored this approach in hybrid quantum-classical microservices systems, outlining both its advantages and inherent development challenges. \textcolor{black}{To address issues related to backend heterogeneity, Garcia-Alonso et al. \cite{garcia2021quantum} introduced quantum API gateways, which dynamically route requests to the appropriate quantum backends, thus improving system flexibility and interoperability.} Additionally, Beisel et al. \cite{beisel2022patterns} emphasized the importance of error correction patterns in mitigating Qubits fragility, ensuring reliable computation even under noise and decoherence, thereby supporting the broader adoption and industrial viability of quantum computing.


\subsection{Decision Models for Selecting Patterns in Software Development}\label{DMS_SE}
Architectural decision-making is a fundamental aspect of software development, as the selection of appropriate patterns and strategies can significantly impact a system's QAs (e.g., scalability, maintainability). However, due to the increasing complexity and domain-specific constraints of modern systems, such as microservices, blockchain, IoT, and quantum software, practitioners often face challenges in identifying suitable architecture patterns. To address this, previous work proposed structured decision models that guide architects and developers in pattern selection by aligning decisions with design goals and contextual constraints. Waseem et al. \cite{waseem2022decision} proposed decision models for microservices architecture, tackling the selection of patterns and strategies across four key design areas, including service decomposition, security, communication, and discovery within microservices systems. Their study utilized a multivocal review to establish a comprehensive base of patterns and strategies. The decision models proposed were then evaluated through interviews with practitioners, highlighting the practical effectiveness of these models in guiding microservices pattern selection. Lewis et al. \cite{lewis2016decision} explored decision-making in the context of cyber-foraging systems, where the challenge lay in extending the capabilities of mobile devices through computational offloading. Their study also investigated the correlation of architectural tactics with both functional and non-functional requirements in their decision model, thereby guiding the development of cyber-foraging systems. \textcolor{black}{Their work offers insights that inform our analysis of decision models for selecting architecture patterns and strategies in quantum software systems.}

Two critical studies focused on decision models in blockchain systems. Liu et al. \cite{liu2023decision} proposed decision models tailored for governance-driven blockchain architectures and emphasized the need for a structured approach to adopting architecture patterns that address governance dimensions such as decision rights, incentives, and accountability. \textcolor{black}{Their proposed decision models facilitate the selection of patterns that enhance governance in blockchain systems, supported by an evaluation involving expert opinions on the usability, correctness, and completeness of these decision models.} Similarly, Xu et al. \cite{xu2021decision} addressed the challenges of selecting patterns in blockchain-based applications. Their study proposed a decision model that assists developers and architects in making informed choices about pattern adoption based on the characteristics of specific use cases and the associated trade-offs. Their work underscores the complexity and necessity of a methodological approach to pattern selection in blockchain applications, validated by expert feedback on the model's effectiveness in guiding architectural decisions. Zimmermann et al. \cite{zimmermann2008combining} proposed a comprehensive design method that integrates pattern languages with reusable architectural decision models. Their method systematically derives architectural decision models in QSE, providing insights into optimizing simulations in the presence of quantum-mechanicality from platform-independent principles (e.g., general architecture patterns) to platform-specific implementation decisions (e.g., technologies tailored to environments such as .NET or J2EE). Their method was validated through applications in enterprise systems and a case study of a Service-Oriented Architecture (SOA) within the finance industry. Jacob et al. \cite{jacob2018software} proposed a model specifically tailored for selecting architecture patterns in Internet of Things (IoT)-based systems. Their model emphasizes the importance of non-functional requirements such as scalability, availability, reliability, security, and heterogeneity, which are crucial for making informed decisions in the IoT domain. Their model analytically validated a structured approach to pattern selection, offering IoT developers a reference for optimizing architectural choices to address both functional and operational requirements. 


\subsection{Decision Models for Architecting Quantum Software Systems}\label{DM_QSA}
\textcolor{black}{In the context of quantum software systems, there is a crucial need for decision models that guide the selection of architecture patterns and strategies. While such patterns and strategies can be selected based on practitioner expertise, decision models provide a systematic approach to guide this selection, particularly in addressing the computational complexity and inherent complexity of quantum technologies.} In the pursuit of architecting efficient quantum software systems, researchers consider key software QAs (e.g., modularity, maintainability, scalability, and reliability) to guide the creation of decision models. Akbar et al. \cite{akbar2023systematic} introduced a systematic framework that identifies, prioritizes, and develops decision-making strategies for QSE. They mapped challenging factors into seven core categories and assessed their criticality using methods such as Interpretive Structure Modeling (ISM) and fuzz Technique for Order of Preference by Similarity to Ideal Solution (TOPSIS). Their approach provided a structured way to tackle challenges in quantum software development and highlighted the decisive influence of resources (e.g., quantum simulators and compilers) on the execution of the QSE process. Similarly, Vietz et al. \cite{vietz2021decision} focused on the support mechanisms for quantum application developers. They categorized and classified available tools, services, and techniques that assist in quantum application development. Their work culminated in a comparison framework that helps developers choose appropriate technologies for specific quantum computing use cases. Nallamothula et al. \cite{nallamothula2020selection} employed a decision tree approach to facilitate the selection of suitable quantum computing architectures. Their study emphasized the need for tailored decision-making strategies based on the characteristics of a quantum project, including the device, algorithm, and programming languages used. Their approach can aid in identifying architectures that could offer a quantum advantage for specific applications. Grurl et al. \cite{grurl2020considering} focused on incorporating decoherence errors into the simulation of quantum circuits and proposed solutions based on decision diagrams that significantly improve simulation performance by accounting for these errors. Their work was pivotal for decision models in QSE, as they provided insights into optimizing simulations in the presence of quantum mechanical instability. In our previous study \cite{aktar2025architecture}, we conducted an empirical study to analyze architecture decisions made in quantum software development using data collected from GitHub repositories and Stack Exchange discussions. Our analysis revealed a range of practical challenges related to maintainability, performance, and architectural design in real-world quantum software development. The findings from our prior work lay the foundation for the current study by pinpointing key design areas where structured decision support is crucial.

\subsection{Conclusive Summary}\label{Conclusive_Summary}
Previous work on decision models in software engineering reveals a critical gap in addressing the unique architectural needs of quantum software systems. Notably, existing studies have primarily focused on conventional software systems, with a recent expansion into emerging domains (e.g., microservices architecture \cite{waseem2022decision}, blockchain architecture \cite{liu2023decision, xu2021decision}, IoT-based systems \cite{jacob2018software}, and cyber-foraging systems \cite{lewis2016decision}). These studies have provided valuable insights into selecting patterns and strategies by employing decision models that leverage reusable knowledge to address specific design challenges. For instance, the idea of structured decision guidance proposed by Zimmermann et al. \cite{zimmermann2008combining} inspires our work to identify and organize architecture patterns and strategies, along with their associated QAs, for architecting quantum software systems. However, when it comes to QSE, architectural design decisions are further complicated due to quantum computing's unique properties (e.g., entanglement, superposition, and decoherence), which render many conventional design practices insufficient \cite{akbar2023systematic, grurl2020considering, vietz2021decision, nallamothula2020selection}. Designing effective quantum software systems requires careful consideration of critical challenges including quantum-classical communication \cite{klymenko2024architectural}, algorithmic decomposition to support hybrid computation \cite{baczyk2024patterns}, processing of entangled or probabilistic data \cite{buhler2023patterns}, maintaining fault tolerance in noisy environments \cite{grurl2020considering}, integrating quantum components with classical systems \cite{klymenko2024architectural}, optimizing resource-limited quantum workloads \cite{perez2024preliminary}, and implementing quantum algorithms across diverse hardware and programming models \cite{vietz2021decision, nallamothula2020selection}. These challenges correspond to six critical design areas that define the core architectural context of quantum software systems: Communication, Decomposition, Data Processing, Fault Tolerance, Integration and Optimization, and Algorithm Implementation. Despite the growing recognition of these challenges in recent literature, there remains a lack of structured, reusable decision support for selecting architecture patterns and strategies, specifically tailored to quantum software systems.

\section{Research Design}\label{ResearchMethodology} 
\begin{figure}[h]
\centering
\includegraphics[width=1\linewidth]{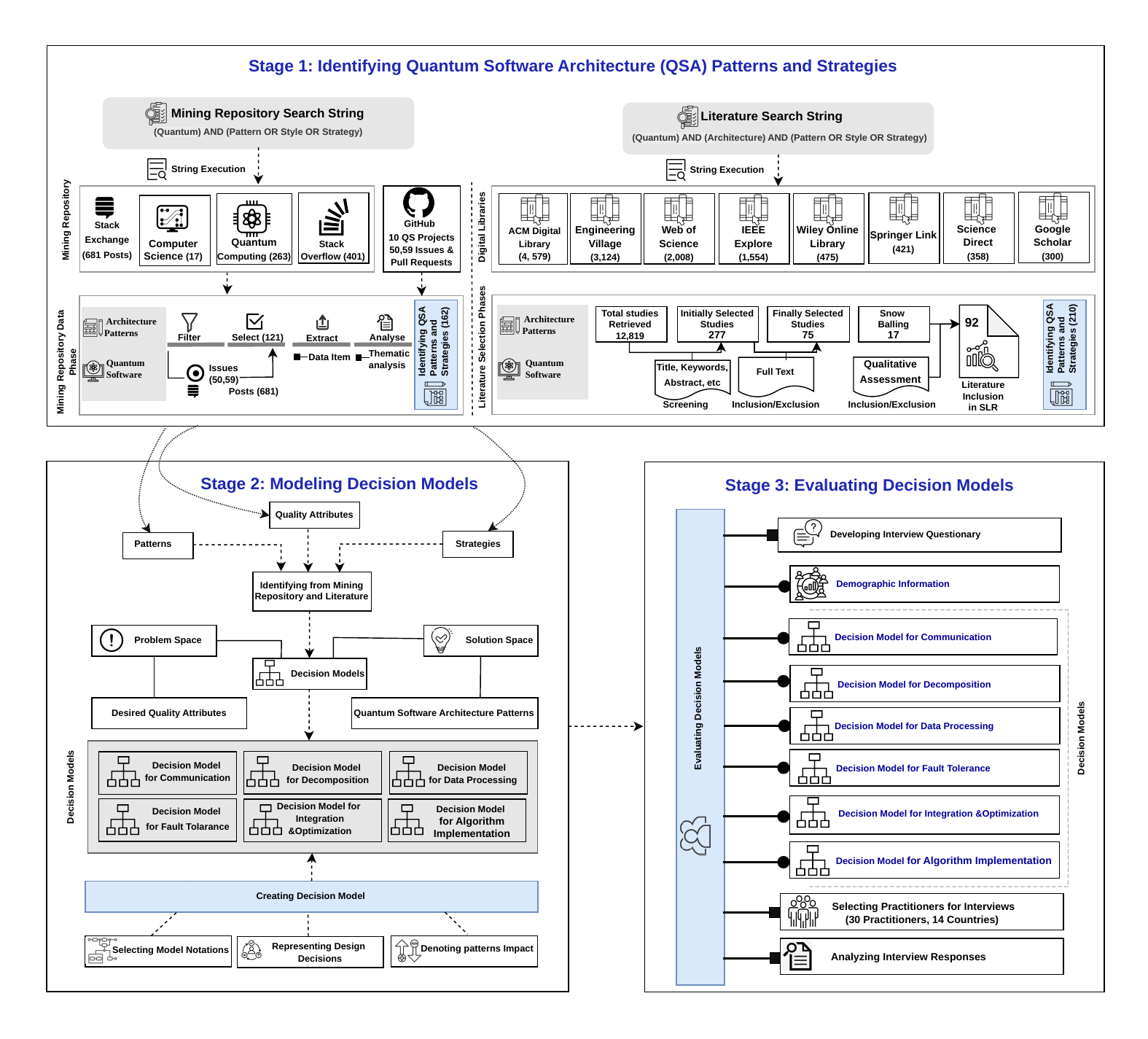}
\caption{Overview of the research process}
\label{ResearchProcess}
\end{figure}

 The research process, illustrated in Figure~\ref{ResearchProcess}, comprises three stages. \textit{Stage 1}: Identifying architecture patterns and strategies through mining GitHub and Stack Exchange data, along with an SLR. \textit{Stage 2}: Modeling six decision models based on the extracted patterns, strategies, and related QAs. \textit{Stage 3}: Evaluating the decision models through interviews with 30 practitioners on quantum software to assess their usefulness, clarity, and completeness.


\subsection{Identifying Architecture Patterns and Strategies}
To identify architecture patterns, strategies, and QAs, and to assess their impact on QAs, we collected data from two sources. First, we conducted a mining study using open-source software projects on GitHub\footnote{\url{https://github.com}} (e.g., issues and pull requests(PRs)) and posts from Stack Exchange sites\footnote{\url{https://stackexchange.com/sites}} (e.g., Stack Overflow and Quantum Computing, focusing on questions and answers related to architecture patterns). Second, we conducted an SLR to identify relevant studies that discuss architectural design decisions in quantum software systems. The selection of GitHub and Stack Exchange sites was inspired by Easterbrook's guidelines for selecting empirical methods in software engineering research \cite{easterbrook2008selecting}. The SLR was conducted following the guidelines by Kitchenham and Charters for performing SLRs in Software Engineering research~\cite{keele2007guidelines}. We explain each of the two studies below and outline the steps to extract relevant data for both the mining study (GitHub and Stack Exchange) and the SLR: 

\subsubsection{Identifying Architecture Patterns and Strategies Using a Mining Study} \label{IdentifyingPatternsMiningStudy}
\textbf{\newline Phase 1 - Data Collection}
\textbf{\newline Step 1 - GitHub Projects:} In this study, we collected data from open-source quantum software projects. This includes both (i) core quantum frameworks (e.g., Qiskit, Cirq, PennyLane) and (ii) supportive tools (e.g., DeepChem, Covalent) that facilitate quantum software development. These projects were initially selected based on a rigorous keyword-driven search and filtering process. We conducted a pilot search using a variety of keywords to determine the most effective terms for identifying relevant quantum-related projects on GitHub. After evaluating different options, we selected ``\textit{quantum}'' and ``\textit{quantum computing}'' as our primary keywords. The search was performed across multiple project components, including names, topics, descriptions, and README files. The keyword ``\textit{quantum}'' returned a total of 74,432 projects, whereas ``\textit{quantum computing}'' yielded significantly fewer results 12,401 projects, many of which were unrelated, such as other learning resources\footnote{\url{https://github.com/desireevl/awesome-quantum-computing}} and books\footnote{\url{https://github.com/JackHidary/quantumcomputingbook}}. Since ``\textit{quantum computing}'' is a subset of the broader term ``\textit{quantum}'', and projects retrieved using ``\textit{quantum}'' also included those related to ``\textit{quantum computing}'', we decided to proceed with ``\textit{quantum}'' as our search keyword. To collect the project data, we utilized GitHub REST API\footnote{\url{https://docs.github.com/en/rest?apiVersion=2022-11-28}}. We applied the following inclusion criteria: (1) the presence of the word ``\textit{quantum}'' in the project metadata (name, description, topics, or README file) following a similar metadata-based project identification approach used in prior work \cite{openja2022technical}, (2) a minimum of 50 stars, and (3) at least 15 forks, to reduce the likelihood of including small-scale or student projects, as suggested by prior research \cite{waseem2021nature}. 


\textcolor{black}{The selection of star and fork thresholds follows established practices in empirical software engineering, where these metrics serve as proxies for repository popularity, community engagement, and maturity \cite{borges2018s, ray2014large, waseem2021nature}. Prior studies on quantum software repositories use similar filtering strategies to ensure dataset quality. For example, Upadhyay et al. \cite{upadhyay2025analyzing} showed that most quantum repositories have very low engagement (3.2 stars and 1.1 forks on average). To exclude inactive or student-level projects, Openja et al. \cite{openja2022technical} required more than one fork, while Li et al. \cite{li2021understanding} used at least two forks. Similarly, Di Rocco et al. \cite{di2020topfilter} demonstrated that stricter thresholds improve dataset quality. Based on this, we set thresholds of 50 stars and 15 forks to exclude low-engagement repositories and ensure a robust dataset.} Based on these criteria, we initially identified 1,226 relevant projects. Subsequently, we manually reviewed each repository’s metadata, including its description, topics, and README file, to further refine our dataset. We excluded repositories that were purely educational (e.g., tutorials or documentation), not written in English, or unrelated to quantum software systems despite containing the keyword ``\textit{quantum}'' (e.g., unrelated tools like FirefoxColor\footnote{\url{https://github.com/mozilla/FirefoxColor}}). This resulted in a curated dataset of 347 quantum software repositories.

\textcolor{black}{After identifying 347 repositories, we piloted a screening to assess their suitability for analyzing architecture patterns via issue and PR discussions. Specifically, we randomly selected 35 repositories for manual inspection. The screening revealed that most repositories contained discussions primarily related to implementation details, bug fixes, or features, which is consistent with QSE findings of low activity and sparse architecture-level discourse \cite{upadhyay2025analyzing}. To increase the likelihood of identifying repositories containing architecture-related discussions, we adopted a ranking-based selection strategy that prioritized repositories with high issue and PR activity, following prior quantum studies \cite{openja2022technical,li2021understanding}. We selected the 10 most active repositories to ensure a rich and relevant dataset while minimizing noise.} From these top 10 projects, we conducted a keyword-based search over the titles and bodies of issues and PRs using terms specifically targeting architecture patterns and strategies:  ``\textit{pattern}'',  ``\textit{patterns}'', ``\textit{style}'', ``\textit{styles}'', ``\textit{strategy}'', and ``\textit{strategies}''. \textcolor{black}{To ensure alignment with our research objective, we deliberately focused on pattern-oriented keywords rather than explicitly labeled using generic terms such as ``architecture'' or ``design''. Including broader keywords could have introduced substantial noise, retrieving discussions unrelated to architecture patterns (e.g., general design advice, implementation details, or high-level conceptual discussions). To mitigate potential omissions, this keyword-based retrieval was complemented by a rigorous content-based filtration process (see Section \ref{dataFiltration}), including predefined inclusion and exclusion criteria and inter-rater validation.} This search retrieved 5,059 issues and PRs containing the keywords. The ten projects, along with their corresponding retrieved and related issues and PRs, are detailed in Table \ref{gitHubProject2}.


\begin{table}[]
\caption{\textcolor{black}{GitHub projects with the number of retrieved issues \& PRs and related issues \& PRs}}
\label{gitHubProject2}
\small
\setlength{\tabcolsep}{4pt}
\renewcommand{\arraystretch}{1.1}
\begin{tabular}{
>{\centering\arraybackslash}p{.8cm}
>{\centering\arraybackslash}p{1.5cm}
>{\centering\arraybackslash}p{2.2cm}
>{\centering\arraybackslash}p{2.2cm}
>{\centering\arraybackslash}p{2.2cm}
>{\centering\arraybackslash}p{1cm}
>{\centering\arraybackslash}p{1cm}}
\hline
\textbf{\#} & \textbf{GitHub Project}  & \textbf{\# of Issues \& PRs}  & \textbf{\# of Retrieved Issues \& PRs}  & \textbf{\# of Related Issues \& PRs} & \textbf{\# of Stars} & \textbf{\# of Forks}\\\hline
\textbf{GP1}        &  Qiskit                       & 3345             & 1169         & 30        &4126       &2153 \\\hline	
\textbf{GP2}        &  Cirq                         & 1855             & 703          & 25        &4017       &945  \\\hline		
\textbf{GP3}        &  ARTIQ                        & 1410             & 375          &  8        &388        &183 \\\hline	
\textbf{GP4}        &  deepchem                     & 1212             & 806          &  1        &4801       &1551\\\hline	
\textbf{GP5}        &  qmcpack                      & 978              & 433          &  2        &272        &132\\\hline	
\textbf{GP6}        &  PennyLane                    & 857              & 441          & 31        &2008       &514\\\hline	
\textbf{GP7}        &  QuTiP                        & 743              & 171          &  7        &1510       &640\\\hline	
\textbf{GP8}        &  Covalent                     & 743              & 171          &  1        &1510       &640\\\hline	
\textbf{GP9}        &  Mitiq                        & 682              & 602          & 14        &303        &640\\\hline	
\textbf{GP10}       &  pyquil                       & 551              & 286          &  2        &1400       &339\\\hline	
\textbf{Total}      &                               & \textbf{11957}            & \textbf{5059}         & 121      & &\\\hline	 
\end{tabular}
\end{table}

\textbf{Step 2 - Stack Exchange Sites}: We searched through questions and answers related to quantum architecture patterns on Stack Exchange sites (i.e., Stack Overflow, Quantum Computing, Computer Science, and Software Engineering). We selected the general terms ``pattern'' (i.e., ``\textit{pattern}'', ``\textit{patterns}'', ``\textit{style}'', ``\textit{styles}'', ``\textit{strategy}'', and ``\textit{strategies}''). Stack Exchange supports the use of wildcard (*) searches, and we defined ``\textit{pattern*}'', ``\textit{style*}'', ``\textit{strateg*}'' as the initial search terms related to architecture patterns. However, no relevant posts were found on the Software Engineering site, and our final dataset was derived from the first three Stack Exchange sites. Table \ref{stackExchageName2} lists the search terms used and the number of posts retrieved from the three Stack Exchange sites. In total, we identified 681 posts. More specifically, 401 from Stack Overflow, 263 from Quantum Computing, and 17 from Computer Science.
\begin{table}[]
\centering
\caption{\textcolor{black}{Stack Exchange sites with the number of retrieved posts and related posts}}
\label{stackExchageName2}
\small
\setlength{\tabcolsep}{4pt}
\renewcommand{\arraystretch}{1.1}
\begin{tabular}{>{\centering\arraybackslash}p{1cm} >{\centering\arraybackslash}p{3.5cm} >{\centering\arraybackslash}p{2.5cm} >{\centering\arraybackslash}p{2cm} >{\centering\arraybackslash}p{2cm}}
\hline
\textbf{\#}& \textbf{Stack Exchange Sites}      & \textbf{Search Terms }            & \textbf{\# of Retrieved Posts}        & \textbf{\# of Related Posts}\\\hline
\textbf{SEN1}   & Stack Overflow                & ``\textit{quantum pattern*}''      & 114                                       & 2 \\\hline
\textbf{SEN2}   & Stack Overflow                & ``\textit{quantum style*}''        & 241                                       & 0 \\\hline
\textbf{SEN3}   & Stack Overflow                & ``\textit{quantum strateg*}''      & 46                                        & 0 \\\hline
\textbf{SEN4}   & Quantum Computing             & ``\textit{quantum pattern*}''      & 91                                        & 5 \\\hline
\textbf{SEN5}   & Quantum Computing             & ``\textit{quantum style*}''        & 56                                        & 3 \\\hline
\textbf{SEN6}   & Quantum Computing             & ``\textit{quantum strateg*}''      & 116                                       & 2 \\\hline
\textbf{SEN7}   & Computer Science              & ``\textit{quantum pattern*}''      & 8                                         & 0 \\\hline
\textbf{SEN8}   & Computer Science              & ``\textit{quantum style*}''        & 4                                         & 1 \\\hline
\textbf{SEN9}   & Computer Science              & ``\textit{quantum strateg*}''      & 5                                         & 0 \\\hline 
\textbf{Total}  &                               &                                    & 681                                       & 13 \\\hline
\end{tabular}
\end{table}

\label{dataFiltration}\textbf{Phase 2 - Data Filtration}: We found issues and posts containing the terms ``\textit{pattern}'', ``\textit{style}'', and ``\textit{strategy}'' but not in the context of software architecture and design, such as hardware (e.g., error pattern\footnote{\url{https://tinyurl.com/mr367anu}}), mathematical concepts (e.g., mathematical pattern\footnote{\url{https://tinyurl.com/b3myzxds}}), and coding style documentation (Docstrings style\footnote{\url{https://tinyurl.com/4wtawp7b}}), when describing their concerns in the GitHub issues and Stack Exchange posts. We also found that both Stack Overflow and Computer Science Stack Exchange featured posts with the term ``\textit{quantum}'' but not in the context of a quantum software system, such as a browser (e.g., Firefox Quantum\footnote{\url{https://tinyurl.com/5b562yb6}}) and Geographic Information System (GIS) software (e.g., Quantum GIS\footnote{\url{https://tinyurl.com/ywfdt4dx}}). To analyze the collected data effectively, we filtered 5,059 issues and PRs from 10 quantum software systems, along with 681 Stack Exchange posts (i.e., 401 from Stack Overflow, 263 from Quantum Computing, and 17 from Computer Science), and excluded those posts that were not related to software architecture patterns and strategies in quantum software systems. This process involved conducting a content analysis and utilizing predefined inclusion and exclusion criteria (see Table \ref{dataExtraction2}) to systematically eliminate unrelated issues and posts. 

\textcolor{black}{To ensure the reliability of the filtration process and address potential researcher bias, we implemented a two-stage screening approach with inter-rater validation. Before the formal screening process involving manual review, we performed an initial selection based on predefined inclusion and exclusion criteria (see Table \ref{Tab: InclusionExclusionCriteriaForMining}). This preliminary phase aimed to identify issues from GitHub projects and posts on Stack Exchange related to architecture patterns and strategies in quantum software systems. During this step, the first author carefully read 250 GitHub issues and PRs, and 50 Stack Exchange posts. Each data point was labeled as ``related'', ``not related'' or ``doubtful''. Among the GitHub issues, 11 were labeled as ``related'', 222 as ``not related'', and 17 as ``doubtful''. For the Stack Exchange posts, 3 were marked as ``related'', 46 as ``not related'', and 1 as ``doubtful''. To ensure reliability and reduce bias, all labeling decisions, including related, not related, and doubtful cases, were reviewed collaboratively by the first four authors using the same set of 300 items, with discrepancies resolved through a negotiated agreement approach \cite{campbell2013coding}.}

\textcolor{black}{Following the initial calibration, the first author conducted the formal data screening of all 5,740 items (5,059 GitHub items and 681 Stack Exchange posts) using the predefined inclusion and exclusion criteria in Table \ref{Tab: InclusionExclusionCriteriaForMining}. To further reduce potential selection bias, we conducted a post hoc reliability check: the second author independently re-screened a statistically representative random sample of 380 items, comprising 316 GitHub issues/PRs and 64 Stack Exchange posts. Inter-rater agreement between the first and second authors was assessed using Cohen's kappa coefficient, which yielded K = 0.89, indicating ``almost perfect'' agreement according to the interpretation guidelines proposed by Landis and Koch et al. \cite{landis1977measurement}. Disagreements occurred in 12 cases and were resolved collaboratively through discussion among the first four authors, resulting in the reinclusion of four items that met the inclusion criteria upon re-evaluation. The formal screening process ultimately identified 121 GitHub issues and PRs (Table \ref{gitHubProject2}) and 13 Stack Exchange posts (Table \ref{stackExchageName2}) as ``related'' to architecture patterns and strategies in quantum software systems, representing 2.3\% of the initially collected data.}


\textbf{Phase 3 - Data Extraction}: We defined a set of data items (see Table \ref{dataExtraction2}) to be extracted from the discussion of the 121 GitHub issues \& PRs and 13 Stack Exchange posts. To check the reliability of the extracted data items, the first author conducted a pilot data extraction on 10 issues, 5 PRs, and 5 posts, and the second, third, and fourth authors evaluated the extracted data. After evaluating the extracted data items, the first author used a revised set of data items to formally extract data from the selected issues, PRs, and posts. The first four authors then discussed the extracted data to reduce potential bias and ambiguity. Data items (D1-D3) are used to extract the general information of the selected issues, PRs, and posts. The remaining data items (D4-D9) are used to identify patterns, strategies, QAs, and impact. Finally, we used spreadsheets to record the extracted data.

\begin{table}[h!]
\caption{\textcolor{black}{Inclusion and exclusion criteria to manually identify GitHub issues, PRs, and Stack Exchange posts}}
\label{Tab: InclusionExclusionCriteriaForMining}
\small
\setlength{\tabcolsep}{4pt}
\renewcommand{\arraystretch}{1.1}
\begin{tabular}{p{2cm}p{11.2cm}}
\hline
\textbf{Inclusion}      & \textbf{I1.} We include an issue, PR, or post if it addresses both architecture patterns and quantum software systems. \\
    \hline
    \textbf{Exclusion}  & \textbf{E1.} We exclude an issue, PR, or post if it is only related to quantum software systems and not to architecture patterns. \newline
                          \textbf{E2.} We exclude an issue, PR, or post if it is only related to the architecture pattern and not the quantum software system. \\
    \hline
\end{tabular}
\end{table}

\textbf{Phase 4 - Data Analysis}: \textcolor{black}{We used thematic analysis \cite{braun2006using} to systematically analyze the qualitative data extracted from GitHub and Stack Exchange (as detailed in Phase 3). This method is well-suited for identifying, analyzing, and reporting patterns (themes) within qualitative data. Our process was predominantly inductive, allowing themes to emerge from the data itself, rather than fitting the data into a pre-existing framework. This approach aligns with established practices in software engineering research, as demonstrated in recent studies analyzing developer discussions (e.g.,~\cite{kashif2025developers}). The analysis was conducted collaboratively by the authors to enhance reliability. Thematic analysis is not a linear but an iterative process. Our thematic analysis proceeded through five main steps:}

\textcolor{black}{
\begin{itemize}
    \item  \textbf{Familiarizing with the Data}: We repeatedly read the selected GitHub issues, PRs, and Stack Exchange posts, noting all mentions relevant to architecture patterns, architecture strategies, related QAs, and references to QA impacts. This step ensured deep engagement with the context and language of developer discussions.
    \item \textbf{Generating Initial Codes}: Based on this familiarization, we generated an initial set of codes that capture recurring architectural features identified in the raw data, including architectural decisions, design rationale, abstraction mechanisms, reusable components, modular decomposition, interface design, and QA implications. For example, in \href{https://github.com/PennyLaneAI/pennylane/pull/710}{PennyLane Pull Request \#710}, the discussion on introducing the ApproxTimeEvolution template and alternative approaches for Hamiltonian representation and wire management was coded as ``reusable template'', ``modular decomposition'', ``component reuse'', ``interface consistency'', and ``Hamiltonian abstraction''.
    \item \textbf{Searching for Themes}: We examined relationships among the initial codes and grouped related codes into broader candidate themes representing architecture patterns or strategies. For example, the codes ``reusable template'', ``modular decomposition'', ``component reuse'', ``interface consistency'', and ``Hamiltonian abstraction'' collectively reflected organizing quantum functionality into reusable layers with clearly separated responsibilities. These related codes were therefore grouped into the Layered Architecture Pattern, representing a higher-level architectural theme that emerged from the combined evidence rather than from any single code.
    \item \textbf{Reviewing Themes}: The candidate themes were iteratively reviewed and refined by all authors. The first author led the coding process by continuously evaluating whether the coded extracts coherently represented each candidate theme and whether the themes accurately reflected the overall dataset. The remaining authors independently reviewed the coding results, discussed and resolved any disagreements, and refined, merged, or removed themes as appropriate to improve consistency and reliability.
    \item \textbf{Defining and Naming Themes}: Following iterative refinement, each theme was assigned a descriptive name that accurately reflected its underlying architectural concept and its associated QA impacts. For instance, the evidence extracted from \href{https://github.com/PennyLaneAI/pennylane/pull/710}{PennyLane Pull Request \#710} was ultimately categorized as the Layered Architecture Pattern because it represented a recurring architectural decision to encapsulate time-evolution functionality within a reusable layer, thereby improving modularity, reusability, and maintainability through clear separation of architectural responsibilities.
\end{itemize}}

\textcolor{black}{By following this process, we consolidated a total of 162 patterns and strategies from 121 GitHub issues and 13 Stack Exchange posts. Table \ref{dataExtraction2} summarizes these extracted data items and their coding categories.}

\begin{table}[]
\caption{\textcolor{black}{Data items extracted from the selected GitHub issues, PRs, and Stack Exchange posts}}
\label{dataExtraction2}
\small
\setlength{\tabcolsep}{4pt}
\renewcommand{\arraystretch}{1.1}
\begin{tabular}{p{1cm}p{4.2cm}p{7cm}}
\hline
 \textbf{\#}&\textbf{Data item}&\textbf{Description}\\\hline
    D1  &   Issues, PRs, \& posts ID                                         & A ID of the issues, PRs, \& posts ID that were discussed on GitHub and Stack Exchange in the context of the quantum software architecture pattern and strategy \\\hline
    D2  &   Issues, PRs, \& posts title                                      & A title of the issues, PRs, \& posts ID from a developer that describes what the quantum software architecture pattern and strategy is all about \\\hline 
    D3	&   Issues, PRs, \& posts link                                       & The URL address of the issues, PRs, \& posts ID\\\hline
    D4  &   Type of quantum software architecture patterns and strategies              & Identification of quantum software architecture patterns and strategies type based on key points\\\hline
    D5  &   Pattern and  strategy related to quantum software architecture design area & Identification of quantum software architecture patterns and strategies design area \\\hline
    D6  &   Pattern and  strategy key point                                            & Identification of quantum software architecture patterns and strategies key points \\\hline
    D7	&   Related QAs                                                 & Related QAs of patterns and strategies that occur in quantum software systems\\\hline
    D8	&   Positive impact QAs                                         & Positively impacted QAs of patterns and strategies that occur in quantum software systems\\\hline
    D9	&   Negative impact QAs                                         &  Negatively impacted QAs of patterns and strategies that occur in quantum software systems\\\hline
\end{tabular}
\end{table}

\subsubsection{Identifying Architecture Patterns and Strategies Using an SLR}\label{IdentifyingArchitecturePatternsSLR}

\textbf{\newline 1) Data Collection}

\textbf{Phase 1 - Defining Data Collection Strategy}: This phase focuses on preparing for the SLR by identifying relevant data sources, formulating a search strategy, and defining inclusion and exclusion criteria for study selection.

\textbf{Step 1- Identify Data Sources}: Electronic Data Sources (EDS) are pivotal in systematic review studies, where they enable automated searches through predefined or customized queries. These automated searches facilitate the identification of relevant literature in a particular domain \cite{van2021automation,chen2010towards}. In our study, we adhered to the guidelines for performing systematic literature reviews in software engineering as proposed by Kitchenham et al. \cite{KITCHENHAM20097}. \textcolor{black}{We selected eight widely recognized EDS: IEEE Xplore, ACM Digital Library, Wiley Online Library, ScienceDirect, SpringerLink, Engineering Village, Web of Science, and Google Scholar, based on their proven effectiveness and relevance to computing, software engineering, and software architecture research \cite{zhang2011empirical, chen2010towards}.}

\textbf{Step 2- Formulate Search Strategy}: To define the search strategy for the SLR, the authors collaboratively identified key terms aligned with the study objective of identifying software architecture patterns and strategies in quantum software systems. The search was structured around three core concepts: quantum computing, software architecture, and architectural solutions. Accordingly, the final search string was defined as: \textit{(Quantum) AND (Architecture) AND (Pattern OR Style OR Strategy)}.

The Boolean operator ``AND'' ensures that retrieved studies lie at the intersection of quantum computing and software architecture, while ``OR'' captures synonymous architectural terms. A pilot search using IEEE Xplore and ACM Digital Library was conducted to refine the search string. During this process, broader terms such as design and model were excluded, as they introduced many irrelevant results related to implementation-level aspects (e.g., circuit design, algorithm design, or software process models) rather than architectural-level concerns. Instead, the search focuses on patterns, styles, and strategies, which better represent reusable architectural solutions and high-level design decisions. The final search string was iteratively refined and agreed upon by all authors to balance comprehensiveness and precision.

\textbf{Step 3 - Defining Inclusion and Exclusion Criteria}: We defined a set of inclusion and exclusion criteria for selecting relevant studies. These criteria were applied to filter out irrelevant, redundant, non-English, and duplicate studies, especially those appearing in multiple venues or overlapping due to shared libraries or cross-referencing. Table \ref{Tab:InclusionExclusionCriteriaStudy} lists our inclusion and exclusion criteria. The inclusion and exclusion process was followed by a quality assessment step to evaluate each included study’s relevance and quality. Studies that did not meet the minimum quality score of 1.5 were excluded from the review (as detailed in Section~\ref{step3} Phase 2 - Step 3).

\textbf{Step 4 - Perform Quality Assessment}: This quality assessment ensured that only high-quality, relevant studies were included in the final analysis.


\begin{table}[ht]
\centering
\caption{Inclusion and exclusion criteria for study selection}
\label{Tab:InclusionExclusionCriteriaStudy}
\small
\setlength{\tabcolsep}{4pt}
\renewcommand{\arraystretch}{1}
\begin{tabular}{p{1.1cm}p{5cm}p{1.1cm}p{5cm}}
\hline
\textbf{Code} & \textbf{Inclusion Criteria} & \textbf{Code} & \textbf{Exclusion Criteria} \\ \hline
I1            & Studies that focus on quantum software architecture patterns, strategies, styles, or QAs.                          & E1         & Studies unrelated to quantum software architecture (e.g., studies focused on quantum computing applications in chemistry, biology, or physics). \\ \hline
I2            & Studies that provide empirical data or conceptual frameworks related to the design and development of quantum software systems.   & E2         & Studies that do not include explicit information on architecture patterns, strategies, or design principles in quantum software systems. \\ \hline
I3            & Studies published in peer-reviewed journals, conferences, or reputable repositories.                                              & E3         & Studies not published in peer-reviewed sources or not available in English. \\ \hline
I4            & Studies available in the English language.                                                                                        & E4         & Duplicate or secondary studies based on the search results. \\ \hline
\end{tabular}
\end{table}

\textbf{Phase 2 - Conducting the Review}: This phase involved executing the search, screening the results through multiple levels of review, and applying inclusion, exclusion, and quality criteria to identify the final set of primary studies.

\textbf{Step 1 - Selecting Primary Studies}: We used the following steps to select the primary studies.

\begin{itemize}
    \item \textbf{String Execution}: The identification of primary studies began by querying the selected EDS with the predefined search string (outlined in Section \ref{IdentifyingArchitecturePatternsSLR}). The initial search process retrieved a total of 12,819 studies aggregated across all EDS, which were further manually filtered by the first author based on the studies' titles, keywords, and abstracts against the inclusion and exclusion criteria (see Table \ref{Tab:InclusionExclusionCriteriaStudy}) to identify studies that were potentially relevant for further evaluation. For example, we used the advanced search functionality in IEEE Xplore to apply the search string and retrieve relevant studies from the ``Full Text \& Metadata''. This search returned 1,584 results, most of which focused on quantum systems in general, with particular emphasis on quantum hardware. To narrow down the overwhelming number of results and assess the scope of relevant entries, we modified the search parameter from ``Full Text \& Metadata'' to ``Abstract''. This refinement yielded 259 studies, but the abstract-only search missed several relevant articles that the broader full-text search captured, including known flagship studies in the field. To ensure the validity of our search strategy, we manually reviewed the results and verified that key benchmark studies were included, reinforcing the need for a comprehensive scanning process. Therefore, we decided to retain the more comprehensive search setting and proceeded with a manual examination of the 1,584 results. After applying IEEE Xplore-specific filtering to exclude non-peer-reviewed content such as books and magazines, we refined the dataset to 1,554 candidate studies. We manually examined the remaining retained studies that were directly related to architecture patterns, styles, or strategies within the domain of quantum software systems. Based on this repository-specific strategy, we extracted these studies for further screening, applying inclusion and exclusion criteria, and conducted a qualitative assessment, as illustrated in Figure~\ref{ResearchProcess}.
    \item \textbf{Title, Abstract, and Keywords Based Screening}:  After the initial search, we screened all retrieved studies based on their titles, abstracts, and keywords. The first author independently reviewed the studies, labeling each as ``relevant'', ``irrelevant'', or ``doubtful''. Doubtful studies were discussed among the first four authors to reach a consensus on their relevance. This process ensured that only the most relevant studies progressed to the next phase. Ultimately, 277 studies were selected based on title, abstract, and keyword filtering (see Figure~\ref{ResearchProcess}), and these studies were considered for the next stage of full-text screening.
    \item \textbf{Full Text Screening:} The 277 studies that passed the abstract and topic-based selection were subjected to a more thorough full-text evaluation. The first author applied the inclusion and exclusion criteria, as outlined in Table \ref{Tab:InclusionExclusionCriteriaStudy} (I1 to I4 for inclusion and E1 to E4 for exclusion), to ensure that the selected studies met the necessary quality and relevance standards. After applying the inclusion and exclusion criteria, we selected 75 primary studies for the final analysis. To ensure the consistency and accuracy of the selection process, the second, third, and fourth authors independently reviewed the list of selected studies and verified the results.
\end{itemize}

\textbf{Step 2 - Snowballing}: To mitigate the risk of overlooking important studies during the automated search, we applied snowballing, following the method outlined by Wohlin et al. \cite{wohlin2014guidelines}. This process involved both forward snowballing, where we gathered studies citing the selected ones, and backward snowballing, where we reviewed the references of the 75 selected studies. Initially, this method returned a total of 451 studies based on the titles. After removing duplicates and conducting a relevance check based on the abstracts, keywords, and conclusions of 189 studies, we proceeded with a detailed full-text screening of these papers. We applied a multi-level screening process, incorporating both general inclusion and exclusion criteria and specific quality assessment guidelines (see Table~\ref{tab:qualityAssessment}). Following this rigorous evaluation, a final set of 17 relevant studies was identified and incorporated into our primary dataset through snowballing.

\begin{table}[h]
\caption{Quality assessment criteria for primary studies}
\label{tab:qualityAssessment}
\small
\setlength{\tabcolsep}{4pt}
\renewcommand{\arraystretch}{1.1}
\centering
\begin{tabular}{p{1.2cm}p{12.5cm}}
\hline
\textbf{Code} & \textbf{Generic Quality Criteria} \\ \hline
G1 & Does the study explicitly define its research objectives and articulate the motivation behind the work? \\ \hline
G2 & \textcolor{black}{Is the context of the study, such as the domain, environment, or technology stack described?} \\ \hline
G3 & \textcolor{black}{Is the research methodology detailed and logically justified, including data sources, techniques, and analytical procedures?} \\ \hline
G4 & \textcolor{black}{Are the key findings and interpretations presented, well-supported by data, and thoroughly discussed?} \\ \hline
G5 & Are the study’s limitations, assumptions, or threats to validity transparently acknowledged and explained? \\ \hline
\textbf{Code} & \textbf{Domain-Specific Criteria for Quantum Software Architecture} \\ \hline
S1 & Does the study contribute to quantum software architecture by discussing or applying patterns, styles, strategies, or decision models tailored for quantum software systems? \\ \hline
S2 & \textcolor{black}{Does the study address at least one of the six key design areas (Communication, Decomposition, Data Processing, Fault Tolerance, Integration \& Optimization, Algorithm Implementation)? These six areas were not established a priori as a predefined taxonomy; rather, they were derived inductively from the patterns identified during the data collection phase. No additional design areas were observed beyond these six categories during the data screening and extraction process.} \\ \hline
S3 & Does the study provide practical insights, empirical evaluations, or real-world use cases that inform architecture design decisions in quantum software systems? \\ \hline
\end{tabular}
\end{table}

\textbf{Step 3 - Quality Assessment of Primary Studies}\label{step3}: To ensure the rigor and relevance of the studies included in this SLR, we conducted a structured quality assessment based on established guidelines. Our assessment combined a general methodological evaluation with domain-specific criteria tailored to the context of quantum software architecture. The criteria were adapted from the methodological frameworks proposed by Kitchenham and Charters et al. \cite{keele2007guidelines}, Brereton et al. \cite{brereton2007lessons}, Waseem et al. \cite{waseem2020systematic} and Ahmad et al. \cite{ahmad2016software}. The final framework comprised eight evaluation items, five generic (G1–G5) and three domain-specific (S1–S3), as listed in Table \ref{tab:qualityAssessment}. Each study was evaluated based on whether it fully addressed, partially addressed, or did not address each criterion. A score of 1 was assigned if the criterion was fully satisfied, 0.5 if partially satisfied, and 0 if not satisfied. \textcolor{black}{The sum of scores from the specific criteria was multiplied by a factor of three to emphasize the relevance of domain-specific insights in the context of quantum software architecture. This weighting approach was inspired by the work of Waseem et al. \cite{waseem2020systematic}, who similarly placed greater weight on domain-specific factors in their quality assessment framework for microservices architecture in DevOps. They found that domain-specific relevance significantly influences the quality and applicability of studies, especially in specialized fields like quantum software systems. The final quality score for each study was computed using the following formula:}

\[
\text{Quality Score} = \left( \sum_{i=1}^{5} G_i + 3 \times \sum_{j=1}^{3} S_j \right)
\]

Studies that received a total quality score of 1.5 or higher were considered suitable for inclusion in the final dataset \cite{waseem2020systematic}. The initial assessments were conducted by the first author and were independently verified by the second, third, and fourth authors to ensure consistency and reduce evaluation bias. All 92 studies (75 from string execution and 17 from snowballing)  satisfied the inclusion criterion of a minimum quality score of 1.5 and were therefore included in the final dataset.

\textbf{Phase 3 - Data Extraction and Analysis for Primary Studies}

For this study, we defined a set of data extraction items to systematically collect and synthesize relevant information from the selected primary studies. These data items (see Table~\ref{tab:data_items}) were designed to capture both general bibliographic information and specific content relevant to quantum software architecture patterns and strategies.

\textbf{Step 1 - Data Extraction Process for Primary Studies}:
The data extraction process was conducted in two phases. Initially, a pilot extraction was performed by the first author on a subset of ten studies to assess the clarity and applicability of the predefined inclusion and exclusion criteria (see Table \ref{Tab:InclusionExclusionCriteriaStudy}). Based on the feedback and discussions among the first four authors, minor refinements were made to the extraction framework to improve consistency and remove ambiguities. Following this, the formal extraction process was conducted, with the first four authors equally distributing the selected studies among themselves based on their research expertise and interests. \textcolor{black}{To ensure reliability and reduce bias, the extracted data were reviewed collaboratively by the first four authors, with discrepancies resolved through a negotiated agreement approach \cite{campbell2013coding}.} All extracted data were systematically recorded in Excel sheets to support traceability and further analysis.


\begin{table}[ht]
\centering
\caption{Data items extracted from the selected primary studies}
\label{tab:data_items}
\small
\setlength{\tabcolsep}{4pt}
\renewcommand{\arraystretch}{1.1}
\begin{tabular}{p{1cm} p{4cm} p{7cm}}
\hline
\textbf{Code} & \textbf{Data Item} & \textbf{Description} \\
\hline
D1  & Index                                      & ID of the study \\\hline
D2  & Study title                                & Title of the study \\\hline
D3  & Author(s) list                             & Full names of the authors \\\hline
D4  & Year                                       & Publication year \\\hline
D5  & Venue                                      & Publishing venue (e.g., journal, conference) \\\hline
D6  & Publication type                           & Type of publication (e.g., journal, conference) \\\hline
D7  & Authors affiliation                        & Academia, industry, or both \\\hline
D8  & Pattern and strategy related to quantum software architecture design area & Identification of quantum software architecture patterns and strategies design area \\\hline
D9  & Pattern and strategy key point             & Key points summarizing the identified patterns and strategies \\\hline
D10 & Related QAs                 & QAs associated with the patterns and strategies \\\hline
D11 & Positive impact QAs         & QAs positively influenced by the patterns and strategies \\\hline
D12 & Negative impact QAs         & QAs negatively influenced by the patterns and strategies \\\hline
\end{tabular}
\end{table}

\textcolor{black}{
\textbf{Step 2 - Data Analysis Process for Primary Studies}: After data extraction, we conducted both descriptive and thematic analysis of the collected data. The general information (D1--D7) was analyzed using descriptive statistics to provide an overview of publication trends, author affiliations, and dissemination venues. The content-specific items (D8–D12) were analyzed qualitatively through thematic analysis, following the well-established guidelines proposed by Braun and Clarke \cite{braun2006using}. This approach is widely adopted in software engineering and systematic literature reviews to systematically identify, analyze, and report recurring patterns within qualitative data \cite{braun2006using,kashif2025developers}. The analysis was conducted iteratively through the following steps:}
\textcolor{black}{
\begin{itemize}
    \item \textbf{Familiarizing with the Data}: We carefully reviewed all selected studies, focusing on extracting relevant information such as pattern names, architectural context, related QAs, and their impacts.
    \item \textbf{Generating Initial Codes}: Key data from each study were coded to capture recurring concepts. For example, from Study [S92] (Data Encoding Patterns for Quantum Computing, PLoP 2021), the extracted data include: Pattern Name: \textit{Basis Encoding}; Related to: \textit{Data Processing}; Key Point: \textit{Representing data in a quantum computer for computation}; Related QAs: \textit{Functional Suitability, Performance, Scalability}. These data were initially coded as ``data representation strategy'', ``quantum data encoding'', and ``QA trade-off''.
    \item \textbf{Searching for Themes}: Related codes were grouped into broader themes. In this case, the codes derived from Study [S92] were grouped under the theme Quantum Data Encoding Patterns, which represents a class of architectural solutions for handling data in quantum systems.
    \item \textbf{Reviewing Themes}: The identified themes were iteratively reviewed and refined by all authors to ensure internal consistency and accurate representation of the data. Redundant or overlapping themes were merged where necessary.
    \item \textbf{Defining and Naming Themes}: Each theme was clearly defined and named to reflect its role in architectural decision-making. For instance, the theme derived from Study [S92] was formalized as a Basis Encoding Pattern within the Data Processing decision model, highlighting its role in influencing functional suitability and performance, while potentially negatively impacting scalability. 
\end{itemize}
}

\textcolor{black}{Through this process, the SLR contributed a total of 210 patterns and strategies from 92 primary studies, which were organized into the six decision models. The example of Study [S92] demonstrates how raw study data is transformed into actionable architectural knowledge, directly supporting the development of decision models.}



\subsubsection{Integration and Synthesis of Findings}
\textbf{\newline} To comprehensively support the development of decision models for quantum software architecture, we systematically integrated findings from both the mining study (GitHub issues and Stack Exchange posts) and the SLR. This integration process followed a rigorous multi-stage methodology to ensure traceability from raw data to synthesized patterns.

\textcolor{black}{\begin{itemize}
    \item \textbf{Data Extraction and Coding}: From the mining study, we extracted 162 patterns/strategies (145 from 121 GitHub issues \& PRs, 17 from 13 Stack Exchange posts), while the SLR contributed 210 patterns from 92 studies. Each identified item was initially coded with: (1) source identifier, (2) raw terminology used, (3) architectural design area (Communication, Decomposition, Data Processing, Fault Tolerance, Integration and Optimization, or Algorithm Implementation), and (4) reported QA impacts.
    \item \textbf{Conceptual Unification}: During integration, we observed that different studies and repositories often use varying terminologies to describe similar architecture patterns and strategies, despite sharing comparable underlying intent and behavior. For example, ``\textit{Quantum Key Distribution Chain Patterns}'' was categorized under the broader design area of ``\textit{Quantum Key Distribution (QKD) Protocols}''. To ensure consistency across heterogeneous sources, we applied a qualitative coding approach inspired by constant comparison and thematic synthesis principles to achieve conceptual unification of semantically similar patterns. Specifically, the first and second authors independently performed open coding of each pattern by carefully reading its description in the original source (e.g., GitHub issues, pull request discussions, or SLR studies) and extracting three key elements: (1) architectural intent (what the pattern aims to achieve), (2) functional behavior (how it operates), and (3) reported QA effects. Based on these elements, each researcher assigned an initial standardized descriptor that best represents the underlying architectural concept rather than the original terminology used in the source. After independent coding, the first and second authors conducted a comparison phase where inconsistencies in assigned descriptors were identified and discussed. In cases of disagreement, the final descriptor was determined through re-examination of the original evidence and agreement on the dominant architectural intent and behavior. This iterative process resulted in a unified and semantically consistent set of standardized descriptors across all sources.
    \item \textbf{Thematic Synthesis and Deduplication}: We conducted iterative thematic analysis following Braun and Clarke's guidelines \cite{braun2006using}. This involved: (a) familiarization with normalized pattern descriptions, (b) generating initial themes based on architectural purposes, (c) reviewing and refining themes through researcher triangulation, (d) consolidating patterns that represented identical architectural concepts under unified themes, and (e) resolving discrepancies through consensus meetings with a third senior researcher. Through this process, we consolidated the initial 372 items into 62 unique architecture patterns and strategies.
    \item \textbf{Quality Attribute Impact Analysis}: For each unified pattern, we systematically documented its reported effects on the QAs of quantum software. Our analysis revealed that patterns could have both positive and negative impacts. For instance, a pattern might improve fault tolerance while potentially degrading performance. This nuanced understanding became foundational to the decision models, in which we explicitly denote both beneficial (+) and detrimental (-) QA effects.
    \item \textbf{Preliminary Model Structuring}: The 62 unified patterns were organized within their six architectural design areas. Specifically, we identified 17 Communication, 7 Decomposition, 12 Data Processing, 8 Fault Tolerance, 9 Integration and Optimization, and 9 Algorithm Implementation patterns and strategies. Within each area, we began identifying logical relationships and decision sequences based on how patterns were discussed in the literature, for example, whether certain patterns were typically implemented before others, served as alternatives, or complemented each other.
\end{itemize}}

\textcolor{black}{Overall, this multi-stage integration ensures that the resulting decision models are systematically derived, empirically grounded, and capable of supporting informed architectural decision-making in quantum software systems.}

\subsection{Model Notation and Representation}
\begin{figure}[h!]
\centering
\includegraphics[width=1 \linewidth]{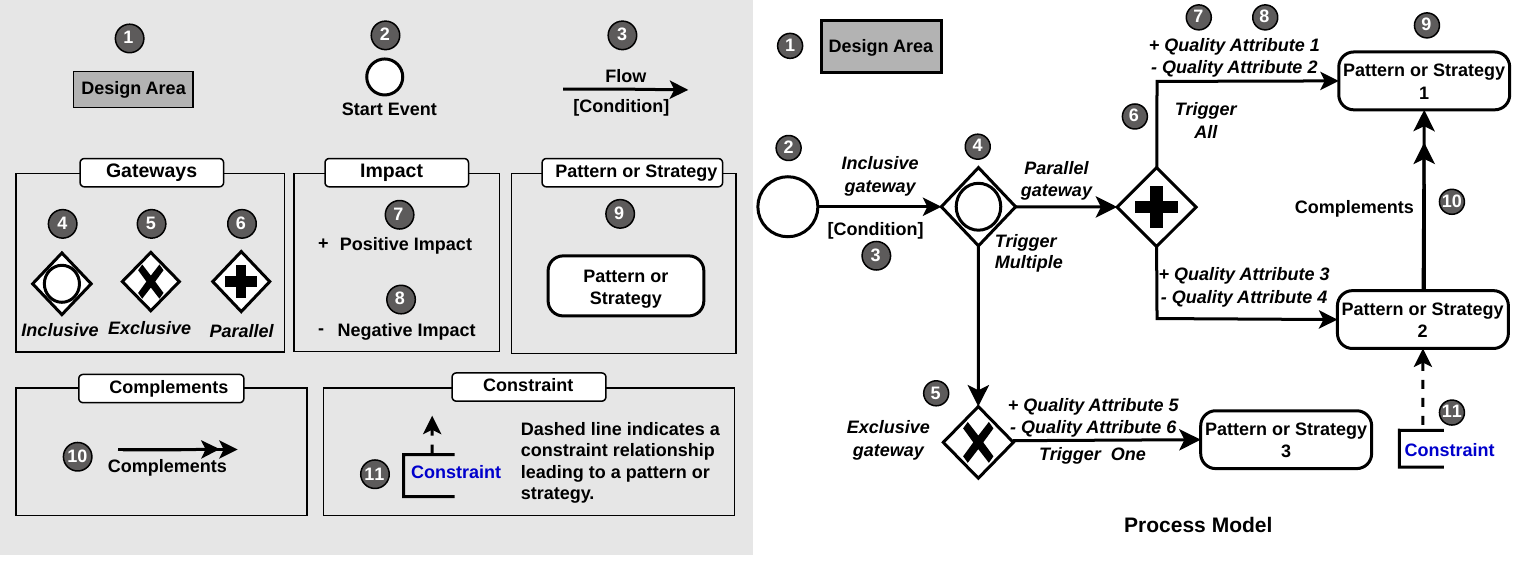}
\caption{Notations used in the decision models}
\label{Modeling_Decision_Model}
\end{figure}

\textcolor{black}{Figure \ref{Modeling_Decision_Model} illustrates the notations employed within the decision models proposed in this study. To represent decision flows, we used the Inclusive, Exclusive, and Parallel gateways from the Business Process Model and Notation. Specifically, \blackcirclednum{1} a gray box indicates a \textit{design area} within quantum software systems; \blackcirclednum{2} a circle signifies the \textit{starting point} of a decision process; \blackcirclednum{3} a single-headed arrow indicates a \textit{conditional flow} between patterns; \blackcirclednum{4} an inclusive gateway may activate \textit{multiple outgoing paths}; \blackcirclednum{5} an exclusive gateway allows only \textit{one outgoing path}; \blackcirclednum{6} a parallel gateway initiates \textit{all outgoing paths} simultaneously; \blackcirclednum{7} a \textit{QA that is improved} by a pattern is indicated with a plus sign (+); \blackcirclednum{8} \textit{a QA that is degraded} by a pattern is indicated with a minus sign (-); \blackcirclednum{9} \textit{architecture patterns} and \textit{strategies} are depicted using rounded rectangles; \blackcirclednum{10} a line with a double-headed arrow between two patterns represents a \textit{complements} relationship; and \blackcirclednum{11} \textit{constraints} associated with a pattern are shown using an open rectangle connected to the pattern by a dashed arrow.}

\textcolor{black}{
\subsection{Constructing Decision Models from Synthesized Findings}
The 62 consolidated architecture patterns/strategies, their associated QA impacts, and identified inter-pattern relationships serve as the primary inputs for constructing the decision models. The construction process consists of three steps:
\begin{itemize}
    \item \textbf{Problem-Solution Space Mapping}: For each design area, we mapped the consolidated patterns (solution space elements) to the QAs they influence (problem space elements). This mapping, derived from the impact analysis in Phase 3, established the foundational relationships between architectural decisions and quality outcomes.
    \item  \textbf{Decision Flow Derivation}: We analyzed the original sources to identify logical sequences and dependencies among patterns. \textcolor{black}{For example, in the Communication design area, the sources indicate that selecting a quantum communication protocol depends on the system's communication requirements, such as whether communication occurs between quantum and classical components or between distributed quantum nodes. Based on these requirements, different architecture patterns or strategies become applicable.} These observed dependencies were translated into decision flows using BPMN gateway notations to represent choice points (exclusive gateways), parallel considerations (parallel gateways), and optional combinations (inclusive gateways).
    \item \textbf{Relationship Identification}: Through qualitative analysis of patterns co-occurring in the same architectural contexts, we identified two types of inter-pattern relationships: (1) complement relationships, where patterns frequently appear together to address related concerns, and (2) conditional relationships, where the applicability of one pattern depends on prior selection of another. These relationships were validated by cross-referencing multiple sources and, when necessary, through follow-up consultations with practicing quantum software architects (as detailed in Section \ref{EvaluatingDecision}).
\end{itemize}}

\subsection{Evaluating Decision Models}\label{EvaluatingDecision}
\textcolor{black}{In our study, we conducted a series of semi-structured interviews with quantum software practitioners to refine and evaluate our proposed decision models. The evaluation was performed at the model level, assessing practitioners' familiarity with the decision models as well as their understandability, completeness, and usefulness. This reflects how multiple architecture patterns are typically considered collectively during real-world architectural decision-making. While this enables a holistic assessment, it may limit pattern-level insights, which we identify as an avenue for future work. The interview guide, adapted from prior studies (e.g., \cite{waseem2022decision}, \cite{lewis2016decision}, \cite{xu2021decision}, \cite{wang2025decision}), includes predefined questions and follow-up prompts designed to elicit focused responses. Although the final question was open-ended, it was applied in a semi-structured manner to capture actionable insights. To mitigate respondent fatigue, participants received detailed supporting materials with practical examples in advance, and the interview flow was carefully designed to maintain clarity and engagement throughout the session. The questionnaire was conducted via Google Docs and is publicly available online\footnote{\url{https://tinyurl.com/bdcvzjjb}}, with supporting materials shared prior to the interview\footnote{\url{https://tinyurl.com/2fwcbkmh}}.}


\subsubsection{Developing Interview Questionnaire}
\textbf{\newline}The interview questionnaire was developed into seven sections: (i) participant demographic information (6 questions), and evaluation of the decision models for (ii) Communication (8 questions), (iii) Decomposition (9 questions), (iv) Data Processing (8 questions), (v) Fault Tolerance (8 questions), (vi) Integration and Optimization (9 questions), and (vii) Algorithm Implementation (9 questions). The questionnaire included both closed-ended and open-ended questions. 

\subsubsection{Participant Selection}
\textbf{\newline Phase 1 - Identifying Practitioners}: We followed the process outlined below to identify practitioners who are actively working on quantum software.

\textbf{1) Quantum Software Practitioners on GitHub}: Our study targeted individuals who actively contributed to quantum software projects on GitHub. We identified potential participants from the 347 quantum software repositories selected in our mining study (see Section \ref{IdentifyingPatternsMiningStudy}). We prioritized projects directly related to quantum software development and extracted publicly available contact information for 1,050 contributors. Invitations were sent via email to these contributors, providing them with access to two key documents: (i) an interview questionnaire titled ``\textit{Decision Models for Selecting Architecture Patterns and Strategies in Quantum Software Systems}'' and (ii) a supplementary document containing detailed descriptions and practical examples titled ``\textit{Architecture Patterns \& Strategies in Quantum Software Systems with Practical Examples}''. These materials were shared in advance to familiarize the participants with the scope and context of the interview and to facilitate meaningful discussion during the sessions.

\textbf{2) Social and Professional Platforms}:
To further expand our participant pool, we were actively engaged with various quantum computing communities on platforms such as LinkedIn and Facebook. These groups bring together quantum software practitioners from diverse geographical regions to discuss challenges, share insights, and exchange expertise. We posted invitations in several professional groups, providing links to the interview questionnaire with the supplementary document describing the architecture patterns and strategies. Participants were encouraged to review these materials prior to the interview sessions. The details of the groups where the invitations were shared are listed in our replication package~\cite{dataset2}.

\textbf{3) Other Contacts}:
In addition to public platforms, we leveraged our professional networks to recruit qualified participants. Email invitations were sent to individuals involved in quantum software organizations and to authors of industrial-track research papers and online blogs focused on quantum software development. We also encouraged recipients to forward the invitation to their colleagues or collaborators with relevant expertise, thus extending our reach through snowball sampling.

\textbf{Phase 2 - Obtaining a Valid Sample}: To ensure the validity and reliability of the interview data, we applied a set of inclusion and exclusion criteria, detailed in Table \ref{tab:InclusionExclusionCriteriaSelectingValidInterviewResponses}. Participants were included if they met criterion I1, namely, having professional responsibility for the development, design, or operation of quantum software systems. Eligibility was assessed through demographic questions on role and experience in software and quantum software development. To maintain data quality, criterion E1 excluded interviews with inconsistent, random, or meaningless responses, while criterion E2 excluded participants who provided such responses to more than five key questions. For example, one participant repeatedly answered ``\textit{I do not know}'' to questions on familiarity, usefulness, and completeness of the decision models and was excluded under E1 and E2. This process helped ensure that only knowledgeable and engaged participants contributed to evaluating the decision models.


\begin{table}[ht]
\centering
\caption{Inclusion and exclusion criteria for selecting valid interview responses}
\label{tab:InclusionExclusionCriteriaSelectingValidInterviewResponses}
\small
\setlength{\tabcolsep}{4pt}
\renewcommand{\arraystretch}{1.1}
\begin{tabular}{p{4cm}p{8cm}}
\hline
\textbf{Inclusion Criterion} & \textbf{Description} \\
\hline
I1 & \textcolor{black}{The participant must have professional experience in the development, software architecture, software design, or architecture-related decision-making of quantum software systems.} \\
\hline
\textbf{Exclusion Criterion} & \textbf{Description} \\
\hline
E1 & An interview exhibiting inconsistent, random, or meaningless responses is excluded. \\
\hline
E2 & If more than five core interview questions are answered inconsistently, randomly, or without meaning, the entire response is excluded. \\
\hline
\end{tabular}
\end{table}

\textbf{Phase 3 - Filtering Interview Data}: Before initiating the formal round of interviews, we conducted pilot interviews involving five quantum software practitioners. These pilot participants represent diverse geographic and professional backgrounds: three were based in Bangladesh, one in China, and one in Senegal. Their roles include researchers and developers working in domains such as quantum cryptography, simulation and optimization, education, machine learning, and quantum software frameworks. In terms of experience, all participants had between 2 and 5 years of software development experience. In contrast, the majority of the participants (three out of five) had between 2 and 5 years of experience, while two participants had less than 1 year of experience in quantum software development. \textcolor{black}{The pilot interview results are included because the responses remain valid for the final analysis and no substantive changes were made to the interview questions after the pilot interview phase.} The primary objective of this phase was to preliminarily assess the clarity, structure, and practical relevance of the proposed decision models across the six architectural design areas: Communication, Decomposition, Data Processing, Fault Tolerance, Integration and Optimization, and Algorithm Implementation. The pilot interviews also helped us refine the interview questionnaire and validate whether the proposed decision models were understandable and applicable in real-world QSE contexts. 


Based on pilot participant feedback, we revised the Communication, Data Processing, and Fault Tolerance decision models by adding new architecture patterns (e.g., Quantum Teleportation Protocol, Amplitude Encoding, and Error Correction) and refining decision flows to distinguish fault detection from fault correction. Accordingly, we updated the corresponding questionnaire items Q14, Q31, and Q39, which ask participants, ``\textit{Which of the following architecture patterns have you implemented in your quantum software systems?}'' for these three decision models, by expanding the multiple-choice options to include the newly incorporated patterns and strategies. The overall question structure and wording remained unchanged. These revisions improved alignment between the refined decision models and the evaluation instrument while enhancing visual presentation and clarifying technical terminology for better participant understanding.

\textcolor{black}{The pilot interviews were included in the final analysis because the refinements to the interview protocol after the pilot phase were incremental and did not alter the underlying concepts being evaluated. To mitigate potential bias, transcripts of the pilot interviews were initially analyzed separately and later independently re-analyzed by the second and third authors. The refinements mainly added new response options without altering existing ones, ensuring comparability between pilot and formal interviews. The same exclusion criteria (Table \ref{tab:InclusionExclusionCriteriaSelectingValidInterviewResponses}) were applied consistently across both datasets, and no pilot interview responses were excluded. We also verified that no pilot interview responses relied on concepts that were later modified or removed. These measures ensure consistency, comparability, and objectivity of the interview data across the pilot and formal interview phases, allowing pilot interview data to be included in the final analysis of the interview responses without compromising the validity of the study.}

Following this refinement, we proceeded with the formal interview phase, during which \textcolor{black}{we interviewed 25 quantum software practitioners. Participants were invited via professional email contacts, relevant online groups (e.g., LinkedIn, Facebook), and personal outreach. A total of 31 confirmations were initially received.} One participant’s response was later excluded due to inconsistency or lack of meaningful engagement, as per the exclusion criteria defined in Table \ref{tab:InclusionExclusionCriteriaSelectingValidInterviewResponses}. \textcolor{black}{This resulted in 30 valid interviews, including 5 from the pilot interview phase, which were used for the final analysis.} All interviews were conducted with informed consent, recorded, transcribed, and translated into English by the first and second authors. The interview results were independently reviewed by the second and third authors to remove irrelevant content and ensure alignment with the study objectives. MS Excel was used to support data organization and categorization.

\textbf{Phase 4 - Interview Data Analysis}: We employed two complementary analysis techniques for the interview data. First, descriptive statistics were applied to analyze closed-ended responses regarding participant demographics, familiarity with decision models, perceived understandability, and completeness. Second, we utilized open coding \cite{seaman1999qualitative} and constant comparison techniques \cite{grove1988analysis} from Grounded Theory methodology \cite{stol2016grounded} to analyze qualitative data obtained from open-ended questions. In the open coding phase, raw textual data were segmented into discrete codes representing key ideas expressed by participants, focusing on aspects such as perceived correctness, sufficiency, usefulness, suggestions for improvement, and real-world usage of the decision models. These responses were manually coded by the first author and verified by the second author. Key phrases and statements were extracted from the interview transcripts and assigned initial codes, such as ``\textit{The decision model correctly defines all decision paths and flows}'', ``\textit{Extremely helpful in guiding us toward the appropriate patterns and strategies}'', ``\textit{The model clearly outlines the trade-offs involved in different communication patterns}'', and ``\textit{I have a suggestion to enhance the model: we could introduce resource-aware processing approaches}''. These initial codes were then grouped into higher-level concepts and iteratively refined into broader analytical categories, including Model Correctness, Evaluation Usefulness, and Suggestions for Improvement. This systematic approach facilitated a rigorous, structured interpretation of the qualitative feedback, enabling the refinement of the decision models.

\section{Decision Models}
\label{DecisionModelsSection}
In software architecture, decision models connect elements of the problem space with aspects of the solution space. The problem space encompasses both functional and non-functional requirements, while the solution space encompasses the design components \cite{lewis2016decision}. In this study, the problem space is defined through a collection of QAs for quantum software systems. The solution space comprises various architecture patterns and strategies for quantum software systems. These decision models were crafted for six critical areas of quantum software design: Communication, Decomposition, Data Processing, Fault Tolerance, Integration and Optimization, and Algorithm Implementation. \textcolor{black}{It is important to note that some architecture patterns, such as the Decorator Design Pattern, appear in multiple decision models because they address different architectural concerns depending on the context. In the Fault Tolerance Decision Model, the Decorator Design Pattern is used to dynamically integrate error mitigation mechanisms into quantum algorithms to improve reliability, whereas in the Integration and Optimization Decision Model, it enables adding new functionalities to quantum operations without modifying their core structure. Therefore, the Decorator Design Pattern is not duplicated but represents various architectural usages across decision models.}


\subsection{Communication Decision Model} \label{Model:Communication}
\begin{figure}[h!]
\centering
\includegraphics[width=.8\linewidth]{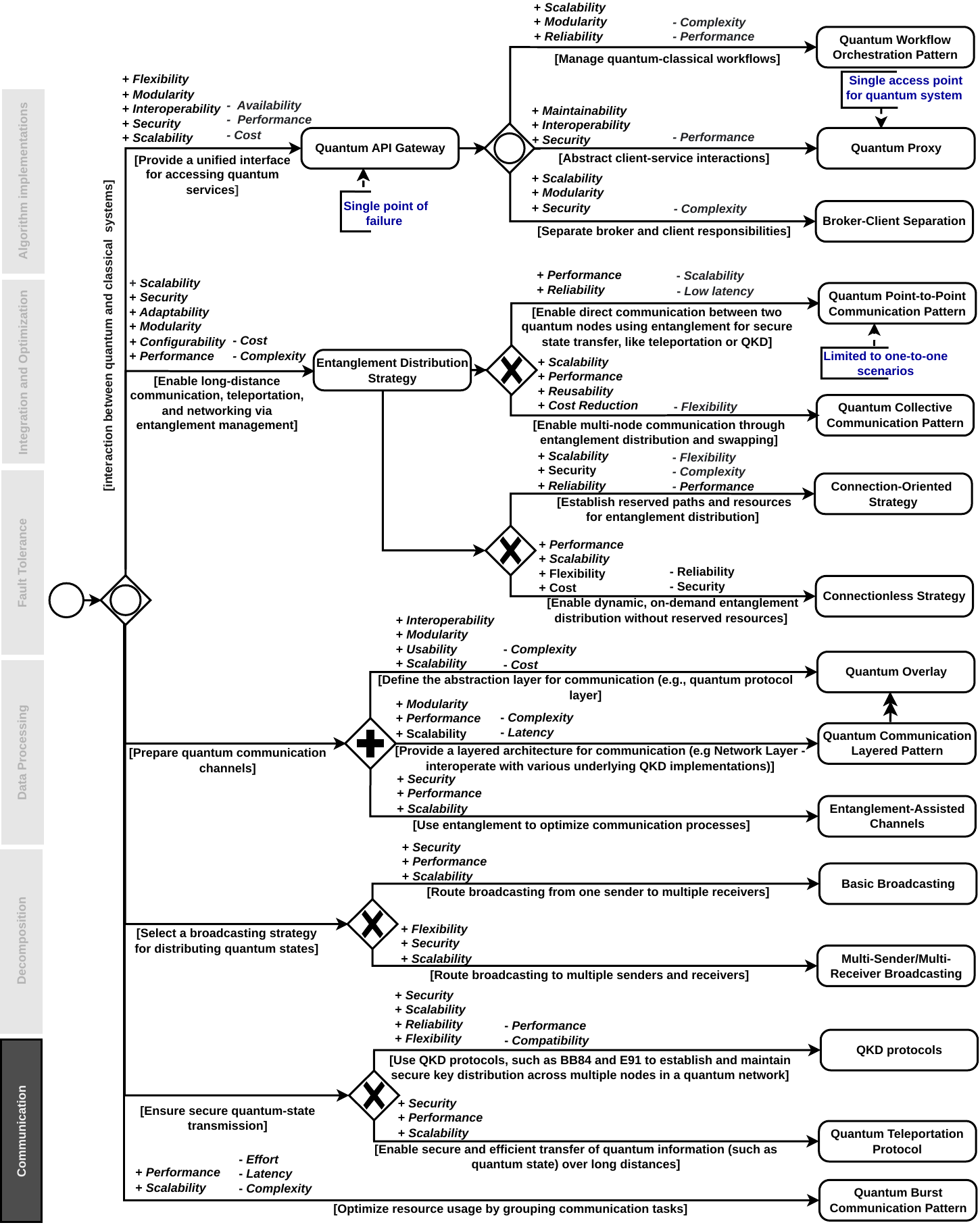}
\caption{Decision model for selecting architecture patterns and strategies for communication}
\label{Fig:Communication}
\end{figure}

The Communication Decision Model provides a structured framework to guide practitioners in selecting appropriate architecture patterns and strategies based on specific conditions and communication impacts in quantum software systems. The decision model guides the selection of communication patterns for both quantum-to-quantum and quantum-to-classical communication. It covers various architecture patterns and strategies, each associated with clearly outlined positive and negative QAs, which are informed by findings from the mining study and the SLR on quantum software communication. It is structured through various decision gateways: Inclusive, Exclusive, and Parallel, representing the decision-making flows required for architecting quantum communication systems. Table \ref{Tab: Communication} lists the patterns and strategies covered by the Communication Decision Model (see Figure \ref{Fig:Communication}). 

\begin{table}[h!]
\centering
\caption{Architecture patterns and strategies for communication}
\label{Tab: Communication}
\small
\setlength{\tabcolsep}{4pt}
\renewcommand{\arraystretch}{1.1}
\resizebox{\textwidth}{!}{%
\begin{tabular}{lp{10cm}}
\hline
\textbf{Pattern Name} & \textbf{Summary} \\ \hline
Quantum API Gateway                      & Provide a unified interface for accessing quantum services, optimizing deployment, and resource selection dynamically. \\ \hline
Quantum Workflow Orchestration           & \textcolor{black}{Coordinates the execution of hybrid quantum-classical workflows by managing task sequencing, resource allocation, and synchronization across distributed components.} \\ \hline
Quantum Proxy                            & Abstract client-service interactions to enhance maintainability and interoperability with secure communication. \\ \hline
Broker-Client Separation                 & Clearly separates broker and client responsibilities, enhancing modularity, scalability, and security at the cost of complexity. \\ \hline
Entanglement Distribution Strategy       & Enable long-distance quantum communication via repeaters, purification, and swapping techniques. \\ \hline
Quantum Point-to-Point Communication     & Facilitate secure direct communication between two quantum nodes through entanglement (e.g., QKD). \\ \hline
Quantum Collective Communication         & Enable efficient multi-node quantum communication leveraging entanglement distribution and swapping. \\ \hline
Connection-Oriented Strategy             & Use dedicated paths and resources to manage stable quantum entanglement distribution. \\ \hline
Connectionless Strategy                  & Dynamically manages on-demand quantum entanglement distribution without fixed resources. \\ \hline
Quantum Overlay                          & Define abstraction layers for quantum communication, standardizing interactions across quantum protocols. \\ \hline
Quantum Communication Layered            & Adopt a layered architecture, facilitating interoperability among different quantum protocols and implementations. \\ \hline
Entanglement-Assisted Channels           & Optimize communication by leveraging entanglement, enhancing security, performance, and scalability. \\ \hline
Basic Broadcasting                       & \textcolor{black}{Enable distribution of quantum information from one sender to multiple receivers through entanglement-assisted or classical coordination mechanisms, acknowledging the constraints imposed by the no-cloning theorem.} \\ \hline
Multi-Sender/Multi-Receiver Broadcasting & \textcolor{black}{Extend virtual broadcasting to scenarios involving multiple senders and receivers, supporting flexible communication patterns through coordinated entanglement distribution and classical control, rather than true quantum state replication.} \\ \hline
QKD Protocols                            & Implement secure quantum key distribution protocols (e.g., BB84, E91) to ensure secure node communication. \\ \hline
Quantum Teleportation Protocol           & \textcolor{black}{Facilitate secure, efficient quantum state transfers across extensive distances.} \\ \hline
Quantum Burst Communication Pattern             & \textcolor{black}{Optimizes quantum network resources by aggregating multiple discrete communication tasks (e.g., entangled pair distribution, remote gate applications) into a single, coordinated batch transmission (burst), where tasks are executed together within a short time interval rather than individually.} \\ \hline
\end{tabular}%
}
\end{table}

For interactions between quantum and classical systems, an \textbf{Inclusive Gateway} initiates the decision process by evaluating various quantum communication requirements. \textcolor{black}{Given the inherently hybrid nature of quantum applications, effective coordination between classical and quantum components requires well-defined interface mechanisms. In this context, patterns such as Quantum API Gateway, Quantum Proxy, and Quantum Workflow Orchestration collectively fulfill the role of a classical–quantum interface by enabling communication, synchronization, and data exchange between classical clients and quantum backends.} Specifically, \textcolor{black}{\textbf{Quantum API Gateway} provides a unified interface for interactions between classical clients and quantum backends. Classical clients refer to external classical systems, applications, or workflow managers that submit quantum jobs and control execution via classical interfaces. For example, Qunicorn represents a legacy, decentralized runtime architecture, whereas IBM's Qiskit Runtime provides a centralized gateway that manages job submissions from classical clients and supports modern hybrid quantum-classical workflows.} It improves \textit{Flexibility} and \textit{Modularity} by encapsulating diverse services under a single access point, and enhances \textit{Interoperability} and \textit{Security} through centralized control. It also supports \textit{Scalability} by efficiently managing increasing client requests. However, it introduces a single point of failure, which affects its \textit{Availability}. Additionally, the need for extra processing degrades \textit{Performance} and increases \textit{Cost} due to management overhead. \textcolor{black}{Quantum applications are inherently hybrid, involving both classical and quantum components that must be executed in a coordinated manner. This hybrid nature necessitates explicit orchestration of execution flows, which can be achieved through an Inclusive Gateway leading to \textbf{Quantum Workflow Orchestration}. This pattern governs the sequencing and synchronization of quantum and classical tasks, enabling efficient execution across heterogeneous environments. It enhances \textit{Scalability} by distributing and coordinating tasks, and improves \textit{Modularity} and \textit{Reliability} through well-defined execution logic. However, the introduction of orchestration layers may increase system \textit{Complexity} and can negatively impact \textit{Performance} in tightly coupled or latency-sensitive workflows. Alternatively, \textbf{Quantum Proxy} is chosen to abstract client-service interactions. It enhances \textit{Maintainability} by isolating clients from implementation changes, strengthens \textit{Interoperability} by managing protocol translations, and improves \textit{Security} by controlling access. It may slightly reduce \textit{Performance} due to request forwarding overhead. \textbf{Broker-Client Separation} supports the separation of concerns between service consumers and communication brokers. It boosts \textit{Modularity} and \textit{Security} by isolating messaging components and improves \textit{Scalability} by decoupling senders and receivers. However, it introduces \textit{Complexity} due to additional configuration and coordination logic.}


To enable long-distance communication, teleportation, and networking via entanglement management, \textbf{Entanglement Distribution Strategy} is used with techniques such as repeater purification and entanglement swapping. This strategy enhances \textit{Scalability} by enabling broader communication ranges, \textit{Security} through controlled entanglement paths, \textit{Adaptability} and \textit{Configurability} by allowing flexible routing, \textit{Modularity} by encapsulating the communication logic as a separate module, and \textit{Performance} due to efficient qubit transfer, albeit with increased \textit{Complexity} from entanglement management overhead. Depending on the communication scenario, an \textbf{Exclusive Gateway} selects either \textbf{Quantum Point-to-Point Communication Pattern} or \textbf{Quantum Collective Communication Pattern}. \textbf{Point-to-Point Pattern} improves \textit{Performance} and \textit{Reliability} by providing direct, dedicated channels between nodes, but introduces \textit{Latency} and limits \textit{Scalability} as it only supports one-to-one links. \textbf{Collective Communication Pattern} supports multi-node entanglement-based messaging, which increases \textit{Scalability}, \textit{Performance}, and \textit{Reusability} by sharing communication logic and resources, and reduces \textit{Cost} through shared infrastructure. However, it limits \textit{Flexibility} due to predefined communication topologies. The decision model also includes \textbf{Connection-Oriented Strategy}, which reserves fixed paths for communication. This ensures \textit{Scalability}, \textit{Security}, \textit{Reliability}, and \textit{Performance} through stable connections, but reduces \textit{Flexibility} and increases \textit{Cost} due to pre-established resource allocation.  In contrast, \textbf{Connectionless Strategy} enables dynamic, on-demand communication. It enhances \textit{Performance}, \textit{Scalability}, \textit{Flexibility}, and \textit{Cost-Efficiency} by avoiding fixed routing. \textcolor{black}{However, it compromises \textit{Reliability} due to the lack of guaranteed delivery paths and acknowledgments, and introduces \textit{Security} vulnerabilities, such as exposure to packet interception, spoofing, and data integrity loss.}

For channel preparation, the decision model uses \textbf{Parallel Gateways} to evaluate multiple configuration paths. \textbf{Quantum Overlay} introduces an abstraction layer between network components, which improves \textit{Interoperability} by standardizing interactions, \textit{Modularity} by isolating layers, \textit{Usability} through simplified interfaces, and \textit{Scalability} by supporting layered expansion. However, it also increases \textit{Complexity} due to layered architecture and adds \textit{Cost} due to additional infrastructure needs. Alternatively, \textbf{Quantum Communication Layered Pattern} organizes communication functions into dedicated layers, enhancing \textit{Modularity} and \textit{Performance} by isolating responsibilities and optimizing each layer. This layered approach supports \textit{Scalability} across different hardware or protocol types but introduces \textit{Complexity} in managing dependencies and \textit{Latency} due to multi-step message processing. For performance optimization, \textbf{Entanglement-Assisted Channels} utilize pre-shared entanglement to streamline quantum communication, resulting in improved \textit{Security}, \textit{Performance}, and \textit{Scalability}, as entanglement can reduce the need for repeated key exchanges and boost throughput. 

\textcolor{black}{For distributing quantum information to multiple nodes, an \textbf{Exclusive Gateway} is used to select between alternative communication strategies. Due to the constraints imposed by the no-cloning theorem, true quantum state broadcasting is not physically realizable. Therefore, \textbf{Basic Broadcasting} (Virtual/Approximate) enables the distribution of quantum information from one sender to multiple receivers through entanglement-assisted or classical coordination mechanisms. This approach improves \textit{Scalability} by supporting multi-node communication, enhances \textit{Security} through quantum-safe distribution, and improves \textit{Performance} via parallel or coordinated delivery to multiple receivers, although it may introduce coordination overhead. \textbf{Multi-Sender/Multi-Receiver Broadcasting} (Virtual/Approximate) extends this concept by allowing multiple senders and receivers, thereby improving \textit{Flexibility} in network configuration, enhancing \textit{Security} through distributed and coordinated communication, and increasing \textit{Scalability} across distributed systems.}

To establish secure quantum communication channels, the decision model evaluates \textbf{Quantum Key Distribution (QKD) Protocol} through an \textbf{Inclusive Gateway}. \textbf{QKD Protocol} supports secure key exchange across quantum network nodes, thereby enhancing \textit{Security} and \textit{Reliability}. Its \textit{Scalability} stems from its applicability to multi-node networks, and \textit{Flexibility} is supported by its use in various quantum architectures. However, it may lead to reduced \textit{Performance} due to key negotiation overhead and face \textit{Compatibility} issues with some classical systems. Similarly, \textbf{Quantum Teleportation Protocol} is selected to enable the transfer of quantum information over long distances. It improves \textit{Security} and \textit{Scalability} by eliminating the need for direct quantum links between nodes and enhances \textit{Performance} through instantaneous state transfer, although this depends on the quality of entanglement and classical signaling. 

\textcolor{black}{For identifying optimal qubit-node interaction patterns and strategies, the decision model includes an  \textbf{Inclusive Gateway}, which leads to the selection of the \textbf{Quantum Burst Communication Pattern}. This pattern groups multiple communication tasks into a coordinated batch (burst) for joint execution within a short time window. This approach improves \textit{Performance} and \textit{Scalability} by reducing idle time and optimizing channel utilization. However, it may increase \textit{Latency}, \textit{Effort}, and \textit{Complexity} due to the additional coordination and scheduling required for burst execution.}

The Communication Decision Model offers a structured approach for practitioners to select communication patterns and strategies in quantum software systems, aligning system communication requirements with trade-offs among various QAs, including security, scalability, and performance.

\subsection{Decomposition Decision Model}\label{Model:Decomposition}
\begin{figure}[h!]
\centering
\includegraphics[width=0.8\linewidth]{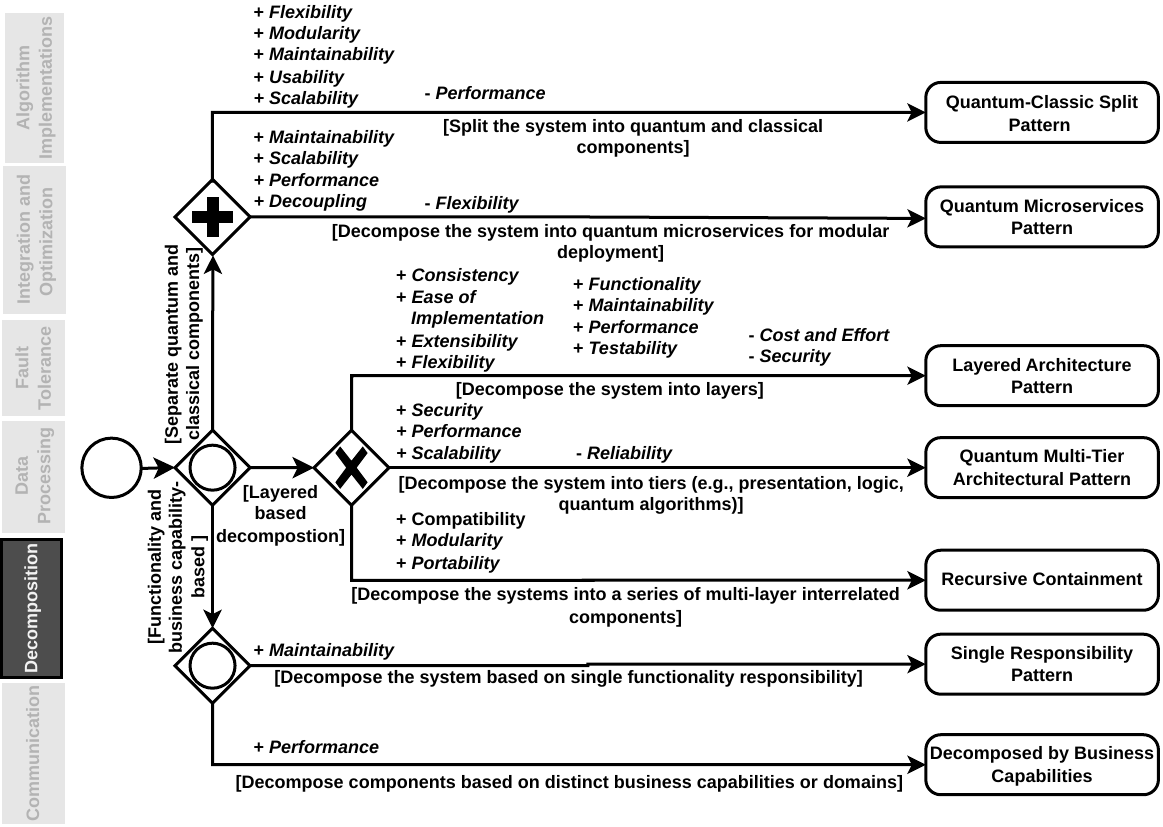}
\caption{Decision model for selecting architecture patterns and strategies for decomposition}
\label{Fig: Decomposition}
\end{figure}

The Decomposition Decision Model provides structured guidance to practitioners in selecting appropriate decomposition patterns for quantum software systems. This decision model is designed to evaluate various decomposition strategies, each triggered by specific conditions and affecting distinct QAs. The decision model begins with an \textbf{Inclusive Gateway}, which triggers multiple decision paths depending on whether the system benefits more from separating quantum and classical components, adopting architectural layering, or aligning decomposition with functional or business capabilities. Table \ref{Tab: Decomposition} lists the patterns and strategies covered by the Decomposition Decision Model (see Figure~\ref{Fig: Decomposition}). 

\begin{table}[h!]
\centering
\caption{Architecture patterns and strategies for decomposition}
\label{Tab: Decomposition}
\small
\setlength{\tabcolsep}{4pt}
\renewcommand{\arraystretch}{1.1}
\resizebox{\textwidth}{!}{%
\begin{tabular}{lp{11cm}}
\hline
\textbf{Pattern Name} & \textbf{Summary} \\ \hline
Quantum-Classic Split Pattern            & Split the system into quantum and classical components, enabling flexibility, modularity, and scalability. Facilitates the integration of NISQ devices and classical controllers to leverage both computation paradigms. Applied in Quantum Computing as a Service (QCaaS) and hybrid systems. \\ \hline

Quantum Microservices Pattern            & \textcolor{black}{Decompose the hybrid quantum-classical system into modular, loosely coupled quantum microservices. Each service encapsulates a specific quantum algorithm or circuit, along with its necessary classical pre- and post-processing logic, for distributed deployment.} \\ \hline

Layered Architecture Pattern             & Decompose the system into hierarchical layers (e.g., hardware, middleware, application), improving consistency and extensibility. Supports the layering of error correction, quantum-classical coupling, and facilitates incremental development of quantum software. Common in compiler infrastructures (e.g., XACC) and design flows. \\ \hline

Quantum Multi-Tier Architectural Pattern & \textcolor{black}{Divide the system into multiple tiers (physical, logical, application), integrating quantum algorithms within the application tier. This approach improves security, scalability, and resource management by effectively orchestrating quantum and classical components in a hybrid system.} \\ \hline

Recursive Containment                    & Structure the system as multi-layer interrelated components, promoting compatibility, modularity, and portability. Suitable for complex systems where each layer abstracts certain functionalities. \\ \hline

Single Responsibility Pattern            & Decompose based on single functionality responsibilities, maximizing maintainability. Applied to isolate concerns and simplify unit testing in quantum-classical software modules. \\ \hline

Decomposed by Business Capabilities      & \textcolor{black}{Divide the system components based on business domains or distinct capabilities, improving performance and aligning quantum solutions to domain-specific requirements (e.g., finance, logistics). This decomposition strategy is vital for ensuring the system's alignment with business needs.}  \\\hline
\end{tabular}%
}
\end{table}

For splitting quantum and classical logic, a \textbf{Parallel Gateway} activates two concurrent paths. \textbf{Quantum-Classic Split} is selected for dividing the system into distinct quantum and classical parts. This enhances \textit{Flexibility} and \textit{Modularity} by isolating concerns, supports \textit{Maintainability} and \textit{Usability} by simplifying component responsibilities, and improves \textit{Scalability} through parallel development. However, integration overhead may reduce \textit{Performance}. In contrast, \textbf{Quantum Microservices} is selected for modular deployment using isolated quantum services. This increases \textit{Maintainability} and \textit{Scalability} by enabling independent service management and boosts \textit{Performance} through parallel execution. However, it reduces \textit{Flexibility} as interactions between tightly scoped services require predefined coordination.

For selecting layered decomposition approaches, an \textbf{Exclusive Gateway} offers three alternative paths based on system structuring needs. \textbf{Layered Architecture} is selected when the system requires separation of responsibilities into horizontal layers. This improves \textit{Maintainability} by isolating layer-specific logic, enhances \textit{Performance} by enabling efficient execution within each layer, and supports \textit{Testability} through clear interface boundaries. However, it increases development effort due to the need for multiple abstraction layers and reduces \textit{Security} as each exposed layer interface becomes a potential vulnerability. \textbf{Quantum Multi-Tier Architecture} is selected for systems that interact with diverse clients or user interfaces. It enhances \textit{Compatibility} by decoupling the presentation and logic tiers, improves \textit{Modularity} by enabling independent updates across tiers, and supports \textit{Portability} by adapting easily to different platforms. At the same time, it reduces \textit{Reliability} by increasing inter-tier dependencies, which complicate failure handling. \textbf{Recursive Containment} is selected for designs requiring nested structural relationships among components. This increases \textit{Security} by encapsulating internal operations, improves \textit{Performance} through localized processing within each nested unit, and enhances \textit{Scalability} by allowing recursive extension of system components. However, it reduces \textit{Reliability} as tightly coupled dependencies across nested layers make fault isolation more difficult.

Lastly, for selecting functionality or business capability-based decomposition, the process employs another \textbf{Inclusive Gateway}, enabling two alternative paths. \textbf{Single Responsibility} is selected when the system is decomposed into singular functional responsibilities, thereby enhancing \textit{Maintainability} by isolating each task within a dedicated unit. \textcolor{black}{Alternatively, the \textbf{Decomposed by Business Capabilities} strategy is selected when system components must align with distinct business domains, thereby improving \textit{Performance} through a focus on domain-specific logic. For example, in a quantum application involving finance or logistics, the system is divided into business-relevant components to optimize performance while ensuring that the quantum system remains adaptable to evolving business needs.}

The Decomposition Decision Model provides a structured approach for practitioners to select decomposition patterns and strategies in quantum software systems, aligning system decomposition needs with trade-offs across various QAs, including flexibility, scalability, maintainability, and performance.

\subsection{Data Processing Decision Model}
\begin{figure}[h!]
\centering
\includegraphics[width=0.8\linewidth]{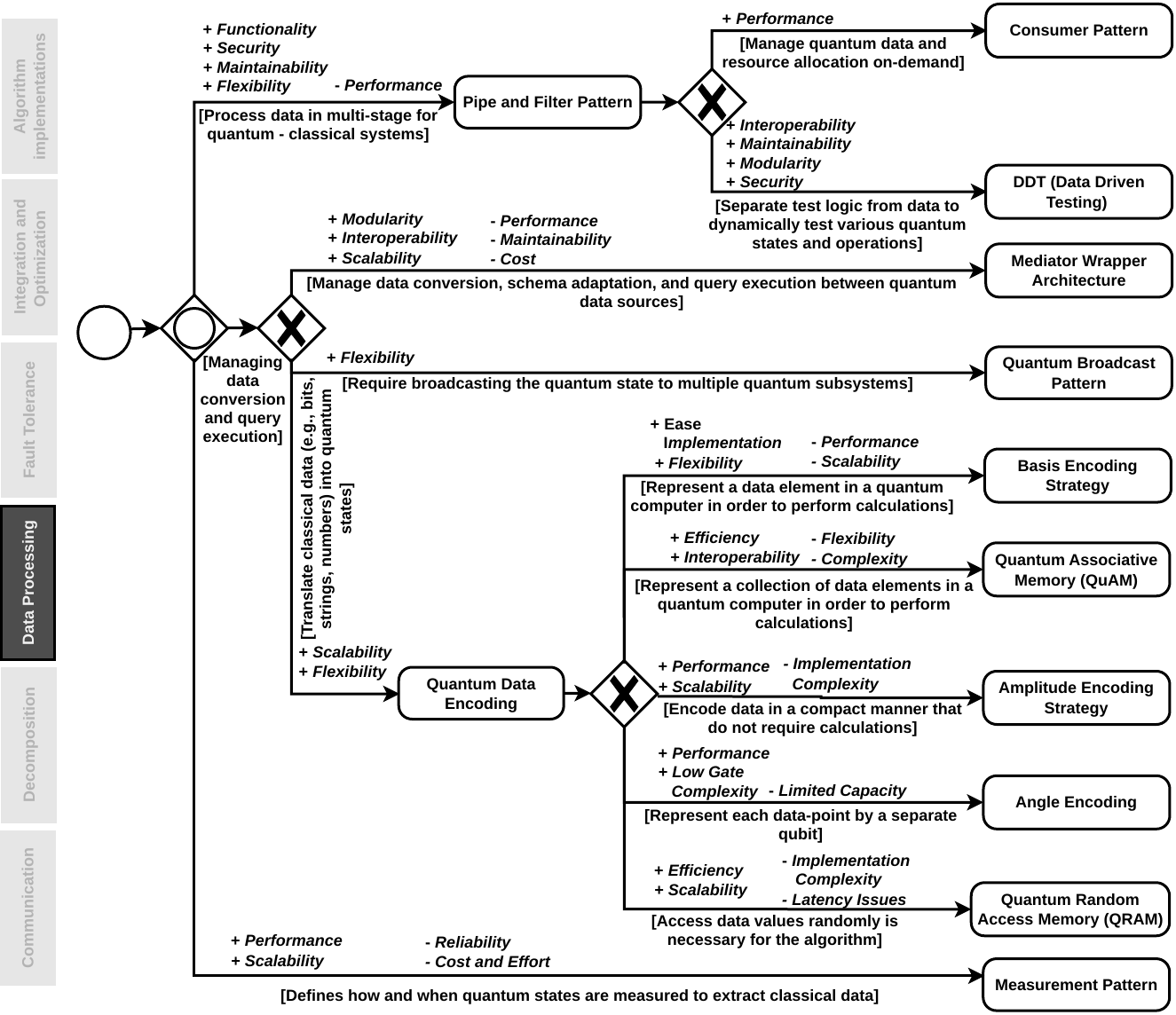}
\caption{Decision model for selecting architecture patterns and strategies for data processing}
\label{Fig: DataProcessing}
\end{figure}

\textcolor{black}{The Data Processing Decision Model provides a structured flow that leverages gateways to guide the selection of appropriate architecture patterns and strategies, based on data transformation needs, workload characteristics, and QAs trade-offs.} The decision flow begins with an \textbf{Inclusive Gateway}, enabling multiple paths to proceed simultaneously. Table \ref{Tab: DataProcessing} lists the patterns and strategies covered by the Data Processing Decision Model (see Figure~\ref{Fig: DataProcessing}).

\begin{table}[h!]
\centering
\caption{Architecture patterns and strategies for data processing}
\label{Tab: DataProcessing}
\small
\setlength{\tabcolsep}{4pt}
\renewcommand{\arraystretch}{1.1}
\resizebox{\textwidth}{!}{%
\begin{tabular}{lp{10cm}}
\hline
\textbf{Pattern Name} & \textbf{Summary} \\
\hline
Pipe and Filter Pattern             & \textcolor{black}{Process quantum data through multi-staged transpilation, transformation, and reconciliation pipelines to support structured quantum-classical workflow execution.} \\\hline
Consumer Pattern                    & \textcolor{black}{Dynamically manage quantum data and allocate resources on-demand. It optimizes resource allocation and data handling in quantum software systems, particularly for scenarios requiring on-demand processing.} \\\hline
Data Driven Testing (DDT)           & Separate test logic from data, enabling dynamic testing of diverse quantum states and operations for adaptability and efficiency. \\\hline
Quantum Mediator Wrapper            & Handle data conversion, schema adaptation, and query execution between varied quantum data sources, supporting system flexibility and scalability. \\\hline
Quantum Broadcast           & \textcolor{black}{Coordinate the application of common operations, control instructions, or dissemination of classical measurement outcomes across selected qubits or subsystems, improving coordination and flexibility while respecting the constraints imposed by quantum mechanics.} \\\hline
Quantum Data Encoding               & \textcolor{black}{Convert classical data (bits, strings, numbers) into quantum states, facilitating quantum computation readiness. This process aligns with the Initialization Pattern,
which is responsible for preparing quantum states before computation begins.} \\\hline
\textcolor{black}{Basis Encoding}            & Represent classical data elements as quantum computational basis states, allowing direct quantum calculations. \\\hline
\textcolor{black}{Quantum Associative Memory (QuAM)}   & Store and retrieves collections of data elements in quantum memory structures, enabling efficient quantum-based data processing. \\\hline
Amplitude Encoding         & Encode data compactly using amplitudes, minimizing computational requirements for efficient quantum data representation. \\\hline
Angle Encoding                      & Represent each data point with separate Qubits through angle encoding, enabling flexible data mapping to quantum states. \\\hline
\textcolor{black}{Quantum Random Access Memory (QRAM)} & Provide random access to quantum data values as required by algorithms, enhancing data retrieval flexibility in quantum computations. \\\hline
Measurement Pattern                 & \textcolor{black}{Define measurement protocols (projective, weak, and destructive) for extracting classical data from quantum states, enabling integration of quantum outputs into classical systems while accounting for measurement-induced collapse.} \\\hline
\end{tabular}%
}
\end{table}

\textcolor{black}{For multi-stage quantum data processing, the \textbf{Pipe and Filter} pattern is employed to structure the workflow. It processes quantum information through multiple sequential stages, supporting complex quantum-classical workflows with a modular and scalable architecture. This approach enables structured data flow, supporting \textit{Functionality} through modular processing, \textit{Security} by isolating components, \textit{Maintainability} due to the separation of stages, and \textit{Flexibility} for easy replacement or reordering of filters. \textbf{Pipe and Filter} is particularly useful in quantum software systems, such as Quantum Key Distribution (QKD), where secure communication and efficient data transmission are critical. However, it may affect \textit{Performance} due to processing overhead, as the system must handle data flow through multiple stages. Within the same decision flow, an  \textbf{Exclusive Gateway} identifies further paths. If the quantum system requires on-demand data handling and dynamic resource allocation, the \textbf{Consumer Pattern} can be selected. In this context, it improves \textit{Performance} by enabling efficient retrieval and processing of quantum-related data based on real-time demands, thereby optimizing computational resources and execution time. While originally a classical pattern, its adaptation in quantum software systems, evidenced by its use in repositories such as QMCPACK, demonstrates its relevance in supporting hybrid quantum workloads.} If the goal is to isolate test logic from data, \textbf{Data-Driven Testing (DDT)} is selected. It supports \textit{Interoperability} by allowing flexible data input formats, \textit{Maintainability} through centralized test data, \textit{Modularity} by decoupling test logic, and \textit{Security} by enabling controlled test environments.

For multi-faceted quantum data handling, an \textcolor{black}{\textbf{Inclusive Gateway}} selects a suitable path based on system needs. When managing data conversion, schema adaptation, and query execution across diverse quantum data sources, \textbf{Quantum Mediator Wrapper} is selected. It improves \textit{Modularity} by isolating adaptation logic from core functionalities, facilitates \textit{Interoperability} by translating between heterogeneous data formats, and supports \textit{Scalability} by decoupling services for easier integration of new sources. However, it may reduce \textit{Maintainability} due to the complexity of managing adapters, impact \textit{Performance} owing to added translation overhead, and increase \textit{Cost} due to required middleware infrastructure. \textcolor{black}{In contrast, when quantum information processing requires multiple subsystems to act in a coordinated manner, \textbf{Quantum Broadcast} can be selected. Rather than broadcasting or copying unknown quantum states, this pattern facilitates the dissemination of classical measurement outcomes, control instructions, or entanglement-assisted correlations to relevant subsystems. This enables coordinated processing with minimal coupling between receivers while remaining consistent with the no-cloning theorem.}


\textcolor{black}{For translating classical data into quantum states, the \textbf{Quantum Data Encoding} strategy is employed. This strategy aligns with the Initialization Pattern, facilitating the preparation of quantum states necessary for quantum computation. It contributes to \textit{Scalability} by supporting the encoding of large and high-dimensional data across Qubits, and to \textit{Flexibility} by enabling the use of various encoding schemes tailored to different computational needs. Importantly, once quantum states are established, direct computations can be performed not only with basis encoding but also with a variety of encoding schemes, including amplitude, angle, associative memory, and QRAM-based approaches. This capability enhances \textit{Scalability} by supporting the encoding of large and high-dimensional data across qubits, and \textit{Flexibility} by enabling encoding schemes tailored to different computational needs.} In one scenario, the decision flow uses an \textbf{Exclusive Gateway} to trigger one path based on specific conditions. \textbf{Basis Encoding} is selected when a straightforward mapping of classical bits to quantum states is sufficient. It enhances \textit{Ease of Implementation} through its simplicity and promotes \textit{Flexibility} in data-processing experiments, though it can negatively affect \textit{Performance} and \textit{Scalability} because its naive structure neither compresses data nor minimizes qubit usage. In contrast, \textbf{Quantum Associative Memory (QuAM)} is selected when associative retrieval is needed in quantum computations. It boosts \textit{Efficiency} through rapid pattern search and improves \textit{Interoperability} by integrating different types of quantum data. However, its rigid structure may reduce \textit{Flexibility}, and its complex architecture increases overall \textit{Complexity} and implementation difficulty. Another path activates \textbf{Amplitude Encoding}, which compactly encodes classical data into amplitude values of Qubits. This enhances \textit{Performance} and supports \textit{Scalability} by reducing the number of Qubits required for large datasets. Yet, it often results in high Implementation \textit{Complexity} due to precision control during state preparation, and increased \textit{Error} rates due to susceptibility to noise. \textcolor{black}{When representing individual data features via rotational angles, \textbf{Angle Encoding} is chosen. It enables high \textit{Performance} through minimal gate usage, requiring only a constant number of single-qubit rotations, and contributes to \textit{Low Gate Complexity}. However, it faces challenges with \textit{Limited Capacity} because it requires one qubit per input feature, resulting in linear scaling of qubit count with data dimensionality.} Lastly, \textbf{Quantum Random Access Memory (QRAM)} is selected when fast, non-sequential access to quantum-stored data is required. It improves \textit{Efficiency} by enabling parallel data access in superposition and aids \textit{Scalability} for large datasets. However, it introduces significant Implementation \textit{Complexity} due to the need for specialized quantum hardware and suffers from potential \textit{Latency} Issues during quantum-classical transitions.

\textcolor{black}{Finally, for tracking and interpreting quantum state collapse during execution, \textbf{Measurement Pattern} is selected. This pattern considers three measurement types: projective measurement (deterministic outcomes with full state collapse), weak measurement (partial disturbance enabling sequential observations with lower fidelity), and destructive measurement (consumes the quantum state, suitable for final readout). Projective measurement improves \textit{Performance} by providing clear data at defined checkpoints and supports \textit{Scalability} through concurrent state evaluation. However, it increases \textit{Cost and Effort} due to repeated state preparation and may reduce \textit{Reliability}, as measurement inherently collapses the quantum state and premature measurement can disrupt the computation. Weak measurement reduces this risk but requires additional post-processing effort, while destructive measurement is primarily used for final output to avoid unintended state loss.}


The Data Processing Decision Model guides practitioners in selecting suitable architecture patterns and encoding strategies based on system-specific data needs and QA trade-offs, offering a structured decision flow that balances modularity, performance, scalability, and implementation complexity in quantum software systems.

\subsection{Fault Tolerance Decision Model}
\begin{figure}[h!]
\centering
\includegraphics[width=0.8\linewidth]{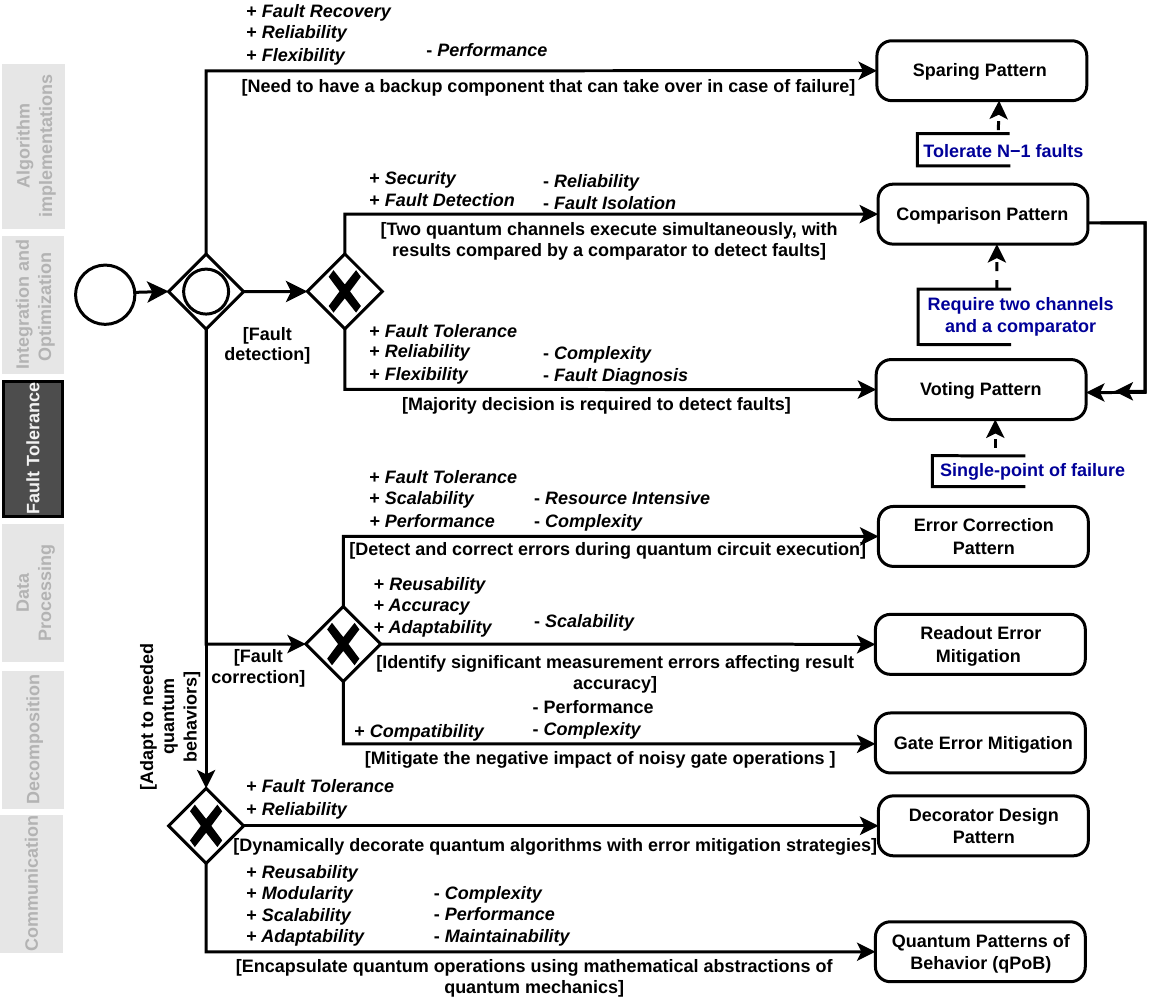}
\caption{Decision model for selecting architecture patterns and strategies for fault tolerance}
\label{Fig: FaultTolerance}
\end{figure}

The Fault Tolerance Decision Model is designed to assist in selecting appropriate architecture patterns that enhance fault detection, correction, and tolerance in quantum systems. This decision model begins with an \textbf{Inclusive Gateway}, enabling the simultaneous evaluation of multiple fault-tolerance requirements. Table \ref{Tab: FaultTolerance} lists the patterns and strategies covered by the Fault Tolerance Decision Model (see Figure~\ref{Fig: FaultTolerance}). 

\begin{table}[h!]
\centering
\caption{Architecture patterns and strategies for fault tolerance}
\label{Tab: FaultTolerance}
\small
\setlength{\tabcolsep}{4pt}
\renewcommand{\arraystretch}{1.1}
\resizebox{\textwidth}{!}{%
\begin{tabular}{lp{10cm}}
\hline
\textbf{Pattern Name} & \textbf{Summary} \\
\hline
Sparing Pattern                     & \textcolor{black}{Introduce redundant quantum components that take over upon fault detection (e.g., via Witness checks), ensuring continuity during faults in quantum software systems.} \\\hline
Comparison Pattern                  & \textcolor{black}{Executing two quantum channels (e.g., qrChannels) in parallel and forwarding their respective measurement results (probability distributions) to a comparator component.} \\\hline
Voting Pattern                      & \textcolor{black}{Uses a majority decision mechanism to detect and handle faults by comparing outputs from multiple components. Adapted for quantum systems (e.g., combining outputs from parallel quantum channels) to mitigate measurement errors and improve reliability.} \\\hline
Error Correction Pattern            & \textcolor{black}{Dynamically integrate error mitigation into quantum algorithms to allow real-time adaptability during execution.} \\\hline
Readout Error Mitigation Pattern    & Reduce the impact of measurement errors on quantum computations to improve accuracy. \\\hline
Gate Error Mitigation Pattern       & Mitigate errors arising from noisy gate operations in quantum circuits. \\\hline
Decorator Design Pattern            & \textcolor{black}{Dynamically integrate error mitigation into quantum algorithms to allow real-time adaptability during execution.} \\\hline
Quantum Patterns of Behavior (qPoB) & Provide high-level abstraction by encapsulating quantum operations mathematically, supporting complex quantum behaviors and error mitigation strategies. \\\hline
\end{tabular}%
}
\end{table}

\textcolor{black}{For ensuring operational continuity during component failures in quantum software systems, \textbf{Sparing Pattern} is selected. It introduces standby quantum components that automatically take over when a failure occurs. Sparing Pattern is particularly relevant for quantum systems, where error detection mechanisms, such as using Witnesses or Certificates, are employed to validate quantum measurements. Sparing Pattern improves \textit{Fault Recovery} by providing immediate redundancy, \textit{Reliability} by reducing downtime risk, and \textit{Flexibility} through seamless switchover. However, it may slightly reduce \textit{Performance} due to synchronization overhead.}


\textcolor{black}{For fault detection, the decision model follows a \textbf{Exclusive Gateway} to explore detection mechanisms. When simultaneous execution and comparison of two quantum channels are required, \textbf{Comparison Pattern} is selected. This pattern uses a comparator component to measure the divergence between the two output distributions using a metric such as the Kullback-Leibler Divergence (KLD). If the measured divergence exceeds a configurable threshold, the output is rejected and triggers countermeasures. This pattern strengthens \textit{Security} by identifying channel mismatches, indicating potential tampering or corruption. It supports \textit{Fault Detection} via real-time comparison of measurement outcomes and provides a clear rejection criterion. However, the need for synchronized operations, the probabilistic nature of quantum measurements, and the choice of an appropriate divergence threshold may reduce overall \textit{Reliability}. Additionally, it can hinder \textit{Fault Isolation}, as errors may propagate across dependent processes or be masked by noise, complicating the identification of discrepancies. If a system benefits from redundancy through collective decision-making, \textbf{Voting Pattern} is selected. This pattern leverages multiple quantum channels (qrChannels) running in parallel on quantum computers, whose outputs are combined using a voter (or combiner) that selects the majority outcome. Although classical in origin, it is effectively adapted for hybrid quantum-classical systems to address quantum-specific challenges such as measurement errors and transient decoherence. By employing this majority decision mechanism, the pattern improves \textit{Fault Tolerance} by using a majority rule to mask faulty results, supports \textit{Reliability} by minimizing single-point failures, and offers \textit{Flexibility} to adapt to various fault scenarios. Yet, maintaining multiple replicas increases \textit{Complexity} and makes fault tracking harder, which can reduce the effectiveness of \textit{Fault Diagnosis}.}


For fault correction, the decision model proceeds through an \textbf{Exclusive Gateway} to explore correction mechanisms. If the system requires dynamic detection and correction of errors during quantum circuit execution, \textbf{Error Correction Pattern} is implemented. It improves \textit{Fault Tolerance} by actively identifying and correcting errors at runtime, supports \textit{Scalability} by ensuring the system remains reliable even as it grows in size or complexity, and boosts \textit{Performance} by preventing failures from propagating. However, the need for additional ancilla Qubits and control logic adds significant \textit{Complexity} to circuit design and resource management. If substantial measurement errors are detected that compromise output accuracy, \textbf{Readout Error Mitigation} is applied. It enhances \textit{Reusability} by calibrating corrections across multiple experiments, improves \textit{Accuracy} through statistical post-processing of erroneous measurements, and increases \textit{Adaptability} by optimizing correction techniques for specific hardware. However, as the system size increases, the \textit{Scalability} of the chosen error mitigation strategy may be limited by the growing complexity of error models and correction matrices. Alternatively, if noisy gate operations present a major challenge, \textbf{Gate Error Mitigation} is selected. It supports \textit{Compatibility} by aligning correction techniques with diverse quantum hardware and gate sets. Yet, it may negatively affect \textit{Performance} by requiring repeated circuit executions or calibration, and increase \textit{Complexity} due to the extra layers of error modeling and compensation logic.


\textcolor{black}{For dynamic adaptation to quantum behaviors, the decision model navigates to an \textbf{Exclusive Gateway}. When the system benefits from seamlessly integrating error correction into quantum algorithms, \textbf{Decorator Design Pattern} is selected. This pattern enhances \textit{Fault Tolerance} by allowing additional behaviors (e.g., correction or validation) to be layered onto core operations without altering the algorithm itself. It also improves \textit{Reliability} by making the system more responsive to changing fault conditions during runtime.} If the system requires a higher level of abstraction by encapsulating quantum operations within structured mathematical constructs, \textbf{Quantum Patterns of Behavior (qPoB)} is selected. It supports \textit{Reusability} by enabling standardized quantum behaviors across different contexts, and enhances \textit{Modularity} by separating high-level behavior from low-level operations. It facilitates \textit{Scalability} by abstracting complexities and supports \textit{Adaptability} through interchangeable behavior compositions. On the downside, its layered abstraction can increase \textit{Complexity}, reduce \textit{Performance} due to interpretative overhead, and pose challenges for \textit{Maintainability}, especially for developers unfamiliar with abstract models of quantum operations or system behaviors, which are generalized, high-level representations used to design and understand quantum systems but may lack the detail necessary for direct implementation.

The Fault Tolerance Decision Model systematically addresses fault tolerance requirements and balances QAs to recommend suitable architecture patterns and strategies for quantum software systems.

\subsection{Integration and Optimization Decision Model}
\begin{figure}[h!]
\centering
\includegraphics[width=0.9\linewidth]{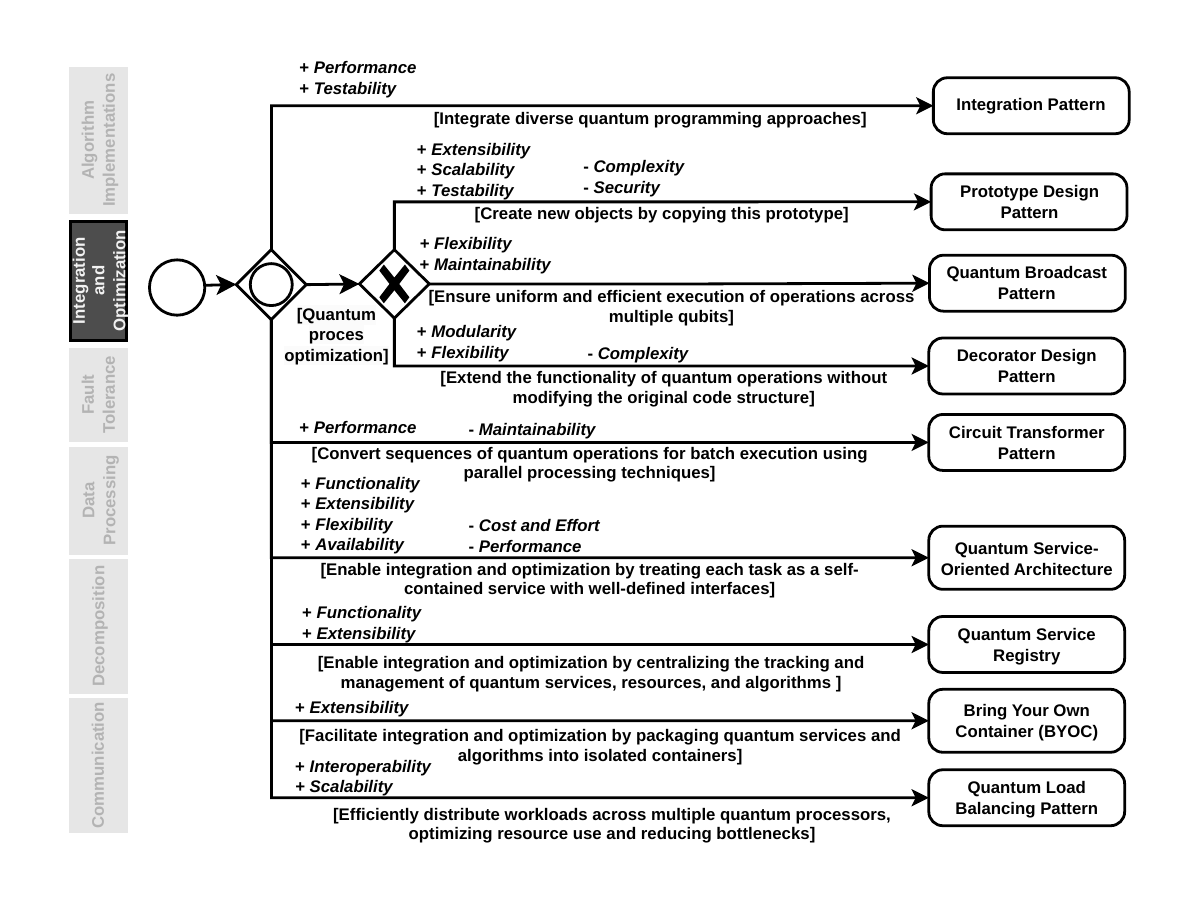}
\caption{Decision model for selecting architecture patterns and strategies for integration and optimization}
\label{Fig: Integration and Optimization}
\end{figure}

The Integration and Optimization Decision Model for quantum software systems follows a structured decision-making process using \textbf{Gateways} to guide the selection of appropriate patterns and strategies based on specific conditions. Table \ref{Tab: IntegrationOptimization} lists the patterns and strategies covered by the Integration and Optimization Decision Model (see Figure~\ref{Fig: Integration and Optimization}). 

\begin{table}[h!]
\centering
\caption{Architecture patterns and strategies for integration and optimization}
\label{Tab: IntegrationOptimization}
\small
\setlength{\tabcolsep}{4pt}
\renewcommand{\arraystretch}{1.1}
\resizebox{\textwidth}{!}{%
\begin{tabular}{lp{10cm}}
\hline
\textbf{Pattern Name} & \textbf{Summary} \\
\hline
Integration Pattern                           & Enable the combination of diverse quantum programming approaches into a unified software workflow. \\\hline
Prototype Design Pattern                      & Support creating new quantum software objects by cloning existing prototypes, enabling faster development and reuse. \\\hline
Quantum Broadcast                     & \textcolor{black}{Facilitate the application of identical quantum operations uniformly across multiple Qubits, enabling logically synchronized execution without violating the no-cloning theorem.} \\\hline
Decorator Design Pattern                      & Allow adding new functionalities to quantum operations dynamically without altering their core structure. \\\hline
Circuit Transformer Pattern                   & \textcolor{black}{Provide a reusable architectural abstraction for transforming quantum circuits or operation sequences into executable representations, supporting pipeline-based processing such as compilation, optimization, and batching in quantum software systems.} \\\hline
Quantum Service-Oriented Architecture & \textcolor{black}{Structure quantum tasks as loosely coupled, independently deployable services using a combination of service decomposition, interface-based communication, and orchestration patterns to support modular integration.}  \\\hline

Quantum Service Registry                      & Centralize the cataloging and discovery of quantum services, resources, and algorithms to simplify integration. \\\hline
Bring Your Own Container (BYOC)       & Allow packaging quantum services and custom libraries into isolated, reusable containers for flexible deployment. \\\hline
Quantum Load Balancing Pattern                & Distribute quantum computational tasks dynamically across multiple providers or systems to optimize resource usage. \\
\hline
\end{tabular}%
}
\end{table}

To select patterns and strategies for integrating diverse quantum programming models, this decision model begins with an \textbf{Inclusive Gateway} that offers parallel paths to address different concerns. If seamless coordination of hybrid components is required, \textbf{Integration} is selected. It enhances \textit{Performance} by reducing overhead in inter-component communication and supports \textit{Testability} by establishing well-defined interface boundaries between classical and quantum parts.

For selecting patterns and strategies when quantum process optimization is the focus, the decision flow proceeds through an \textbf{Exclusive Gateway} that activates a specific strategy based on the optimization needs. If creating new configurations by replicating existing templates is beneficial, \textbf{Prototype Design} is applied. It supports \textit{Extensibility} by allowing reuse of established structures, improves \textit{Scalability} through quick duplication of logic, and enhances \textit{Testability} by enabling controlled variation of object states. However, it may reduce \textit{Compatibility} when reused prototypes do not align with hardware constraints, and compromise \textit{Security} if sensitive logic is copied without isolation. \textcolor{black}{\textbf{Quantum Broadcast} is selected when uniform execution of quantum operations across multiple qubits is required. It allows for synchronized updates across qubits, improving \textit{Flexibility} by enabling uniform changes. This pattern also supports \textit{Maintainability} by simplifying the propagation of updates. However, applying identical operations across all qubits may reduce \textit{Precision} in systems where individual qubits require specific handling, and may introduce challenges in \textit{Compatibility} if reused prototypes do not align with hardware constraints. Additionally, there may be \textit{Security} concerns if sensitive logic is not properly isolated, although no quantum state cloning occurs in this process.} Alternatively, if extending functionality without altering core components is necessary, \textbf{Decorator Design} is used. It promotes \textit{Modularity} by attaching features dynamically and enhances \textit{Flexibility} by allowing layered behavior changes. However, it increases \textit{Complexity} by making layered decorators harder to track, complicating debugging and logic tracing.

\textcolor{black}{For selecting patterns and strategies that support integration and optimization in quantum software systems, the decision model returns to an \textbf{Inclusive Gateway} to evaluate system-wide concerns such as modularity, service orchestration, and performance tuning. If the objective is to transform quantum circuits or sequences of quantum operations into execution-ready representations, \textbf{Circuit Transformer Pattern} is adopted. It provides a reusable architectural abstraction for circuit transformation within the quantum software pipeline, supporting compilation-related processing, optimization, and controlled execution. The pattern improves \textit{Performance} by enabling more efficient execution structures and reducing redundant operations and circuit overhead. However, the additional transformation layer introduces complexity that may reduce \textit{Maintainability}, particularly in debugging and adaptation of transformation logic.} \textcolor{black}{To integrate diverse quantum tasks as independently managed services, \textbf{Quantum Service-Oriented Architecture} is considered. It improves \textit{Functionality} by allowing services to encapsulate specific logic, supports \textit{Extensibility} through service reuse, increases \textit{Flexibility} by enabling component reconfiguration, and boosts \textit{Availability} through service redundancy. However, the need for service interface definitions, orchestration, and infrastructure setup increases \textit{Cost and Effort} and can degrade \textit{Performance} due to communication overhead between services.} To manage and orchestrate these services more efficiently, \textbf{Quantum Service Registry} is selected. It improves \textit{Functionality} by maintaining a centralized catalog of available services, endpoints, and metadata. It also supports \textit{Extensibility} by allowing new services to be dynamically registered and discovered, promoting modular growth of the system. \textcolor{black}{If packaging services and algorithms into isolated environments is required for deployment flexibility, \textbf{Bring Your Own Container (BYOC) Pattern} can be adopted \cite{tsymbalista2025toward}. This pattern is grounded in container-based deployment strategies commonly used in quantum runtime architectures, where custom execution environments are packaged as containers and deployed across distributed quantum-classical infrastructures. It enhances \textit{Extensibility} by enabling developers to integrate custom runtime dependencies and libraries without affecting system-wide configurations. However, it increases system-level \textit{Complexity} by introducing additional architectural components and management concerns, such as container isolation, scheduling, and deployment coordination.} To distribute the computational load across multiple quantum processors, \textbf{Quantum Load Balancing} is selected. It improves \textit{Interoperability} by enabling smooth coordination between different hardware backends and execution queues.  It also enhances \textit{Scalability} by enabling the system to adjust the number of active quantum processors based on workload demands, thereby supporting efficient scaling of quantum services in hybrid quantum-classical environments.


The Integration and Optimization Decision Model offers a structured approach for selecting architecture patterns and strategies in quantum software systems, aligning integration and optimization needs with trade-offs among key QAs, including performance, scalability, flexibility, and testability.

\subsection{Algorithm Implementation Decision Model}
\begin{figure}[h!]
\centering
\includegraphics[width=0.9\linewidth]{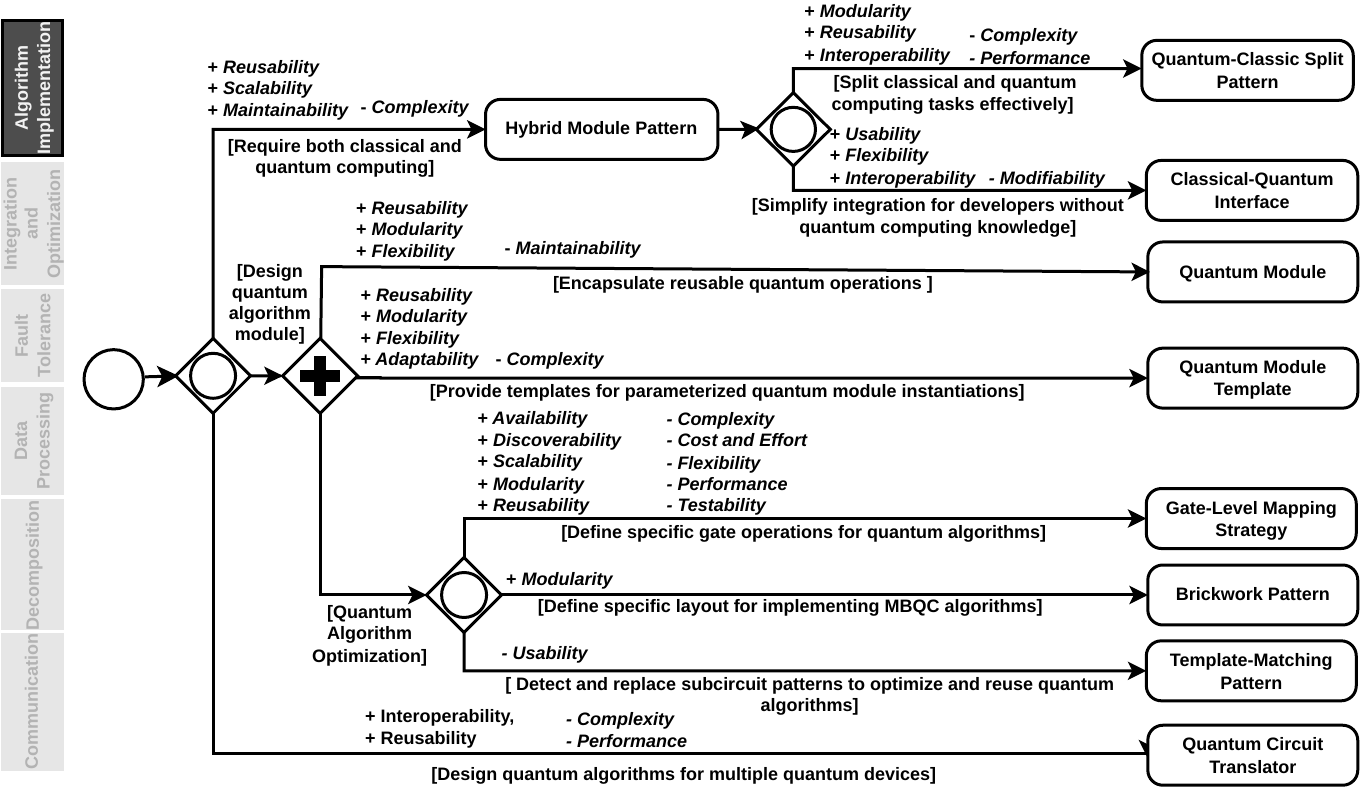}
\caption{Decision model for selecting architecture patterns and strategies for algorithm implementation}
\label{Fig: Algorithm Implementation}
\end{figure}

The Algorithm Implementation Decision Model is designed to guide practitioners through systematically selecting patterns and strategies for implementing quantum algorithms, ensuring that solutions align with specific project requirements and QAs. The decision model initiates with an \textbf{Inclusive Gateway}, allowing multiple paths to be explored simultaneously. Table \ref{Tab: AlgorithmImplementation} lists the patterns and strategies covered by the Algorithm Implementation Decision Model (see Figure~\ref{Fig: Algorithm Implementation}). 

\begin{table}[h!]
\centering
\caption{Architecture patterns and strategies for algorithm implementation}
\label{Tab: AlgorithmImplementation}
\small
\setlength{\tabcolsep}{4pt}
\renewcommand{\arraystretch}{1.1}
\resizebox{\textwidth}{!}{%
\begin{tabular}{lp{12cm}}
\hline
\textbf{Pattern Name} & \textbf{Summary} \\
\hline
Hybrid Module Pattern         & Package both quantum and classical parts of an algorithm into a single module with control flow to manage their orchestration. \\\hline
Quantum-Classic Split Pattern & Separate the execution of classical and quantum tasks into distinct components to support hybrid computation. \\\hline
Classical-Quantum Interface   & Provide a standardized interface that conceals quantum implementation details and converts problem-specific inputs for quantum processing. \\\hline
Quantum Module Pattern        & Encapsulate reusable quantum code that generates quantum circuits based on provided inputs to promote modularity. \\\hline
Quantum Module Template       & Define generic templates for quantum algorithm components that can be instantiated with varying inputs or submodules. \\\hline
Gate-Level Mapping Strategy           & \textcolor{black}{Specify the selection, composition, and sequencing of fundamental quantum gate operations (e.g., single-qubit and two-qubit gates) applied to qubits for building and optimizing quantum circuits.} \\\hline
Brickwork Pattern             & Arrange quantum operations in a grid-like structure supporting measurement-based quantum computing (MBQC). \\\hline
Template-Based Circuit Rewriting Strategy     & \textcolor{black}{Detect predefined subcircuit patterns and rewrite them using equivalent implementations to enable reusable algorithm construction and optimization.}
\\\hline
Quantum Circuit Translator    & Convert quantum circuits into target programming languages and transpile them to supported hardware instruction sets. \\
\hline
\end{tabular}%
}
\end{table}

To select patterns and strategies that support seamless coordination between classical and quantum components, the decision model starts with the \textbf{Hybrid Module}. It improves \textit{Reusability} by encapsulating classical-quantum interactions in shared modules that can be reused across different applications. It supports \textit{Scalability} by allowing independent extension of classical or quantum parts without restructuring the entire system. It also enhances \textit{Maintainability} by separating concerns between classical and quantum components. However, managing dependencies across dual components increases \textit{Complexity}, especially when synchronizing their execution flows. Next, an \textbf{Exclusive Gateway} directs the decision flow based on integration needs. If the goal is to distinctly divide classical and quantum responsibilities, \textbf{Quantum-Classic Split} is selected. It enhances \textit{Modularity} by separating the logic into discrete parts, thereby enabling focused development. It improves \textit{Reusability} by allowing each component to be applied independently in other contexts. It also strengthens \textit{Interoperability} by enabling integration with other systems that interact with either the quantum or classical part. However, managing communication between two isolated parts introduces \textit{Complexity} and may reduce \textit{Performance} due to data transfer overhead. Alternatively, if simplifying integration for developers with limited quantum expertise is a priority, \textbf{Classical-Quantum Interface} is selected. It improves \textit{Modularity} by offering defined boundaries between classical and quantum logic. It supports \textit{Reusability} by allowing interface components to be reused across projects with similar interaction needs. It enhances \textit{Interoperability} by providing standard access points to quantum resources. Yet, the abstraction layers and conversion logic add \textit{Complexity} and reduce \textit{Performance} because of translation and coordination delays.

For selecting patterns and strategies for designing quantum algorithm modules, an \textbf{Inclusive Gateway} enables parallel optimization approaches through a \textbf{Parallel Gateway}. One path encapsulates reusable quantum operations using the \textbf{Quantum Module}, which improves \textit{Reusability} by packaging operations into isolated modules, supports \textit{Modularity} by separating concerns into well-defined units, and promotes \textit{Flexibility} by enabling easy substitution of operations. However, managing these modules over time may pose \textit{Maintainability} challenges as dependencies and updates grow more complex. In parallel, \textbf{Quantum Module Template} provides parameterized blackprints for module instantiations. This improves \textit{Reusability} by enabling consistent reuse of parameter-driven modules, enhances \textit{Modularity} through structured construction, increases \textit{Flexibility} by supporting variation in instantiation, and boosts \textit{Adaptability} by allowing dynamic binding to specific use cases. Still, defining and maintaining these templates may add \textit{Complexity} due to template hierarchies and configuration overhead. \textcolor{black}{In optimizing quantum algorithms, another \textbf{Inclusive Gateway} activates additional paths. If defining fine-grained gate-level behavior is critical, the \textbf{Gate-Level Mapping Strategy} is selected. This strategy refers not to a specific gate type but to a reusable design strategy for orchestrating low-level gate operations on qubits. It improves \textit{Availability} by enabling low-level customization, enhances \textit{Discoverability} through explicit operation definitions, supports \textit{Scalability} by aligning with fine-grained control in large quantum systems, and benefits \textit{Modularity} and \textit{Reusability} through composable gate blocks. However, it introduces \textit{Complexity} in quantum software design, increases \textit{Cost and Effort} due to manual configuration, limits \textit{Flexibility} in higher-level quantum programming abstractions, reduces \textit{Performance} owing to control overhead, and complicates \textit{Testability} due to low-level verification needs.} Alternatively, when targeting MBQC, \textbf{Brickwork} is adopted. Its structured grid layout improves \textit{Modularity} by mapping quantum logic to a fixed computational structure that supports predictable configuration. \textcolor{black}{When quantum algorithms require reusable components, the model suggests using the \textbf{Template-Based Circuit Rewriting Strategy}. This strategy involves applying predefined rewrite rules to identify and replace subcircuits in quantum algorithms with equivalent implementations. By automating the pattern-matching process, it enables more efficient construction and optimization of quantum algorithms. However, this strategy may reduce \textit{Usability} by introducing rigid structures that constrain the developers' ability to customize or optimize quantum algorithms.}


For selecting implementation patterns and strategies when algorithms must be compatible with multiple quantum devices, the decision flow proceeds to \textbf{Quantum Circuit Translator}. This improves \textit{Reusability} by allowing circuits to adapt across diverse backends and enhances \textit{Interoperability} by translating device-specific representations. However, it increases \textit{Complexity} due to the need for dynamic translation layers and reduces \textit{Performance} through additional compilation and transformation overhead.

Through this structured decision flow of \textbf{Inclusive}, \textbf{Exclusive}, and \textbf{Parallel Gateways}, the Algorithm Implementation Decision Model systematically supports algorithm implementation by evaluating context-specific requirements and balancing key QAs, such as reusability, interoperability, modularity, and performance, to recommend suitable architecture patterns and strategies for quantum software systems.

\section{Evaluation}\label{EvaluationOfDecisionModels}
To evaluate the proposed decision models, we conducted semi-structured interviews with 30 practitioners using a standardized interview questionnaire\footnote{\url{https://tinyurl.com/bdcvzjjb}}. The results of this evaluation are organized into two parts: (1) participant demographics and professional background, and (2) evaluation of the decision models with respect to familiarity, understandability, and correctness.

\subsection{Participant Demographics and Professional Background}\label{Demographics}
\begin{figure}[h!]
\centering
\includegraphics[width=1.0 \linewidth]{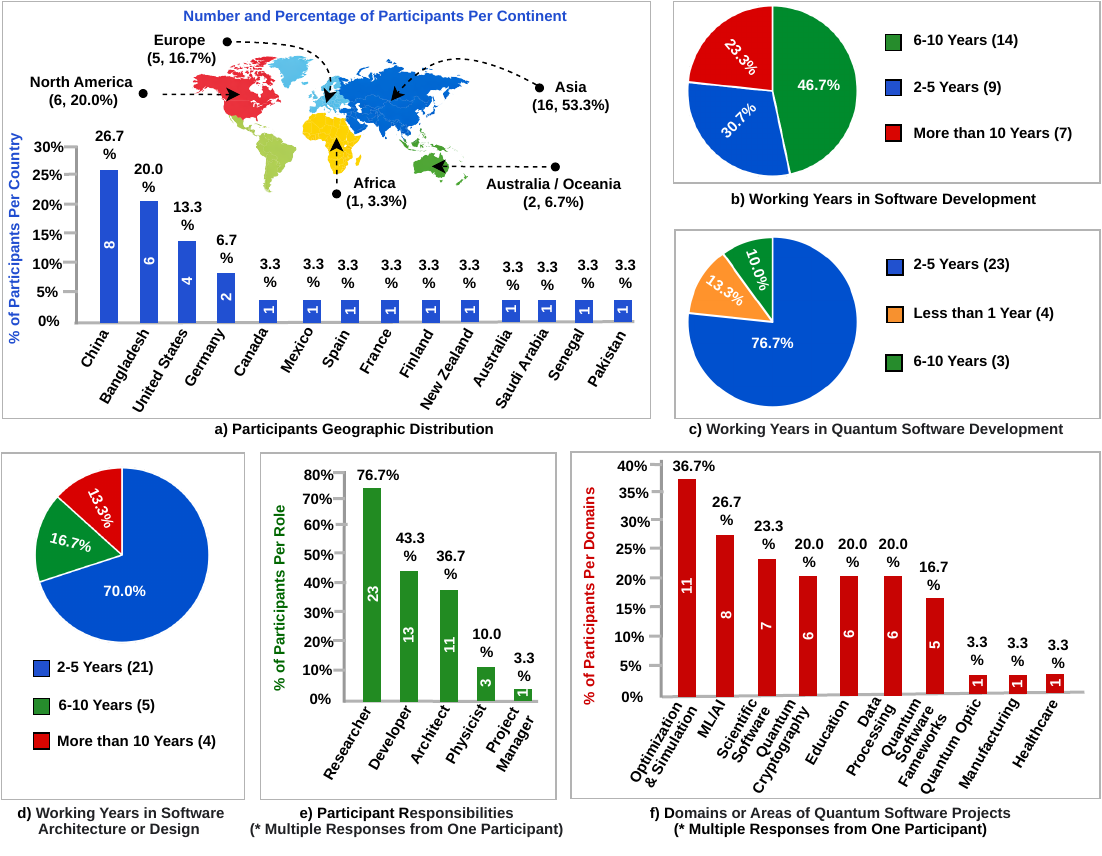}
\caption{Demographic and professional profile of interview participants (n=30)}
\label{DemographyDetails}
\end{figure}

We present an overview of the demographics and professional experience of the 30 interview participants in Figure \ref{DemographyDetails}, 
which provides essential context for interpreting the feedback provided on the decision models.

\textbf{Geographic Distribution}: \textcolor{black}{The participants represent a geographically diverse group across five continents: Asia, North America, Europe, Australia/Oceania, and Africa. The majority are from Asia with 53.3\% (16 participants), followed by North America with 20.0\% (6 participants), and Europe with 16.7\% (5 participants). Australia/Oceania accounts for 6.7\% (2 participants), and Africa accounts for 3.3\% (1 participant). At the country level, China leads with 26.7\% (8 participants), Bangladesh with 20.0\% (6 participants), and the United States with 13.3\% (4 participants), with the remaining participants distributed across several other countries (see Figure \ref{DemographyDetails}(a) Participants Geographic Distribution).}



\textbf{Experience in Software Development}: \textcolor{black}{Most respondents (14 out of 30 participants, 46.7\%) reported having 6--10 years of software development experience, while 30.0\% (9 participants) reported having 2--5 years of experience. Notably, a substantial segment,  representing 23.3\% (7 participants), reported more than 10 years of software development experience. This distribution indicates a strong foundation in established software engineering principles, providing a reliable basis for evaluating practices within the emerging quantum domain (see Figure~\ref{DemographyDetails}(b) Working Years in Software Development).}
 

\textbf{Experience in Quantum Software Development}: \textcolor{black}{The majority of participants (76.7\%, 23 out of 30) report 2--5 years of quantum software experience. An additional 10.0\% (3 participants) bring 6--10 years of experience. The remaining 13.3\% (4 participants) have less than one year of experience. This distribution reflects the field's rapid maturation: since quantum software development began gaining significant industrial and academic traction around 2017–2019, marked by the release of foundational frameworks such as IBM's Qiskit \cite{ibm_qiskit_history}, with 2--5 years of dedicated experience represent a meaningful and substantive portion of the field's history (see Figure~\ref{DemographyDetails}(c) Working Years in Quantum Software Development).}


\textbf{Experience in Software Architecture or Design}: \textcolor{black}{Regarding experience in software architecture or design, 70.0\% (21 participants) reported having 2--5 years of experience, followed by 16.7\% (5 participants) with 6--10 years, and 13.3\% (4 participants) with more than 10 years (see Figure~\ref{DemographyDetails}(d) Working Years in Software Architecture or Design).}

\textbf{Professional Roles}: \textcolor{black}{During the interviews, participants reported holding a variety of professional roles within their organizations, reflecting the interdisciplinary nature of the quantum software field. Researcher was the most frequently reported role, mentioned by 76.7\% of participants (23 respondents), followed by Developer at 43.3\% (13 respondents), Architect at 36.7\% (11 respondents), Physicist at 10.0\% (3 respondents), and Project Manager at 3.3\% (1 respondent) (see Figure \ref{DemographyDetails}(e) Participant Responsibilities). Given that some participants reported involvement in multiple roles, the total percentage and the number of participants exceed 100\% and 30.}


\textbf{Domains of Quantum Software Projects}: \textcolor{black}{The participants reported involvement in various domains of quantum software projects. Optimization \& Simulation is the most frequently mentioned area, with 36.7\% (11 participants), followed by ML/AI with 26.7\% (8 participants) and Scientific Software with 23.3\% (7 participants). Quantum Cryptography, Education, and Data Processing were each mentioned by 20.0\% (6 participants), and Quantum Software Frameworks by 16.7\% (5 participants). Domains such as Quantum Optics, Manufacturing, and Healthcare were each mentioned by 3.3\% (1 participant) (see Figure \ref{DemographyDetails}(f) Domains or Areas of Quantum Software Projects). Given that some participants reported involvement in multiple domains, the total percentage and the number of participants exceed 100\% and 30.}
                                               


\subsection{Practitioner Feedback on Decision Models}\label{ModelFeedback}
Below, we discuss practitioners’ assessments of the proposed decision models. This includes their familiarity with the patterns and strategies, the perceived understandability and ease of use, the correctness and sufficiency of the decision models in supporting architectural decisions, and suggestions for improvement. \textcolor{black}{The evaluation is reported at the decision-model level, as the 62 consolidated architecture patterns and strategies are organized into six design areas. To retain pattern-level insight, participants were also asked: ``\textit{Which of the following architecture patterns or strategies have you implemented in your quantum software systems for each decision model?}''. However, reporting fully disaggregated quantitative results at the individual-pattern level was not feasible due to the large number of patterns and strategies (62 in total), as it would significantly increase analysis complexity, impose a substantial burden on the participants during interviews, and make it difficult to maintain a manageable interview duration. Therefore, the evaluation in this study was designed to assess the overall usefulness and applicability of the proposed decision models, rather than to validate each individual architecture pattern or strategy separately.} The results reflect both the strengths and areas for potential refinement across the six decision models presented in Section \ref{DecisionModelsSection}.

\begin{figure}[h!]
    \centering
    \includegraphics[width=1.0 \linewidth]{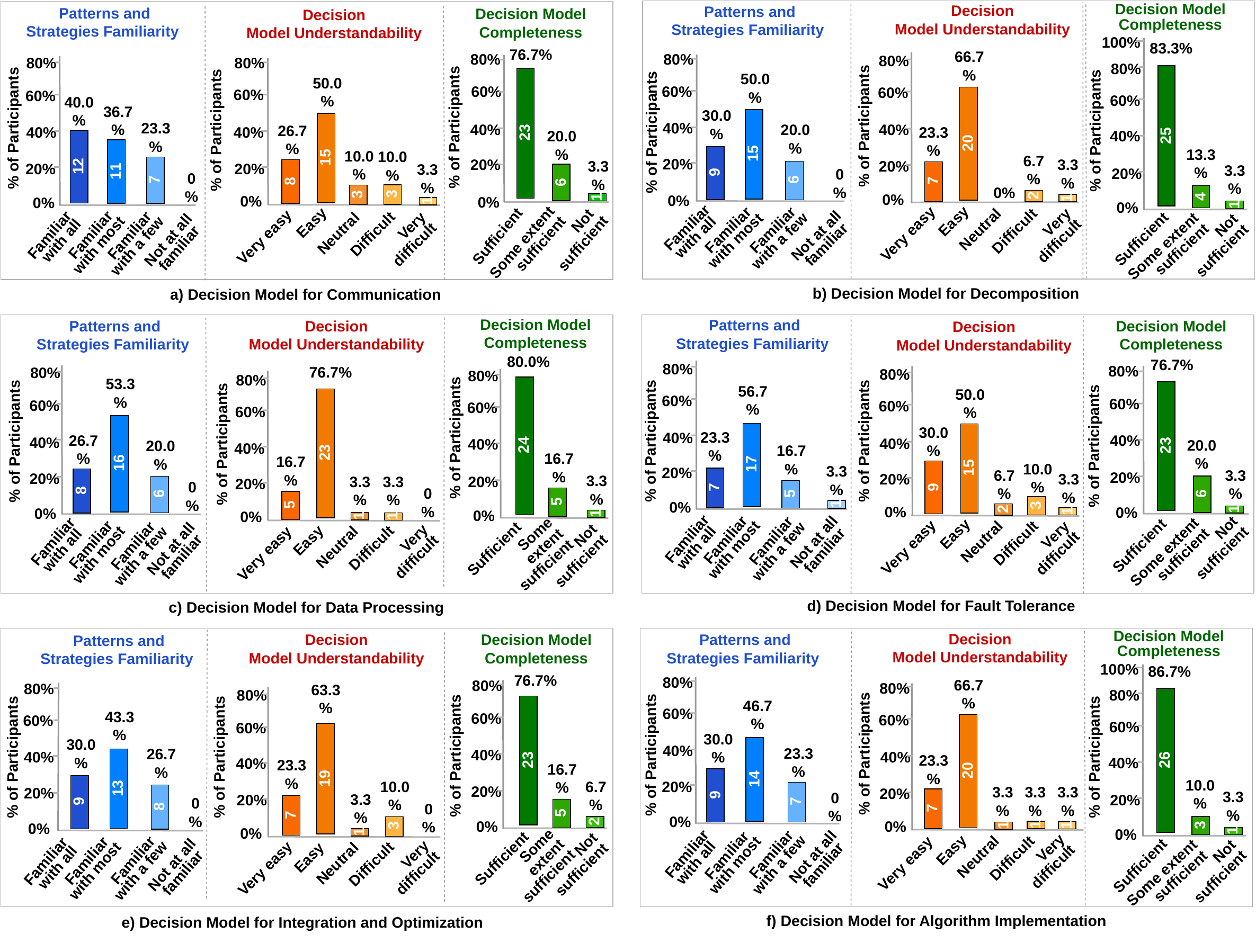}  
    \caption{Overview of practitioners’ responses for familiarity, understandability, and completeness of the six decision models}
    \label{ResultsInterviewQuestions}
\end{figure}

\textcolor{black}{
\textbf{Familiarity with Patterns and Strategies}: During the interviews, we asked participants whether they were familiar with the patterns and strategies used in the six decision models, ``\textit{Are you familiar with the patterns and strategies used in the six decision models?}''. The majority reported a high level of familiarity with the architecture patterns and strategies (Figure \ref{ResultsInterviewQuestions}). Specifically, 80.0\% (24 participants) were familiar with most or all patterns in the Decomposition, Data Processing, and Fault Tolerance Decision models, 76.7\% (23 participants) in the Communication and Algorithm Implementation Decision models, and 73.3\% (22 participants) in the Integration and Optimization Decision model. These results indicate that participants were broadly familiar with all six decision models. A smaller proportion reported familiarity with only a few patterns and strategies, ranging from 16.7\% to 26.7\% across the models, while only one participant (3.3\%) reported no familiarity with the Fault Tolerance Decision model. To support participants with limited familiarity, we provided a detailed document describing the patterns and strategies, including examples and contextual explanations\footnote{\url{https://tinyurl.com/2fwcbkmh}}. These materials helped participants engage with the decision models and suggested that the decision models can serve not only as decision-support tools but also as learning resources for practitioners with varying levels of expertise.}

\textcolor{black}{\textbf{Understandability of Decision Models}: To evaluate the understandability of the proposed decision models, participants were asked: ``\textit{To what extent are the six decision models easy to understand and use?}''. The results indicate strong perceived clarity and usability across all models (see Figure \ref{ResultsInterviewQuestions}). The Data Processing Decision Model receives the highest rating, with 93.4\% (28 participants) rating it as \textit{easy} or \textit{very easy}, followed by Decomposition and Algorithm Implementation (90.0\%, 27 participants each), Integration and Optimization (86.6\%, 26 participants), Fault Tolerance (80.0\%, 24 participants), and Communication (76.7\%, 23 participants). \textit{Neutral} ratings are uncommon, 3.3\%–10.0\%, and \textit{difficult} or \textit{very difficult} ratings are reported by only 3.3\%–13.3\% of participants across the models. Overall, the findings show that a substantial majority of participants found all six decision models \textit{easy} or \textit{very easy} to understand and use, indicating strong clarity and usability.}

\textcolor{black}{\textbf{Completeness of Decision Models}: To assess the sufficiency of the proposed decision models, we asked participants whether each decision model sufficiently supports architectural decision-making, ``\textit{Does the information in the six decision models sufficiently support making architecture decisions about quantum software systems?}''. The results show that all six models were perceived as largely sufficient (Figure \ref{ResultsInterviewQuestions}). The Algorithm Implementation Model receives the highest sufficiency rating, with 86.7\% (26 participants), followed by Decomposition 83.3\% (25 participants), Data Processing 80.0\% (24 participants), and Communication, Fault Tolerance, and Integration and Optimization 76.7\% (23 participants each). Between 10.0\% and 20.0\% of participants considered the decision models sufficient to some extent, while only 3.3\%–6.7\% rated them as not sufficient. Overall, the low proportion of insufficient ratings indicates that the proposed decision models comprehensively cover the key architectural concerns required for decision-making in specific design areas of quantum software systems.}

\textcolor{black}{\textbf{Correctness of decision models}: We assessed the perceived correctness of the proposed decision models by asking the interviewees: ``\textit{Are these decision models correct? If not correct, please indicate the problem(s)}''. The results show strong consensus regarding the correctness of the decision models, with the vast majority agreeing that they align with practical experience in quantum software development. Only 3.3\%–6.7\% of participants considered the decision models partially correct or expressed uncertainty due to limited familiarity with certain patterns and strategies. The participants also identified a few minor issues, including missing decision conditions, unclear complementary relationships, duplicated patterns and strategies, and minor inaccuracies in QA mappings. Based on this feedback, we refined the decision logic, improved relationship representation, removed redundancies, and corrected QA mappings, thereby enhancing the correctness and reliability of the decision models.}

\textcolor{black}{\textbf{Usefulness for Quantum Software Design and Development}: Regarding usefulness, we asked: ``\textit{Are these decision models useful to support making decisions in the design and development of quantum software systems? Why?}''. Overall, the participants agreed that the six decision models provide structured guidance for addressing key architectural challenges in quantum software systems. They emphasized that the decision models help systematically select appropriate patterns and strategies while considering domain-specific constraints such as quantum-classical interactions, data encoding, and system scalability. In particular, the participants highlighted that the decision models support architectural decision-making by making trade-offs explicit and by organizing complex design choices into manageable steps. For instance, P28 noted: \faComment ``\textit{So, it makes sense and is efficient. This communication decision model is indeed useful}''. Similarly, the importance of data-related decisions was emphasized by P25: \faComment ``\textit{Data Processing Decision Model is highly useful. The most critical part is the comparison of quantum encoding patterns (Amplitude, Angle, Basis, QRAM). Choosing the right encoding is fundamental to algorithm performance, and this model provides the necessary decision criteria and trade-offs}''. These responses indicate that the decision models not only guide practitioners in selecting suitable architectural solutions but also improve confidence in making design decisions by clarifying alternatives and their implications.}

\textcolor{black}{\textbf{Usefulness for Architectural Evaluation of Quantum Software Systems}: We further asked: ``\textit{Are these decision models useful in the architectural evaluation of quantum software systems? Why?}''. The majority of participants confirmed that the decision models are valuable for evaluating architectural choices, particularly by linking patterns and strategies with relevant QAs such as performance, scalability, reliability, and efficiency. For example, P27 stated: \faComment ``\textit{Yes. The Communication Decision Model can support architectural evaluation by explicitly linking architecture patterns with QAs such as performance, scalability, and reliability, which are key concerns in quantum software systems}''. P17 highlighted the broader impact of decision models on system-level considerations: \faComment ``\textit{I'd say yes. This decomposition decision strongly affects strategy, a deployment pipeline, and applying quantum of implementation; this model can be used to address ambiguity}''. Overall, the participants indicated that the six decision models facilitate a more systematic and transparent evaluation process by enabling architects to assess how different design choices influence system qualities. This supports more informed architectural decisions and helps identify potential trade-offs and optimization opportunities across the quantum software lifecycle.}

\textbf{Suggestions for Improving Decision Models}: We asked the participants to offer suggestions for improving the decision models, ``\textit{What are your suggestions to improve the decision models?}''. Most participants believed that the decision models are sufficient for guiding the architecture design of quantum software systems, particularly in selecting appropriate architecture patterns and strategies. They attributed this effectiveness to several factors: the presence of constraints on patterns, the necessary conditions that these patterns must fulfill, the trade-offs concerning QAs, and the influence of both patterns and strategies on these QAs. For instance, P13 mentioned that \faComment ``\textit{Yes, I believe all the models are very effective, especially for guiding optimization. These models will be highly useful in practical quantum computing projects}''. Finally, participants also provided several suggestions. These include presenting patterns with code, measuring quantitative values for QAs, and using decision models in industrial quantum software projects. These suggestions address issues such as missing conditions, unclear \textit{complement} relations, duplicate patterns and strategies, bidirectional influence, and incorrect impact of patterns on QAs. We fixed these issues in the updated version of the decision models. For instance, one participant (P2) suggested that \faComment ``\textit{Providing a more detailed breakdown of potential trade-offs for each pattern could help users better understand their impact on performance, system stability, and resource utilization}''. \textcolor{black}{We note that, to support grounded evaluation, all patterns and strategies in our study are derived from empirical evidence, including mined GitHub repositories, Stack Exchange discussions, and the systematic literature review. Additionally, a detailed document describing the architecture patterns and strategies, along with practical examples, was shared with participants prior to the interviews and discussed during the sessions. This ensured that practitioners assessed patterns and strategies based on concrete, real-world instances rather than abstract definitions.}

\section{Discussion} \label{Discussion}
\textcolor{black}{In this section, we discuss the evaluation and implications of the proposed decision models based on practitioner feedback. Specifically, Section \ref{Disscussion:CommunicationDecision Model} to Section \ref{Disscussion:AlgorithmImplementationDecisionModel} discuss each decision model respectively, and Section \ref{InterdependenciesBetweenModels} discusses the interdependencies between decision models, including QA propagation and their combined application in real-world hybrid quantum-classical systems.}

\subsection{Communication Decision Model} \label{Disscussion:CommunicationDecision Model}
\textbf{Decisions related to communication design must account for hardware constraints and evolving protocols}: The evaluation of the Communication Decision Model reveals its strong applicability in supporting architecture-level decisions in quantum software systems, particularly by providing structured guidance through reusable patterns such as Entanglement-Assisted Channels, Quantum API Gateways, and Workflow Orchestration. The majority of practitioners agreed that this decision model enhances modularity and scalability in quantum software design. One interviewee (P11) explained that the model \faComment ``\textit{provides a structured framework for selecting communication patterns, making architectural decisions more informed, consistent, and optimized}''. Another participant (P6) stated that \faComment ``\textit{It clearly outlines trade-offs in performance, scalability, security, and complexity, helping balance system constraints}''. These observations suggest that this decision model has practical value in guiding early design decisions. However, despite these strengths, several practitioners pointed out a crucial limitation: the current decision model does not fully account for the variability and constraints of quantum hardware. For instance, a participant (P13) emphasized that \faComment ``\textit{As hardware implementations vary, the communication model must evolve or adapt to meet the specific requirements and constraints of each platform}''. Practitioners also raised concerns about the absence of universally applicable communication protocols, particularly in hybrid quantum–classical systems, where coherence and fidelity can be compromised. This highlights a significant challenge: \textit{quantum communication patterns are not one-size-fits-all}. As systems become increasingly complex and distributed, the need for context-aware communication strategies that take into account the underlying hardware becomes critical. This presents an important opportunity for future research, which can focus on extending communication decision models with adaptive logic that incorporates quantum hardware specifications (e.g., topology, qubit fidelity, decoherence rate). Future research could explore how to formalize such context-awareness through constraint-based decision models or learning-enhanced decision support. \textcolor{black}{For practitioners, the key takeaway is that applying communication strategies without considering hardware-specific factors may lead to inefficiencies or incompatibilities in system performance. As architecture patterns provide abstract solution templates rather than concrete implementations, practitioners should interpret communication patterns as flexible blackprints instead of fixed recipes, adapting them to system characteristics, hardware constraints, and the evolving demands of quantum–classical integration.}


\subsection{Decomposition Decision Model}\label{Disscussion:DecompositionDecisionModel}
\textbf{Decomposition patterns must enable hardware-awareness and scalable modularity in hybrid quantum systems}: The Decomposition Decision Model was widely perceived as useful for guiding architectural decisions in quantum software systems, particularly in scenarios where hybrid quantum classical integration is prevalent. Many practitioners emphasized that while decomposition is a well-known concept in classical software engineering, quantum decomposition requires unique considerations given the nascent, hardware-sensitive nature of quantum systems. For example, one participant (P1) noted \faComment ``\textit{Quantum computing is highly sophisticated and complex, which makes a one-size-fits-all decomposition approach insufficient. We need to tailor decomposition techniques to specific tasks such as algorithm implementation or fault tolerance}''. Others highlighted that, just as classical systems evolve from monolithic to microservices-based architectures, quantum software will inevitably require structured decomposition to ensure scalability, flexibility, and maintainability. Interestingly, several participants emphasized that decomposition patterns, such as Quantum Classical Split and Quantum Microservices, already facilitate modular development and smoother interaction between classical and quantum layers, an essential requirement for current hybrid architectures. A participant (P10) explained that \faComment ``\textit{Since most quantum applications today still rely on classical quantum hybrid systems, decomposition is essential to ensure smooth interaction between components}''. This aligns with a common theme across interviews: decomposition decisions must be hardware-aware. As one participant (P4) pointed out \faComment ``\textit{If any type of hardware has any kind of issue with the decomposition process, that will create problems when implementing algorithms on that hardware}''. This suggests that decomposition in quantum systems is not only about logical separation, but also about compatibility with physical constraints, e.g., connectivity, qubit coherence, and gate fidelity. While the decision model was largely rated as sufficient and easy to understand by most participants \textcolor{black}{(66.7\% \textit{Easy}; 23.3\% \textit{Very easy})}, several suggestions were raised to enhance its applicability. Participants suggested various approaches, including a cost–benefit analysis framework to weigh trade-offs between system complexity, resource consumption, and latency. Others recommended integrating an evaluation module into the Decomposition Decision Model to assess the compatibility of decomposition patterns and strategies with available hardware resources and performance capabilities, including factors such as resource allocation, hardware availability, and operational efficiency. Additional suggestions include incorporating adaptive decomposition capabilities or using AI-based optimization to dynamically refine decomposition strategies. For researchers, this signals an opportunity to refine decomposition decision models by introducing context-awareness, specifically adapting the models to quantum hardware constraints (e.g., qubit connectivity, coherence times) and system performance conditions (e.g., real-time computational resource usage, latency) that vary depending on the quantum environment. Future work can also explore formal validation frameworks or AI-assisted architecture search techniques to automate the decomposition planning process. For practitioners, the decision model encourages a more structured approach to modularizing quantum systems, moving away from ad hoc, hardware-agnostic decisions. Practitioners are advised to tailor decomposition strategies to their specific quantum hardware capabilities and operational goals, particularly as system scale, error rates, and algorithmic complexity increase. Finally, while the Decomposition Decision Model provides a valuable framework for modular architecture design, its future evolution hinges on its ability to adapt to quantum-specific constraints. Making it hardware-aware, cost-sensitive, and dynamically tunable will significantly improve its utility for both researchers and practitioners building the next generation of quantum software systems.

\subsection{Data Processing Decision Model}\label{Disscussion:Data processing Decision Model}
\textbf{Quantum data processing requires structured and scalable handling of classical-quantum transitions}: The evaluation of the Data Processing Decision Model highlights its effectiveness in supporting architectural decision-making for managing the transition between classical and quantum data, a task fundamental to quantum software design. \textcolor{black}{With 93.4\%} of the participants rating the Data Processing Decision Model as \textit{Easy} or \textit{Very easy} to understand, this decision model demonstrates a clear advantage in clarity and practical applicability. Participants particularly appreciated the structured flow and the inclusion of well-established patterns such as the Pipe-and-Filter Pattern, Quantum Data Encoding, and the Consumer Pattern, which enable the construction of scalable data pipelines and reduce computational load. As one participant (P5) explained \faComment ``\textit{The Quantum Data Processing Decision Model is highly useful. It supports decision-making by incorporating patterns like the Pipe and Filter Pattern and the Quantum Data Encoding Pattern, which enhance scalability and flexibility, enabling quantum software to adapt to growing data processing demands}''. Likewise, another participant (P10) explained \faComment ``\textit{The model provides a clear structural flow for handling quantum data. The processing steps are well defined and systematic. For example, the use of the Consumer Pattern allows for on-demand quantum data management, which helps optimize resource allocation and improve overall system performance}''. These insights confirm the model's practical relevance in guiding pattern selection during architectural design. However, several interviewees also highlighted opportunities for improvement, particularly in supporting real-world applicability and adaptability. One participant (P2) suggested that the decision model should include \faComment ``\textit{More examples specifically, for each pattern, demonstrating how it performs or fails in real-world applications}''. Another participant (P8) proposed that \faComment ``\textit{I suggest adding a benchmarking framework to compare data processing patterns based on their efficiency}'', which would assist practitioners in selecting the most effective strategy according to their context. Additionally, a participant (P12) noted that \faComment ``\textit{I would suggest integrating artificial intelligence. In my view, quantum computing and AI are the leading technologies of this era, and combining them will be very impactful}'', and another participant (P16) recommended \faComment ``\textit{To enhance the (decision) model, it could be designed to accommodate multi-bit architectures and distributed quantum computing systems}''. For researchers, these insights emphasize the need for adaptive and empirically validated extensions to data processing decision models that reflect evolving hardware and software contexts. For practitioners, the takeaway is that while the current Data Processing Decision Model offers valuable scaffolding for architectural thinking, future enhancements should incorporate operational metrics and context-aware data processing recommendations. The ultimate goal is to ensure the decision model remains relevant, usable, and impactful as quantum data processing challenges grow in scale and complexity.

\subsection{Fault Tolerance Decision Model}\label{Disscussion:Fault Tolerance Decision Model}
\textbf{Researchers should advance multi-level fault correction techniques, while practitioners need layered, runtime-aware mechanisms for resilient quantum software systems}: The Fault Tolerance Decision Model plays a central role in guiding quantum software engineers to mitigate quantum-specific risks such as decoherence, error propagation, and gate instability. One participant (P5) emphasized that \faComment ``\textit{The Decorator Design Pattern enables real-time error handling by allowing quantum algorithms to dynamically adapt fault-tolerant strategies based on runtime conditions}''. Likewise, another participant (P8) highlighted the model's contribution to resilient quantum architectures by stating that \faComment ``\textit{It promotes best practices by evaluating fault governance against performance and resource utilization}''. Yet, the interviews also revealed evolving demands for layered decision-making that reflect the operational complexity of quantum systems. Several participants, such as P9, recommended hierarchical fault abstraction \faComment ``\textit{One suggestion would be to enhance the model by categorizing fault tolerance into multiple levels (e.g., hardware level, gate level, circuit level, and algorithm level). This layered classification could make the model more comprehensive and easier to apply in different contexts}''. This suggestion aligns with the unique error models in quantum computing, where fault propagation dynamics differ significantly from those in classical systems. For researchers, these findings highlight a pressing need to develop adaptive and modular fault tolerance frameworks that evolve in tandem with quantum hardware. This includes defining a taxonomy of fault categories, embedding predictive analytics into decision models, and exploring automated reasoning tools that can recommend suitable patterns for emerging error types. Researchers should also investigate cross-layer coordination techniques to ensure that fault mitigation is consistent from the hardware abstraction layer up to algorithm execution. For practitioners, fault tolerance must be treated as a first-class architectural concern. \textcolor{black}{The adoption of layered, runtime-aware fault handling, such as using Sparing, Voting, and Gate Error Mitigation Pattern, should be grounded in the specific operational conditions of the target platform.} Moreover, practitioners are encouraged to establish validation pipelines that include error injection, resilience testing, and resource profiling to ensure that fault tolerance is not only designed but empirically verified throughout the lifecycle of quantum software. The Fault Tolerance Decision Model provides a structured approach for using and combining fault tolerance patterns and strategies. This approach ensures that error handling is effectively designed and integrated into quantum software systems. However, as quantum systems grow in complexity, both research innovation and practitioner discipline will be necessary to ensure resilient, scalable, and high-performing quantum software systems.

\subsection{Integration and Optimization Decision Model}\label{Disscussion:IntegrationandOptimizationDecision Model}
\textbf{Researchers should refine integration methods and modularize optimization strategies, while practitioners should adopt adaptable, service-oriented patterns to build scalable quantum systems}: The Integration and Optimization Decision Model was broadly regarded by practitioners as a highly valuable resource for guiding both design and architectural evaluation in quantum software systems. \textcolor{black}{Across the 30 interviews, practitioners were consistently positive about the decision model's structured approach in supporting the selection of suitable patterns, such as Quantum Load Balancing Pattern, Circuit Transformer Pattern, Prototype Design Pattern, and Quantum Service Registry, for ensuring system adaptability, resource efficiency, and interoperability across hybrid infrastructures.} One participant (P5), for example, highlighted that \faComment ``\textit{The Quantum Load Balancing Pattern ensures efficient workload distribution, preventing bottlenecks and optimizing computational resources across multiple quantum processors}''. Similarly, another participant (P9) emphasized the model's utility in \faComment ``\textit{optimizing execution time and resource allocation, ultimately leading to better performance}''. However, some interviewees noted that the decision model could further improve its practical applicability by offering clearer guidance on pattern prioritization and trade-off analysis. Specifically, one participant (P6) pointed out \faComment ``\textit{The model could be improved by providing more explanation for different problem types, particularly optimization problems. For instance, how constraints are handled should be clearly defined}''. This indicates a need for the decision model to address not just pattern selection, but also the evaluation of trade-offs in integration and optimization decisions, especially as quantum applications scale and face increasingly complex operational environments. \textcolor{black}{For researchers, this opens a significant opportunity to advance the current decision models by incorporating quantitative trade-off evaluation frameworks, constraint handling techniques, and architecture decision support mechanisms. Modular representations of integration and optimization pipelines, especially for multi-qubit orchestration and classical-quantum bridging, could increase generalizability across evolving quantum software ecosystems. For practitioners, the findings reinforce the importance of modularity and service-oriented integration in QSE.} As one participant (P8) noted \faComment ``\textit{The Quantum Service Registry supports centralized tracking and management of quantum services, enhancing interoperability across platforms}''. Practitioners should prioritize patterns that support scalable integration, composability, and ease of future extension. Leveraging well-structured patterns such as Integration Pattern, Service-Oriented Architecture, and BYOC (Bring Your Own Container) can help teams avoid ad hoc architecture decisions, mitigate integration risk, and better prepare for evolving hardware and software toolchains. Moreover, adopting modular, loosely coupled architectures supports the deployment of optimized workflows, especially in distributed quantum environments. While the Integration and Optimization Decision Model is already perceived as sufficient and beneficial, it could become a more powerful tool through enhancements that support trade-off awareness and dynamic adaptability. Researchers should focus on embedding system-level intelligence into the decision model, while practitioners are encouraged to adopt modular, service-driven approaches that align with the rapid evolution of quantum technologies and standards.

\subsection{Algorithm Implementation Decision Model Decision Model}\label{Disscussion:AlgorithmImplementationDecisionModel}
\textbf{Researchers should explore algorithm adaptation and modular integration techniques, while practitioners need reusable components for scalable quantum development}: The Algorithm Implementation Decision Model was widely acknowledged by practitioners as essential for navigating architectural complexity in quantum software systems. It supports a range of architecture patterns, including Quantum Module Templates, Qubit Gate Patterns, and Classical–Quantum Interface Patterns, each aimed at facilitating efficient algorithm design and deployment. \textcolor{black}{With 90.0\% of the participants} found the decision model either \textit{Easy} or \textit{Very easy} to understand, and all agreed on its correctness and sufficiency to guide implementation decisions. One participant (P5) noted that \faComment ``\textit{The model facilitates performance-based decision-making through the use of Qubit Gate Patterns and Brickwork Patterns, allowing developers to balance efficiency, flexibility, and cost}''. This reflects a strong alignment between the decision model and real-world demands for performance optimization, hardware abstraction, and architectural flexibility. In particular, many practitioners emphasized the role of the decision model in promoting modularity and reusability, which are critical for building scalable quantum applications. As another participant (P10) observed: \faComment ``\textit{Patterns such as the Quantum Module and Quantum Module Template promote reusability, modularity, and maintainability. They help make quantum software more adaptable and efficient}''. Other participants, such as P9, pointed out that \faComment ``\textit{The Quantum Circuit Translator ensures that algorithms can be adapted for cross-device compatibility, allowing architectures to remain flexible across different quantum hardware platforms}''. For instance, the inclusion of the Quantum Circuit Translator pattern was seen as vital for achieving cross-device portability and reducing hardware lock-in, a key concern in a rapidly evolving ecosystem. Despite the strengths of this decision model, participants provided valuable insights into areas for improvement. One commonly cited suggestion was the need to provide clearer trade-off analysis for each implementation pattern. As a participant (P2) proposed \faComment ``\textit{Providing a more detailed breakdown of potential trade-offs for each pattern could help users better understand their impact on performance, system stability, and resource utilization}''. Furthermore, as quantum systems increasingly adopt hybrid architectures, practitioners highlighted the need to expand support for classical-quantum coordination patterns. One participant (P6) emphasized \faComment ``\textit{In our case, we’ve used the Classical Quantum Interface pattern, particularly in quantum machine learning. This model supports hybrid algorithm implementation effectively}''. For researchers, these insights suggest the need to further formalize and empirically validate trade-off considerations in algorithm implementation decisions. This includes developing tooling or model extensions that integrate cost-performance metrics, usability constraints, and system-fidelity considerations. Additionally, researchers should investigate how adaptive implementation strategies, such as AI-assisted selection of algorithmic patterns, can support runtime flexibility in emerging quantum-classical hybrid systems. For practitioners, the key takeaway is the value of implementing modular, reusable algorithm components that can evolve with system complexity and hardware transitions. Practitioners are encouraged to adopt the decision model not as a static guide, but as a living framework flexible enough to support experimentation, benchmarking, and the layering of advanced strategies such as template-based circuit rewriting strategy, circuit translation, and hybrid orchestration. In summary, the Algorithm Implementation Decision Model provides a robust foundation for guiding architectural choices; however, its long-term effectiveness depends on continued adaptation to algorithmic complexity, cross-platform compatibility, and the emergence of hybrid quantum-classical computing workflows.

\begin{figure}[h!]
\centering
\includegraphics[width=1\linewidth]{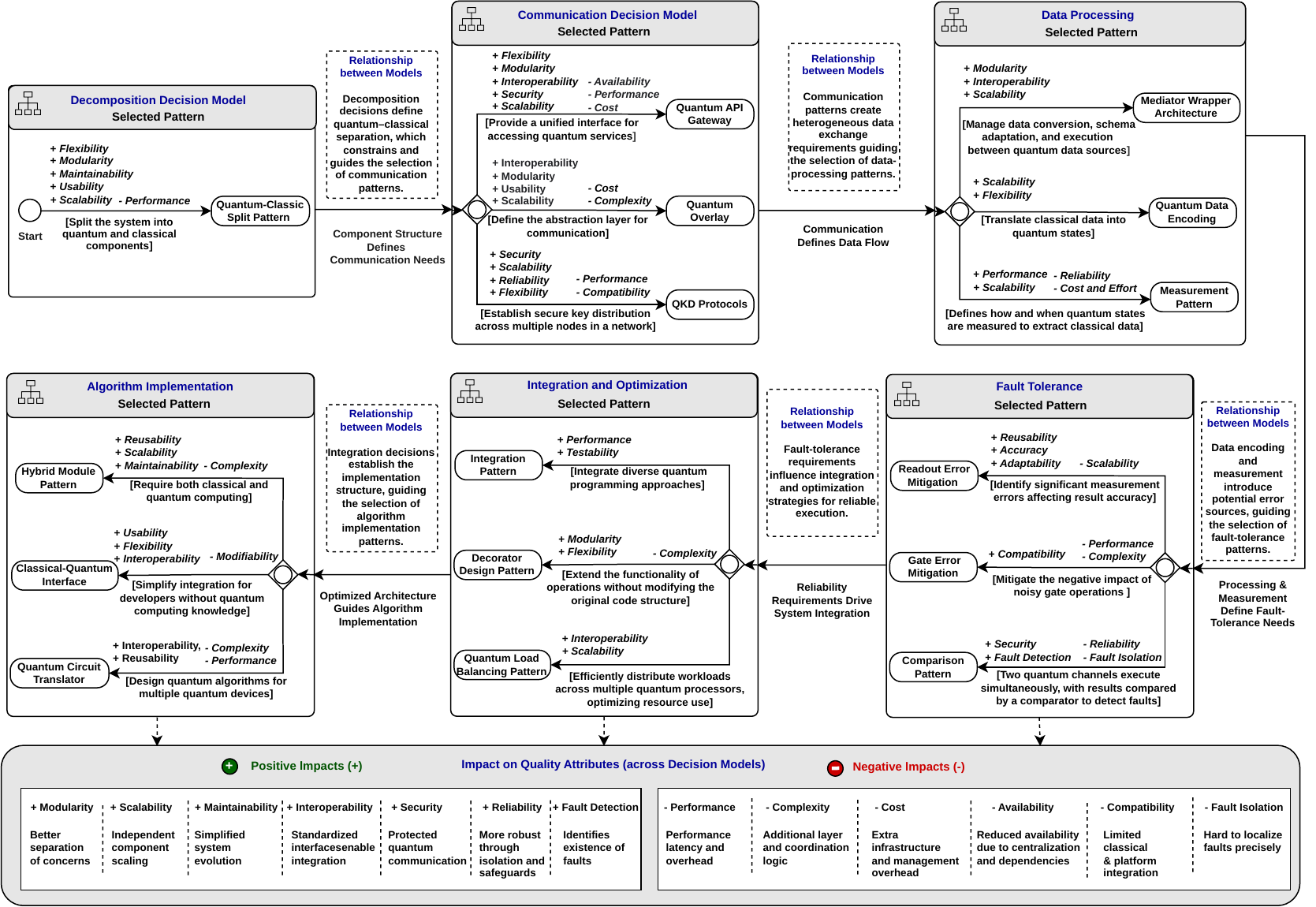}
\caption{An illustrative example of decision propagation and interdependencies among the six decision models}
\label{Fig: InterModelRelationship3}
\end{figure}

\textcolor{black}{
\subsection{Interdependencies between Decision Models}\label{InterdependenciesBetweenModels}} 
\textcolor{black}{Each of the proposed decision models can be used independently to address specific architectural concerns in quantum software systems. However, when solving complex and interdependent design problems, practitioners can combine multiple decision models, where decisions in one model influence and constrain the available choices in subsequent models. To capture these interactions, we adopt a graph-based dependency representation in which nodes represent decision models and selected architecture patterns and strategies, while edges represent dependency relationships such as structural constraints, data-flow dependencies, and QA propagation. To demonstrate the feasibility of this dependency analysis, Figure \ref{Fig: InterModelRelationship3} presents an \textbf{illustrative example} beginning with the selection of \textbf{Quantum-Classic Split Pattern} in \textbf{Decomposition Decision Model}. This example is not intended to exhaustively cover all possible configurations; rather, it illustrates how a single architectural decision propagates across six decision models: Decomposition, Communication, Data Processing, Fault Tolerance, Integration and Optimization, and Algorithm Implementation, and shapes subsequent pattern selections. Each model encapsulates architecture patterns and strategies that address specific concerns in quantum–classical systems.}

\textcolor{black}{The \textbf{illustrative example} begins with the \textbf{Decomposition Decision Model}, where \textbf{Quantum-Classic Split Pattern} establishes the structural separation between quantum and classical components. This structural decision defines the system's architectural boundaries and directly constrains subsequent communication requirements. As a result, the \textbf{Communication Decision Model} selects appropriate interaction mechanisms such as \textbf{Quantum API Gateway}, \textbf{Quantum Overlay}, and \textbf{QKD} protocols, which support secure and scalable coordination between separated components. The communication structure further determines the nature of data exchange, thereby guiding the \textbf{Data Processing Decision Model}. In particular, heterogeneous communication flows necessitate the use of \textbf{Mediator Wrapper Architecture} for data transformation and schema adaptation, the \textbf{Quantum Data Encoding} strategy for converting classical data into quantum states, and \textbf{Measurement Pattern} for extracting classical information from quantum computations. These data-processing operations introduce potential error sources arising from encoding, transformation, and measurement noise, which influence the \textbf{Fault Tolerance Decision Model}. To address these risks, patterns such as \textbf{Readout Error Mitigation}, \textbf{Gate Error Mitigation}, and \textbf{Comparison Pattern} are selected to detect, localize, and mitigate faults. In this context, fault detection identifies the presence of anomalies, while fault isolation highlights the challenges of precisely localizing faults in quantum-classical environments due to entanglement and distributed execution. The reliability requirements emerging from fault-tolerance decisions further influence the \textbf{Integration and Optimization Decision Model}, where system-level coordination is achieved through patterns such as \textbf{Integration Pattern}, \textbf{Decorator Design Pattern}, and \textbf{Quantum Load Balancing Pattern}. These patterns enable efficient orchestration of heterogeneous quantum services while improving scalability and execution efficiency. Finally, these integrated architectural decisions guide the \textbf{Algorithm Implementation Decision Model}, where implementation strategies such as \textbf{Hybrid Module Pattern}, \textbf{Classical–Quantum Interface}, and \textbf{Quantum Circuit Translator} support the construction and deployment of quantum algorithms across diverse hardware platforms.}

\textcolor{black}{\textbf{Impact on Quality Attributes (across Decision Models)} summarizes how architectural decisions propagate quality effects across the six decision models rather than remaining local to individual models. Positive attributes such as modularity, scalability, interoperability, maintainability, security, reliability, and fault detection emerge and are progressively reinforced from decomposition through communication, data processing, fault tolerance, integration, and algorithm implementation. In contrast, negative attributes such as performance overhead, increased complexity, cost, availability limitations, compatibility constraints, and fault isolation challenges accumulate due to layered abstractions, distributed quantum-classical interactions, and error-prone data transformations.}

\textcolor{black}{\textbf{Real World Scenario - Hybrid Quantum Drug Discovery Platform}: To further illustrate the practical applicability of combining decision models, we present an illustrative real-world-inspired scenario. This scenario is intended as a conceptual demonstration rather than empirical validation, showing how the six decision models and their patterns work together cohesively. In a pharmaceutical research platform for drug discovery, scientists process massive molecular datasets and execute hybrid quantum-classical simulations to identify potential compounds efficiently. Recent studies highlight the growing adoption of hybrid quantum pipelines and quantum machine learning techniques for drug discovery applications \cite{duong2026quantum}. The Decomposition Decision Model structures the system using Quantum Microservices \cite{moguel2022quantum} and Layered Architecture \cite{svore2006layered}, separating classical preprocessing, quantum simulation, and result analysis components to support scalability, maintainability, and interoperability. The Communication Decision Model employs a Quantum API Gateway and Quantum Point-to-Point Communication channels to coordinate requests between classical services and distributed quantum backends. Existing studies emphasize API-based orchestration and communication mechanisms for integrating classical and quantum infrastructures. Quantum API Gateway pattern supports service-oriented quantum software systems and coordinates quantum service execution across cloud-based quantum platforms \cite{garcia2021quantum}. Furthermore, scalable multi-core quantum architectures rely on point-to-point and teleportation-based quantum communication mechanisms between distributed quantum cores \cite{palesi2024assessing}. For handling large molecular datasets, the Data Processing Model applies Amplitude Encoding and QRAM Encoding \cite{weigold2021expanding}. Due to the noisy nature of current NISQ hardware, the Fault Tolerance Decision Model integrates Error Correction, Readout Error Mitigation \cite{scheerer2022fault}, and Decorator-based mitigation strategies to improve execution reliability dynamically \cite{mintz2020qcor}. Prior studies demonstrate that hybrid quantum-classical algorithms strongly depend on quantum error mitigation techniques to obtain reliable computation results on noisy hardware. The Integration and Optimization Decision Model then applies Quantum Load Balancing and workflow optimization strategies to distribute workloads across multiple quantum processors and optimize execution efficiency \cite{alvarado2024orchestration}. Finally, the Algorithm Implementation Decision Model executes simulations using Hybrid Modules, Quantum Circuit Translation, and Template-Based Circuit Rewriting Strategy to optimize quantum circuits according to encoding and execution constraints \cite{buhler2023patterns}. Together, the six decision models support scalable, reliable, and efficient drug discovery by coordinating modular decomposition, communication orchestration, quantum data encoding, error mitigation, and execution optimization across hybrid quantum-classical infrastructures.}

\section{Threats to Validity} \label{ResearchThreatsValidity}

\subsection{Construct Validity} \label{ConstructValidity}
In the context of our research on decision models for selecting architecture patterns and strategies in quantum software systems, construct validity refers to the degree to which the study accurately captures and operationalizes its intended theoretical concepts, namely, quantum software architecture patterns, decision-making strategies, and their influence on QAs. Ensuring strong construct validity is essential, as it underpins the credibility, generalizability, and real-world applicability of our findings. To enhance construct validity, we adopted multiple methodological safeguards: 1) \textit{Pilot Search for Terminological Precision}: Before initiating formal data collection, we conducted a pilot search to assess the relevance and appropriateness of the search terms associated with quantum software architecture patterns and strategies. This preliminary step allowed us to refine the search strings, ensuring that the terms used aligned with the study's key theoretical concepts. Specifically, we refined the search terms to capture architecture patterns and decision-making strategies used in quantum software systems. As a result, we maximized the retrieval of conceptually relevant studies and minimized the noise in the dataset. \textcolor{black}{2) \textit{Methodological Triangulation}: To mitigate bias that may arise from relying on a single data source or method, we employed a mixed-methods approach. Quantitative insights were derived from semi-structured interviews with 30 quantum software practitioners across 14 countries and 5 continents, focusing on the perceived familiarity, understandability, completeness, and usefulness of the decision models. In parallel, qualitative data were systematically mined from community-driven repositories such as GitHub and Stack Exchange and thematically analyzed.} This methodological triangulation strengthened the internal consistency of the concepts (i.e., the architecture patterns, decision-making strategies, and their associated QAs). It allowed us to validate emergent themes across independent sources. By combining rigorous terminological refinement with diverse data sources and analytical techniques, our study ensures that the constructs we aim to measure, namely architecture patterns, decision-making strategies, and their impact on QAs, are not only theoretically sound but also empirically grounded. This contributes to the robustness of our conclusions and supports the applicability of the proposed decision models in real-world QSE practices. \textcolor{black}{3) \textit{Selection Bias}: A potential threat to construct validity stems from the bias in selecting and interpreting data used to build the decision models. Our decision models were derived from a combination of mining publicly available repositories (i.e., GitHub and Stack Exchange) and an SLR. Although these sources offer both practical insights into real-world use and academic research on patterns and strategies, they may not reflect the full spectrum of quantum software development practices, particularly those in private or proprietary projects not hosted on GitHub. To mitigate this threat, we explicitly asked practitioners during the interviews to comment on whether any important patterns or strategies had been missed. Their feedback helped validate the comprehensiveness of the patterns and strategies we collected. Moreover, the selected GitHub projects are the most widely adopted quantum frameworks, serving as foundational infrastructure for many proprietary systems \cite{li2021understanding}. Consequently, their architectural decisions may indirectly influence some private-sector developments. Nevertheless, we acknowledge that proprietary systems may take additional considerations into account due to commercial constraints, and future research may validate our findings through industrial case studies.} \textcolor{black}{4) \textit{Data Source Coverage and Temporal Scope}: A potential threat to construct validity concerns the representativeness of our data sources and the absence of temporal filtering in project selection. Regarding temporal filtering, our study aims to capture the current state of architecture patterns and strategies in quantum software systems, rather than analyze their historical evolution. Therefore, we did not apply temporal restrictions, which is consistent with existing empirical QSE studies that prioritize project maturity over temporal constraints when investigating contemporary practices \cite{upadhyay2025analyzing,openja2022technical,li2021understanding}. Our selection criteria (\(\geq 50\) stars, \(\geq 15\) forks) already serve as proxies for project maturity and sustained community engagement, ensuring the analyzed repositories represent meaningful, established systems regardless of their creation or update dates. Notably, during practitioner interviews, no participants indicated that historically earlier patterns were missing from our decision models, further supporting the adequacy of our current-state focus. Regarding source coverage, while our dataset derives exclusively from open-source GitHub projects and Stack Exchange discussions, the selected projects, including Qiskit, Cirq, and PennyLane, are precisely those that dominate both academic research and industrial practice \cite{li2021understanding, openja2022technical}.}

\subsection{Internal Validity}\label{InternalValidity}
Ensuring the internal validity of our research is critical, particularly when developing decision models that support the selection of architecture patterns and strategies in quantum software systems. We identified the following potential threats to internal validity: 1) \textit{Correctness of Decision Models}: A potential threat lies in the correctness and accuracy of the developed decision models. To address this, we adopted a collaborative approach among the authors and incorporated practitioner feedback. Specifically, the first four authors conducted the initial identification of architecture patterns, strategies, and their associated QAs (Stage 1), while others focused on modeling and cross-validating the decision models to ensure internal consistency and alignment with the extracted data (Stage 2) (see Figure \ref{ResearchProcess}). We also included illustrative examples and constraints to clarify usage contexts and minimize misinterpretation. Moreover, the decision models were refined and validated based on the feedback obtained from the semi-structured interviews with practitioners to ensure practical relevance and clarity. 2) \textit{Evaluation Effect}: Another potential threat involves the misinterpretation of the decision models or evaluation criteria by interview participants. Misunderstandings regarding the study's purpose, the terminology used, or the models' structure could affect the quality of the feedback. To mitigate this threat, we provided participants with detailed explanations of the research objectives, terminology, model structure, and rationale, as well as clear instructions for the interview process. This helped ensure that their feedback was informed, accurate, and contextually grounded.

\subsection{External Validity}\label{ExternalValidity}
The external validity of this study, which focuses on the decision models for quantum software systems, may be influenced by the degree to which the findings can be generalized beyond the specific contexts and participants involved. \textit{1) Limited Practitioner Sample and Subjectivity}: A primary concern lies in the limited scope of practitioner validation. While the proposed decision models were evaluated through semi-structured interviews with \textcolor{black}{30 quantum software practitioners from 14 countries across 5 continents}, the relatively small sample size may constrain the generalizability of the results. To mitigate this limitation, we intentionally sought diversity across geographic regions, years of professional experience, organizational roles, and application domains (see Figure~\ref{DemographyDetails}). Despite the modest number of participants, their qualitative feedback played a crucial role in refining the decision models and ensuring they addressed practical concerns in real-world QSE. Practitioners assessed the decision models based on their familiarity, understandability, completeness, and usefulness in realistic quantum software development scenarios. However, the reliance on subjective evaluations introduces potential bias, particularly if the participants’ experiences, while diverse, do not fully represent all possible contexts in which the decision models may be applied. To further strengthen the external validity of the proposed models, future work should evaluate them across a broader range of real-world quantum software scenarios. \textcolor{black}{\textit{2) Impact of AI-Assisted Development Practices}: Another potential threat to external validity arises from recent shifts in developer practices driven by the widespread adoption of generative AI tools, such as GitHub Copilot \cite{song2024impact} and ChatGPT \cite{ahmad2023towards, husain2025exploring, d2024exploring}. Recent studies have reported declining participation and question volumes in online knowledge-sharing communities such as Stack Overflow following the introduction of ChatGPT, particularly among newer users \cite{burtch2024consequences}. Since our mining study relies on data from GitHub and Stack Exchange, the collected dataset may primarily reflect historically established practices rather than AI-assisted development workflows. However, mining these platforms remains critically important for empirical QSE research for several reasons. First, GitHub serves as the main archival source of quantum software evolution, capturing large-scale trends such as rapid repository and contributor growth \cite{upadhyay2025analyzing}, as well as enabling studies on programming purposes and bug classifications \cite{de2022software,yousuf2026bug}. Second, Stack Exchange provides valuable socio-technical insights into developers' reasoning, including challenges in interpreting quantum outputs and bridging classical-quantum gaps \cite{li2021understanding,khan2025mining}. Third, GitHub and Stack Exchange remain useful and traceable sources for studying domain-specific challenges in QSE, such as technical debt~\cite{ishimoto2024empirical}, noise validation~\cite{marquez2025security}, and circuit-level bugs~\cite{yousuf2026bug}. At the same time, the increasing adoption of AI-assisted development workflows may influence how practitioners seek and share knowledge in the future, potentially complementing traditional evidence sources. Finally, GitHub and Stack Exchange offer longitudinal, curated, and reproducible data (e.g., versioned commits, issues, and discussions), which are critical for scientific rigor and cannot be replaced by transient AI-generated outputs. Rather than reducing their relevance, AI increases the need for such ground truth data to validate AI-generated results. Future work should consider AI-assisted development workflows, but these community-driven platforms remain foundational for capturing authentic and traceable developer practices.}

\subsection{Conclusion Validity}\label{ConclusionValidity}
Conclusion validity concerns the extent to which the conclusions drawn from the analysis are credible and accurately reflect the collected data. To strengthen the validity of our study's conclusions, we employed a rigorous multi-stage research methodology grounded in established empirical software engineering practices (e.g., \cite{keele2007guidelines, easterbrook2008selecting, khan2023software}). Our decision models were developed using data derived from a systematic literature review and a mining study of real-world quantum software projects and discussions. To evaluate their reliability and practical relevance, we conducted semi-structured interviews with \textcolor{black}{30 quantum software practitioners from 14 countries}, spanning diverse roles and domains. The interviews assessed the models in terms of familiarity, understandability, completeness, and usefulness. Finally, to support transparency, reproducibility, and future research, we provided a publicly available replication package \cite{dataset2} that includes all the datasets, analysis artifacts, and interview materials of this study.

\section{Conclusions} \label{ResearchConclusion}
This study proposes a set of six decision models that integrate architecture patterns, strategies, and QAs to guide quantum software architects and developers in the architecture design of quantum systems. These decision models are constructed based on the data collected from both a mining study of GitHub and Stack Exchange and an SLR, addressing key design challenges in six vital areas: Communication, Decomposition, Data Processing, Fault Tolerance, Integration and Optimization, and Algorithm Implementation. To validate their practical relevance, we conducted semi-structured interviews with \textcolor{black}{30 practitioners} on quantum software development. \textcolor{black}{The proposed decision models demonstrate strong applicability in guiding architecture-centric development for quantum software systems and provide a structured approach to select reusable architecture patterns and strategies based on multiple quality criteria (e.g., maintainability, performance, scalability), thereby addressing the complexity and abstraction inherent in quantum-classical hybrid environments.} To evaluate the practicality and usability of these decision models, we assessed them across three key dimensions: practitioner familiarity, understandability, and perceived correctness. The findings suggest that the decision models offer substantial practical value and can effectively support both novice and experienced practitioners in architecting and evolving quantum software systems. For researchers, the high ratings in familiarity with the patterns and strategies and understandability provide a strong foundation for further extending these decision models with formal methods, tooling support, and domain-specific adaptations. For practitioners, the decision models serve as a reliable reference for structuring architectural thinking, improving design quality, and ensuring consistency when dealing with the complexities of quantum-classical integration.

In the next step, we plan to expand and refine the decision models to further support the architectural design of quantum software systems in both academic and industrial environments. Building upon this groundwork, our future research will focus on several key directions: \textcolor{black}{(1) Expanding the empirical validation by creating an open-access repository, where we will systematically document detailed mappings between architecture patterns, strategies, QAs, and design constraints, supported by large-scale industrial and longitudinal case studies, thereby complementing and extending existing resources in this domain;} (2) Developing a human-centric recommendation system that integrates contextual metadata (e.g., system requirements, performance metrics, and environmental factors) and AI-driven reasoning to provide adaptive, explainable guidance for pattern selection in quantum software system development; (3) Integration with quantum design tools such as Qiskit, PennyLane, and Cirq, embedding architectural decision-making support directly into the developer workflow. This could include leveraging existing tools such as the Quantum Architecture Description Language (QADL), which facilitates architecture-driven development of quantum software systems by offering a graphical interface, syntactic parsing, and integration with platforms like IBM Qiskit, as demonstrated by Waseem et al. \cite{waseem2025qadl}; and (4) Extension to domain-specific applications, including quantum cryptography, machine learning, and simulation, where domain-specific decision models will address unique architectural challenges like latency sensitivity and data handling.

\section*{Data Availability}
The replication package of this work has been made available at \cite{dataset2}.



\bibliographystyle{ACM-Reference-Format}
\bibliography{basebib}


\begin{thebibliography}{112}


\ifx \showCODEN    \undefined \def \showCODEN     #1{\unskip}     \fi
\ifx \showISBNx    \undefined \def \showISBNx     #1{\unskip}     \fi
\ifx \showISBNxiii \undefined \def \showISBNxiii  #1{\unskip}     \fi
\ifx \showISSN     \undefined \def \showISSN      #1{\unskip}     \fi
\ifx \showLCCN     \undefined \def \showLCCN      #1{\unskip}     \fi
\ifx \shownote     \undefined \def \shownote      #1{#1}          \fi
\ifx \showarticletitle \undefined \def \showarticletitle #1{#1}   \fi
\ifx \showURL      \undefined \def \showURL       {\relax}        \fi
\providecommand\bibfield[2]{#2}
\providecommand\bibinfo[2]{#2}
\providecommand\natexlab[1]{#1}
\providecommand\showeprint[2][]{arXiv:#2}

\bibitem[Ahmad and Babar(2016)]%
        {ahmad2016software}
\bibfield{author}{\bibinfo{person}{Aakash Ahmad} {and} \bibinfo{person}{Muhammad~Ali Babar}.} \bibinfo{year}{2016}\natexlab{}.
\newblock \showarticletitle{Software Architectures for Robotic Systems: A Systematic Mapping Study}.
\newblock \bibinfo{journal}{\emph{Journal of Systems and Software}}  \bibinfo{volume}{122} (\bibinfo{year}{2016}), \bibinfo{pages}{16--39}.
\newblock


\bibitem[Ahmad et~al\mbox{.}(2022)]%
        {ahmad2022towards}
\bibfield{author}{\bibinfo{person}{Aakash Ahmad}, \bibinfo{person}{Arif~Ali Khan}, \bibinfo{person}{Muhammad Waseem}, \bibinfo{person}{Mahdi Fahmideh}, {and} \bibinfo{person}{Tommi Mikkonen}.} \bibinfo{year}{2022}\natexlab{}.
\newblock \showarticletitle{Towards Process Centered Architecting for Quantum Software Systems}. In \bibinfo{booktitle}{\emph{Proceedings of the 1st IEEE International Conference on Quantum Software (QSW)}}. \bibinfo{publisher}{IEEE}, \bibinfo{address}{Barcelona, Spain}, \bibinfo{pages}{26--31}.
\newblock


\bibitem[Ahmad et~al\mbox{.}(2023a)]%
        {ahmad2023towards}
\bibfield{author}{\bibinfo{person}{Aakash Ahmad}, \bibinfo{person}{Muhammad Waseem}, \bibinfo{person}{Peng Liang}, \bibinfo{person}{Mahdi Fahmideh}, \bibinfo{person}{Mst~Shamima Aktar}, {and} \bibinfo{person}{Tommi Mikkonen}.} \bibinfo{year}{2023}\natexlab{a}.
\newblock \showarticletitle{Towards human-bot collaborative software architecting with ChatGPT}. In \bibinfo{booktitle}{\emph{Proceedings of the 27th International Conference on Evaluation and Assessment in Software Engineering (EASE)}}. \bibinfo{publisher}{ACM}, \bibinfo{address}{Oulu, Finland}, \bibinfo{pages}{279--285}.
\newblock


\bibitem[Ahmad et~al\mbox{.}(2023b)]%
        {ahmad2023engineering}
\bibfield{author}{\bibinfo{person}{Aakash Ahmad}, \bibinfo{person}{Muhammad Waseem}, \bibinfo{person}{Peng Liang}, \bibinfo{person}{Mahdi Fehmideh}, \bibinfo{person}{Arif~Ali Khan}, \bibinfo{person}{David~Georg Reichelt}, {and} \bibinfo{person}{Tommi Mikkonen}.} \bibinfo{year}{2023}\natexlab{b}.
\newblock \showarticletitle{Engineering Software Systems for Quantum Computing as a Service: A Mapping Study}.
\newblock \bibinfo{journal}{\emph{arXiv preprint arXiv:2303.14713}} (\bibinfo{year}{2023}).
\newblock


\bibitem[Akbar et~al\mbox{.}(2023)]%
        {akbar2023systematic}
\bibfield{author}{\bibinfo{person}{Muhammad~Azeem Akbar}, \bibinfo{person}{Arif~Ali Khan}, {and} \bibinfo{person}{Saima Rafi}.} \bibinfo{year}{2023}\natexlab{}.
\newblock \showarticletitle{A Systematic Decision-Making Framework for Tackling Quantum Software Engineering Challenges}.
\newblock \bibinfo{journal}{\emph{Automated Software Engineering}} \bibinfo{volume}{30}, \bibinfo{number}{2} (\bibinfo{year}{2023}), \bibinfo{pages}{22}.
\newblock


\bibitem[Akbar et~al\mbox{.}(2024)]%
        {akbar2024genetic}
\bibfield{author}{\bibinfo{person}{Muhammad~Azeem Akbar}, \bibinfo{person}{Arif~Ali Khan}, \bibinfo{person}{Mohammad Shameem}, {and} \bibinfo{person}{Mohammad Nadeem}.} \bibinfo{year}{2024}\natexlab{}.
\newblock \showarticletitle{Genetic Model-Based Success Probability Prediction of Quantum Software Development Projects}.
\newblock \bibinfo{journal}{\emph{Information and Software Technology}}  \bibinfo{volume}{165} (\bibinfo{year}{2024}), \bibinfo{pages}{107352}.
\newblock


\bibitem[Aktar et~al\mbox{.}(2025)]%
        {aktar2025architecture}
\bibfield{author}{\bibinfo{person}{Mst~Shamima Aktar}, \bibinfo{person}{Peng Liang}, \bibinfo{person}{Muhammad Waseem}, \bibinfo{person}{Amjed Tahir}, \bibinfo{person}{Aakash Ahmad}, \bibinfo{person}{Beiqi Zhang}, {and} \bibinfo{person}{Zengyang Li}.} \bibinfo{year}{2025}\natexlab{}.
\newblock \showarticletitle{Architecture Decisions in Quantum Software Systems: An Empirical Study on Stack Exchange and GitHub}.
\newblock \bibinfo{journal}{\emph{Information and Software Technology}}  \bibinfo{volume}{177} (\bibinfo{year}{2025}), \bibinfo{pages}{107587}.
\newblock


\bibitem[Aktar et~al\mbox{.}(2026)]%
        {dataset2}
\bibfield{author}{\bibinfo{person}{Mst~Shamima Aktar}, \bibinfo{person}{Peng Liang}, \bibinfo{person}{Muhammad Waseem}, \bibinfo{person}{Amjed Tahir}, \bibinfo{person}{Mojtaba Shahin}, \bibinfo{person}{Muhammad~Azeem Akbar}, \bibinfo{person}{Arif~Ali Khan}, \bibinfo{person}{Aakash Ahmad}, \bibinfo{person}{Musengamana~Jean de Dieu}, {and} \bibinfo{person}{Ruiyin Li}.} \bibinfo{year}{2026}\natexlab{}.
\newblock \showarticletitle{Replication Package for the Study: Decision Models for Selecting Patterns and Strategies in Quantum Software Systems}. \bibinfo{publisher}{\url{https://github.com/shamimaaktar1/DMQSA}}.
\newblock


\bibitem[Alvarado-Valiente et~al\mbox{.}(2024)]%
        {alvarado2024orchestration}
\bibfield{author}{\bibinfo{person}{Jaime Alvarado-Valiente}, \bibinfo{person}{Javier Romero-{\'A}lvarez}, \bibinfo{person}{Enrique Moguel}, \bibinfo{person}{Jose Garc{\'\i}a-Alonso}, {and} \bibinfo{person}{Juan~M Murillo}.} \bibinfo{year}{2024}\natexlab{}.
\newblock \showarticletitle{Orchestration for quantum services: The power of load balancing across multiple service providers}.
\newblock \bibinfo{journal}{\emph{Science of Computer Programming}}  \bibinfo{volume}{237} (\bibinfo{year}{2024}), \bibinfo{pages}{103139}.
\newblock


\bibitem[Aparicio-Morales et~al\mbox{.}(2024)]%
        {aparicio2024overview}
\bibfield{author}{\bibinfo{person}{{\'A}lvaro~M Aparicio-Morales}, \bibinfo{person}{Enrique Moguel}, \bibinfo{person}{Luis~Mariano Bibbo}, \bibinfo{person}{Alejandro Fernandez}, \bibinfo{person}{Jose Garcia-Alonso}, {and} \bibinfo{person}{Juan~M Murillo}.} \bibinfo{year}{2024}\natexlab{}.
\newblock \showarticletitle{An Overview of Quantum Software Engineering in Latin America}.
\newblock \bibinfo{journal}{\emph{Quantum Information Processing}}  \bibinfo{volume}{23} (\bibinfo{year}{2024}), \bibinfo{pages}{Article No.: 380}.
\newblock


\bibitem[Arute et~al\mbox{.}(2019)]%
        {arute2019quantum}
\bibfield{author}{\bibinfo{person}{Frank Arute}, \bibinfo{person}{Kunal Arya}, \bibinfo{person}{Ryan Babbush}, \bibinfo{person}{Dave Bacon}, \bibinfo{person}{Joseph~C Bardin}, \bibinfo{person}{Rami Barends}, \bibinfo{person}{Rupak Biswas}, \bibinfo{person}{Sergio Boixo}, \bibinfo{person}{Fernando~GSL Brandao}, \bibinfo{person}{David~A Buell}, {et~al\mbox{.}}} \bibinfo{year}{2019}\natexlab{}.
\newblock \showarticletitle{Quantum Supremacy using a Programmable Superconducting Processor}.
\newblock \bibinfo{journal}{\emph{Nature}} \bibinfo{volume}{574}, \bibinfo{number}{7779} (\bibinfo{year}{2019}), \bibinfo{pages}{505--510}.
\newblock


\bibitem[Baczyk and P{\'e}rez-Castillo(2025)]%
        {baczyk2025guidelines}
\bibfield{author}{\bibinfo{person}{Michal Baczyk} {and} \bibinfo{person}{Ricardo P{\'e}rez-Castillo}.} \bibinfo{year}{2025}\natexlab{}.
\newblock \showarticletitle{Guidelines for the Application of Hybrid Software Design Patterns}. In \bibinfo{booktitle}{\emph{Proceedings of the 1st International Conference on Quantum Software (IQSOFT)}}. \bibinfo{publisher}{SCITEPRESS}, \bibinfo{address}{Bilbao, Spain}, \bibinfo{pages}{105--111}.
\newblock


\bibitem[Baczyk et~al\mbox{.}(2024a)]%
        {baczyk2024patterns}
\bibfield{author}{\bibinfo{person}{Michal Baczyk}, \bibinfo{person}{Ricardo P{\'e}rez-Castillo}, {and} \bibinfo{person}{Mario Piattini}.} \bibinfo{year}{2024}\natexlab{a}.
\newblock \showarticletitle{Patterns for Quantum Software Engineering}. In \bibinfo{booktitle}{\emph{Proceedings of 1st Recent Advances in Quantum Computing and Technology (ReAQCT)}}. \bibinfo{publisher}{ACM}, \bibinfo{address}{Budapest, Hungary}, \bibinfo{pages}{1--6}.
\newblock


\bibitem[Baczyk et~al\mbox{.}(2024b)]%
        {baczyk2024towards}
\bibfield{author}{\bibinfo{person}{Michal Baczyk}, \bibinfo{person}{Ricardo P{\'e}rez-Castillo}, {and} \bibinfo{person}{Mario Piattini}.} \bibinfo{year}{2024}\natexlab{b}.
\newblock \showarticletitle{Towards a Framework of Architectural Patterns for Quantum Software Engineering}. In \bibinfo{booktitle}{\emph{Proceedings of the 5th IEEE International Conference on Quantum Computing and Engineering (QCE)}}. \bibinfo{publisher}{IEEE}, \bibinfo{address}{Montreal, QC, Canada}, \bibinfo{pages}{228--233}.
\newblock


\bibitem[Bechtold et~al\mbox{.}(2023)]%
        {bechtold2023patterns}
\bibfield{author}{\bibinfo{person}{Marvin Bechtold}, \bibinfo{person}{Johanna Barzen}, \bibinfo{person}{Martin Beisel}, \bibinfo{person}{Frank Leymann}, {and} \bibinfo{person}{Benjamin Weder}.} \bibinfo{year}{2023}\natexlab{}.
\newblock \showarticletitle{Patterns for Quantum Circuit Cutting}. In \bibinfo{booktitle}{\emph{Proceedings of the 30th Conference on Pattern Languages of Programs (PLoP)}}. \bibinfo{publisher}{The Hillside Group, United States}, \bibinfo{address}{Monticello, IL, USA}, \bibinfo{pages}{1--12}.
\newblock


\bibitem[Beisel et~al\mbox{.}(2022)]%
        {beisel2022patterns}
\bibfield{author}{\bibinfo{person}{Martin Beisel}, \bibinfo{person}{Johanna Barzen}, \bibinfo{person}{Frank Leymann}, \bibinfo{person}{Felix Truger}, \bibinfo{person}{Benjamin Weder}, {and} \bibinfo{person}{Vladimir Yussupov}.} \bibinfo{year}{2022}\natexlab{}.
\newblock \showarticletitle{Patterns for Quantum Error Handling}. In \bibinfo{booktitle}{\emph{Proceedings of the 14th International Conference on Pervasive Patterns and Applications (PATTERNS)}}. \bibinfo{publisher}{XPS}, \bibinfo{address}{Barcelona, Spain}, \bibinfo{pages}{22--30}.
\newblock


\bibitem[Bensoussan et~al\mbox{.}(2025)]%
        {bensoussan2025taxonomy}
\bibfield{author}{\bibinfo{person}{Avner Bensoussan}, \bibinfo{person}{Gunel Jahangirova}, {and} \bibinfo{person}{Mohammadreza Mousavi}.} \bibinfo{year}{2025}\natexlab{}.
\newblock \showarticletitle{A Taxonomy of Real Faults for Hybrid Quantum-Classical Software Architectures}.
\newblock \bibinfo{journal}{\emph{ACM Transactions on Software Engineering and Methodology}} (\bibinfo{year}{2025}).
\newblock


\bibitem[Bi et~al\mbox{.}(2018)]%
        {bi2018architecture}
\bibfield{author}{\bibinfo{person}{Tingting Bi}, \bibinfo{person}{Peng Liang}, {and} \bibinfo{person}{Antony Tang}.} \bibinfo{year}{2018}\natexlab{}.
\newblock \showarticletitle{Architecture Patterns, Quality Attributes, and Design Contexts: How Developers Design with Them}. In \bibinfo{booktitle}{\emph{Proceedings of the 25th Asia-Pacific Software Engineering Conference (APSEC)}}. \bibinfo{publisher}{IEEE}, \bibinfo{address}{Nara, Japan}, \bibinfo{pages}{49--58}.
\newblock


\bibitem[Biamonte et~al\mbox{.}(2017)]%
        {biamonte2017quantum}
\bibfield{author}{\bibinfo{person}{Jacob Biamonte}, \bibinfo{person}{Peter Wittek}, \bibinfo{person}{Nicola Pancotti}, \bibinfo{person}{Patrick Rebentrost}, \bibinfo{person}{Nathan Wiebe}, {and} \bibinfo{person}{Seth Lloyd}.} \bibinfo{year}{2017}\natexlab{}.
\newblock \showarticletitle{Quantum Machine Learning}.
\newblock \bibinfo{journal}{\emph{Nature}} \bibinfo{volume}{549}, \bibinfo{number}{7671} (\bibinfo{year}{2017}), \bibinfo{pages}{195--202}.
\newblock


\bibitem[Borges and Valente(2018)]%
        {borges2018s}
\bibfield{author}{\bibinfo{person}{Hudson Borges} {and} \bibinfo{person}{Marco~Tulio Valente}.} \bibinfo{year}{2018}\natexlab{}.
\newblock \showarticletitle{What’s in a github star? understanding repository starring practices in a social coding platform}.
\newblock \bibinfo{journal}{\emph{Journal of Systems and Software}}  \bibinfo{volume}{146} (\bibinfo{year}{2018}), \bibinfo{pages}{112--129}.
\newblock


\bibitem[Braun and Clarke(2006)]%
        {braun2006using}
\bibfield{author}{\bibinfo{person}{Virginia Braun} {and} \bibinfo{person}{Victoria Clarke}.} \bibinfo{year}{2006}\natexlab{}.
\newblock \showarticletitle{Using Thematic Analysis in Psychology}.
\newblock \bibinfo{journal}{\emph{Qualitative Research in Psychology}} \bibinfo{volume}{3}, \bibinfo{number}{2} (\bibinfo{year}{2006}), \bibinfo{pages}{77--101}.
\newblock


\bibitem[Brereton et~al\mbox{.}(2007)]%
        {brereton2007lessons}
\bibfield{author}{\bibinfo{person}{Pearl Brereton}, \bibinfo{person}{Barbara~A Kitchenham}, \bibinfo{person}{David Budgen}, \bibinfo{person}{Mark Turner}, {and} \bibinfo{person}{Mohamed Khalil}.} \bibinfo{year}{2007}\natexlab{}.
\newblock \showarticletitle{Lessons from Applying the Systematic Literature Review Process within the Software Engineering Domain}.
\newblock \bibinfo{journal}{\emph{Journal of Systems and Software}} \bibinfo{volume}{80}, \bibinfo{number}{4} (\bibinfo{year}{2007}), \bibinfo{pages}{571--583}.
\newblock


\bibitem[B{\"u}hler et~al\mbox{.}(2023)]%
        {buhler2023patterns}
\bibfield{author}{\bibinfo{person}{Fabian B{\"u}hler}, \bibinfo{person}{Johanna Barzen}, \bibinfo{person}{Martin Beisel}, \bibinfo{person}{Daniel Georg}, \bibinfo{person}{Frank Leymann}, {and} \bibinfo{person}{Karoline Wild}.} \bibinfo{year}{2023}\natexlab{}.
\newblock \showarticletitle{Patterns for Quantum Software Development}. In \bibinfo{booktitle}{\emph{Proceedings of the 15th International Conference on Pervasive Patterns and Applications (PATTERNS)}}. \bibinfo{publisher}{XPS}, \bibinfo{address}{Nice, France}, \bibinfo{pages}{30--39}.
\newblock


\bibitem[Burtch et~al\mbox{.}(2024)]%
        {burtch2024consequences}
\bibfield{author}{\bibinfo{person}{Gordon Burtch}, \bibinfo{person}{Dokyun Lee}, {and} \bibinfo{person}{Zhichen Chen}.} \bibinfo{year}{2024}\natexlab{}.
\newblock \showarticletitle{The consequences of generative AI for online knowledge communities}.
\newblock \bibinfo{journal}{\emph{Scientific Reports}} \bibinfo{volume}{14}, \bibinfo{number}{1} (\bibinfo{year}{2024}), \bibinfo{pages}{10413}.
\newblock


\bibitem[Buschmann et~al\mbox{.}(2007)]%
        {buschmann2007pattern}
\bibfield{author}{\bibinfo{person}{Frank Buschmann}, \bibinfo{person}{Kevlin Henney}, {and} \bibinfo{person}{Douglas~C. Schmidt}.} \bibinfo{year}{2007}\natexlab{}.
\newblock \bibinfo{booktitle}{\emph{Pattern-Oriented Software Architecture, Vol. 5: On Patterns and Pattern Languages}}.
\newblock \bibinfo{publisher}{John Wiley \& Sons}.
\newblock


\bibitem[Campbell et~al\mbox{.}(2013)]%
        {campbell2013coding}
\bibfield{author}{\bibinfo{person}{John~L Campbell}, \bibinfo{person}{Charles Quincy}, \bibinfo{person}{Jordan Osserman}, {and} \bibinfo{person}{Ove~K Pedersen}.} \bibinfo{year}{2013}\natexlab{}.
\newblock \showarticletitle{Coding in-Depth Semistructured Interviews: Problems of Unitization and Intercoder Reliability and Agreement}.
\newblock \bibinfo{journal}{\emph{Sociological Methods \& Research}} \bibinfo{volume}{42}, \bibinfo{number}{3} (\bibinfo{year}{2013}), \bibinfo{pages}{294--320}.
\newblock


\bibitem[Carneiro and Schmelmer(2016)]%
        {carneiro2016microservices}
\bibfield{author}{\bibinfo{person}{Cloves Carneiro} {and} \bibinfo{person}{Tim Schmelmer}.} \bibinfo{year}{2016}\natexlab{}.
\newblock \bibinfo{booktitle}{\emph{Microservices from Day One}}.
\newblock \bibinfo{publisher}{Springer}.
\newblock


\bibitem[Chen et~al\mbox{.}(2010)]%
        {chen2010towards}
\bibfield{author}{\bibinfo{person}{Lianping Chen}, \bibinfo{person}{Muhammad~Ali Babar}, {and} \bibinfo{person}{He Zhang}.} \bibinfo{year}{2010}\natexlab{}.
\newblock \showarticletitle{Towards an Evidence-Based Understanding of Electronic Data Sources}. In \bibinfo{booktitle}{\emph{Proceedings of the 14th International Conference on Evaluation and Assessment in Software Engineering (EASE)}}. \bibinfo{publisher}{BCS Learning \& Development}, \bibinfo{address}{Swindon, UK}, \bibinfo{pages}{1--4}.
\newblock


\bibitem[d'Aloisio et~al\mbox{.}(2024)]%
        {d2024exploring}
\bibfield{author}{\bibinfo{person}{Giordano d'Aloisio}, \bibinfo{person}{Sophie Fortz}, \bibinfo{person}{Carol Hanna}, \bibinfo{person}{Daniel Fortunato}, \bibinfo{person}{Avner Bensoussan}, \bibinfo{person}{E{\~n}aut Mendiluze~Usandizaga}, {and} \bibinfo{person}{Federica Sarro}.} \bibinfo{year}{2024}\natexlab{}.
\newblock \showarticletitle{Exploring LLM-driven explanations for quantum algorithms}. In \bibinfo{booktitle}{\emph{Proceedings of the 18th ACM/IEEE International Symposium on Empirical Software Engineering and Measurement (ESEM)}}. \bibinfo{publisher}{ACM}, \bibinfo{address}{Barcelona, Spain}, \bibinfo{pages}{475--481}.
\newblock


\bibitem[De~Stefano et~al\mbox{.}(2022)]%
        {de2022software}
\bibfield{author}{\bibinfo{person}{Manuel De~Stefano}, \bibinfo{person}{Fabiano Pecorelli}, \bibinfo{person}{Dario Di~Nucci}, \bibinfo{person}{Fabio Palomba}, {and} \bibinfo{person}{Andrea De~Lucia}.} \bibinfo{year}{2022}\natexlab{}.
\newblock \showarticletitle{Software engineering for quantum programming: How far are we?}
\newblock \bibinfo{journal}{\emph{Journal of Systems and Software}}  \bibinfo{volume}{190} (\bibinfo{year}{2022}), \bibinfo{pages}{111326}.
\newblock


\bibitem[De~Stefano et~al\mbox{.}(2024)]%
        {de2024quantum}
\bibfield{author}{\bibinfo{person}{Manuel De~Stefano}, \bibinfo{person}{Fabiano Pecorelli}, \bibinfo{person}{Fabio Palomba}, \bibinfo{person}{Davide Taibi}, \bibinfo{person}{Dario Di~Nucci}, {and} \bibinfo{person}{Andrea De~Lucia}.} \bibinfo{year}{2024}\natexlab{}.
\newblock \showarticletitle{Quantum software engineering issues and challenges: Insights from practitioners}.
\newblock In \bibinfo{booktitle}{\emph{Quantum Software: Aspects of Theory and System Design}}. \bibinfo{publisher}{Springer}, \bibinfo{pages}{337--355}.
\newblock


\bibitem[Di~Rocco et~al\mbox{.}(2020)]%
        {di2020topfilter}
\bibfield{author}{\bibinfo{person}{Juri Di~Rocco}, \bibinfo{person}{Davide Di~Ruscio}, \bibinfo{person}{Claudio Di~Sipio}, \bibinfo{person}{Phuong Nguyen}, {and} \bibinfo{person}{Riccardo Rubei}.} \bibinfo{year}{2020}\natexlab{}.
\newblock \showarticletitle{Topfilter: An Approach to Recommend Relevant Github Topics}. In \bibinfo{booktitle}{\emph{Proceedings of the 14th ACM/IEEE International Symposium on Empirical Software Engineering and Measurement (ESEM)}}. \bibinfo{publisher}{ACM}, \bibinfo{address}{Bari, Italy}, \bibinfo{pages}{1--11}.
\newblock


\bibitem[Ding et~al\mbox{.}(2017)]%
        {ding2017classification}
\bibfield{author}{\bibinfo{person}{Kai Ding}, \bibinfo{person}{Andrey Morozov}, {and} \bibinfo{person}{Klaus Janschek}.} \bibinfo{year}{2017}\natexlab{}.
\newblock \showarticletitle{Classification of Hierarchical Fault-Tolerant Design Patterns}. In \bibinfo{booktitle}{\emph{Proceedings of the 15th International Conference on Dependable, Autonomic and Secure Computing (DASC)}}. \bibinfo{publisher}{IEEE}, \bibinfo{address}{Orlando, FL, USA}, \bibinfo{pages}{612--619}.
\newblock


\bibitem[Duong et~al\mbox{.}(2026)]%
        {duong2026quantum}
\bibfield{author}{\bibinfo{person}{Hoang Phi~Yen Duong}, \bibinfo{person}{Syed Muhammad~Abuzar Rizvi}, \bibinfo{person}{Brad McNiven}, \bibinfo{person}{Thanh~Tuan Nguyen}, \bibinfo{person}{Hyundong Shin}, \bibinfo{person}{Octavia Dobre}, {and} \bibinfo{person}{Trung~Q Duong}.} \bibinfo{year}{2026}\natexlab{}.
\newblock \showarticletitle{Quantum machine learning for drug discovery: taxonomy, research challenges, and the road ahead}.
\newblock \bibinfo{journal}{\emph{Comput. Surveys}} \bibinfo{volume}{58}, \bibinfo{number}{8} (\bibinfo{year}{2026}), \bibinfo{pages}{1--36}.
\newblock


\bibitem[Dutta and Bhuyan(2024)]%
        {dutta2024quantum}
\bibfield{author}{\bibinfo{person}{Hrishikesh Dutta} {and} \bibinfo{person}{Amit~Kumar Bhuyan}.} \bibinfo{year}{2024}\natexlab{}.
\newblock \showarticletitle{Quantum Communication: From Fundamentals to Recent Trends, challenges and open problems}.
\newblock \bibinfo{journal}{\emph{arXiv preprint arXiv:2406.04492}} (\bibinfo{year}{2024}).
\newblock


\bibitem[Easterbrook et~al\mbox{.}(2008)]%
        {easterbrook2008selecting}
\bibfield{author}{\bibinfo{person}{Steve Easterbrook}, \bibinfo{person}{Janice Singer}, \bibinfo{person}{Margaret-Anne Storey}, {and} \bibinfo{person}{Daniela Damian}.} \bibinfo{year}{2008}\natexlab{}.
\newblock \showarticletitle{Selecting Empirical Methods for Software Engineering Research}.
\newblock \bibinfo{journal}{\emph{Guide to Advanced Empirical Software Engineering}} (\bibinfo{year}{2008}), \bibinfo{pages}{285--311}.
\newblock


\bibitem[Endrei et~al\mbox{.}(2004)]%
        {endrei2004patterns}
\bibfield{author}{\bibinfo{person}{Mark Endrei}, \bibinfo{person}{Jenny Ang}, \bibinfo{person}{Ali Arsanjani}, \bibinfo{person}{Sook Chua}, \bibinfo{person}{Philippe Comte}, \bibinfo{person}{P{\aa}l Krogdahl}, \bibinfo{person}{Min Luo}, {and} \bibinfo{person}{Tony Newling}.} \bibinfo{year}{2004}\natexlab{}.
\newblock \bibinfo{booktitle}{\emph{Patterns: Service-Oriented Architecture and Web Services}}.
\newblock \bibinfo{publisher}{IBM Corporation}.
\newblock


\bibitem[Fern{\'a}ndez-Osuna et~al\mbox{.}(2025)]%
        {fernandez2025exploring}
\bibfield{author}{\bibinfo{person}{Miriam Fern{\'a}ndez-Osuna}, \bibinfo{person}{Ricardo P{\'e}rez-Castillo}, \bibinfo{person}{Jos{\'e}~A Cruz-Lemus}, \bibinfo{person}{Michal Baczyk}, {and} \bibinfo{person}{Mario Piattini}.} \bibinfo{year}{2025}\natexlab{}.
\newblock \showarticletitle{Exploring design patterns in quantum software: a case study}.
\newblock \bibinfo{journal}{\emph{Computing}} \bibinfo{volume}{107}, \bibinfo{number}{5} (\bibinfo{year}{2025}), \bibinfo{pages}{1--31}.
\newblock


\bibitem[Feynman(2018)]%
        {feynman2018simulating}
\bibfield{author}{\bibinfo{person}{Richard~P Feynman}.} \bibinfo{year}{2018}\natexlab{}.
\newblock \showarticletitle{Simulating Physics with Computers}.
\newblock In \bibinfo{booktitle}{\emph{Feynman and Computation}}. \bibinfo{publisher}{CRC Press}, \bibinfo{pages}{133--153}.
\newblock


\bibitem[Garcia-Alonso et~al\mbox{.}(2021)]%
        {garcia2021quantum}
\bibfield{author}{\bibinfo{person}{Jose Garcia-Alonso}, \bibinfo{person}{Javier Rojo}, \bibinfo{person}{David Valencia}, \bibinfo{person}{Enrique Moguel}, \bibinfo{person}{Javier Berrocal}, {and} \bibinfo{person}{Juan~Manuel Murillo}.} \bibinfo{year}{2021}\natexlab{}.
\newblock \showarticletitle{Quantum Software as a Service Through a Quantum API Gateway}.
\newblock \bibinfo{journal}{\emph{IEEE Internet Computing}} \bibinfo{volume}{26}, \bibinfo{number}{1} (\bibinfo{year}{2021}), \bibinfo{pages}{34--41}.
\newblock


\bibitem[Grove(1988)]%
        {grove1988analysis}
\bibfield{author}{\bibinfo{person}{Richard~W Grove}.} \bibinfo{year}{1988}\natexlab{}.
\newblock \showarticletitle{An Analysis of the Constant Comparative Method}.
\newblock \bibinfo{journal}{\emph{Internation Journal of Qualitative Studies in Education}} \bibinfo{volume}{1}, \bibinfo{number}{3} (\bibinfo{year}{1988}), \bibinfo{pages}{273--279}.
\newblock


\bibitem[Grurl et~al\mbox{.}(2020)]%
        {grurl2020considering}
\bibfield{author}{\bibinfo{person}{Thomas Grurl}, \bibinfo{person}{J{\"u}rgen Fu{\ss}}, {and} \bibinfo{person}{Robert Wille}.} \bibinfo{year}{2020}\natexlab{}.
\newblock \showarticletitle{Considering Decoherence Errors in the Simulation of Quantum Circuits Using Decision Diagrams}. In \bibinfo{booktitle}{\emph{Proceedings of the 39th International Conference on Computer-Aided Design (ICCAD)}}. \bibinfo{publisher}{IEEE}, \bibinfo{address}{Thoothukudi, India}, \bibinfo{pages}{1--7}.
\newblock


\bibitem[Harper and Zheng(2015)]%
        {harper2015exploring}
\bibfield{author}{\bibinfo{person}{K~Eric Harper} {and} \bibinfo{person}{Jiang Zheng}.} \bibinfo{year}{2015}\natexlab{}.
\newblock \showarticletitle{Exploring software architecture context}. In \bibinfo{booktitle}{\emph{Proceedings of the 12th Working IEEE/IFIP Conference on Software Architecture (WICSA)}}. IEEE, \bibinfo{address}{Montreal, QC, Canada}, \bibinfo{pages}{123--126}.
\newblock


\bibitem[Humble and DeBenedictis(2019)]%
        {humble2019quantum}
\bibfield{author}{\bibinfo{person}{Travis~S Humble} {and} \bibinfo{person}{Erik~P DeBenedictis}.} \bibinfo{year}{2019}\natexlab{}.
\newblock \showarticletitle{Quantum Realism}.
\newblock \bibinfo{journal}{\emph{IEEE Computer}} \bibinfo{volume}{52}, \bibinfo{number}{6} (\bibinfo{year}{2019}), \bibinfo{pages}{13--17}.
\newblock


\bibitem[Husain et~al\mbox{.}(2025)]%
        {husain2025exploring}
\bibfield{author}{\bibinfo{person}{Mobashir Husain}, \bibinfo{person}{Muhammad~Sohail Khan}, \bibinfo{person}{Javed~Ali Khan}, \bibinfo{person}{Nek~Dil Khan}, \bibinfo{person}{Arif Khan}, {and} \bibinfo{person}{Muhammad~Azeem Akbar}.} \bibinfo{year}{2025}\natexlab{}.
\newblock \showarticletitle{Exploring Developers Discussion Forums for Quantum Software Engineering: A Fine-Grained Classification Approach Using Large Language Model (ChatGPT)}. In \bibinfo{booktitle}{\emph{Proceedings of the 33rd ACM International Conference on the Foundations of Software Engineering (FSE)}}. \bibinfo{publisher}{ACM}, \bibinfo{address}{New York, NY, USA}, \bibinfo{pages}{1742--1755}.
\newblock


\bibitem[{IBM Quantum}(2026)]%
        {ibm_qiskit_history}
\bibfield{author}{\bibinfo{person}{{IBM Quantum}}.} \bibinfo{year}{2026}\natexlab{}.
\newblock \bibinfo{title}{History of Qiskit}.
\newblock \bibinfo{howpublished}{\url{https://www.ibm.com/quantum/qiskit/history}}.
\newblock
\newblock
\shownote{Accessed: 2026-03-26}.


\bibitem[Ishimoto et~al\mbox{.}(2024)]%
        {ishimoto2024empirical}
\bibfield{author}{\bibinfo{person}{Yuta Ishimoto}, \bibinfo{person}{Yuto Nakamura}, \bibinfo{person}{Ryota Katsube}, \bibinfo{person}{Naoto Sato}, \bibinfo{person}{Hideto Ogawa}, \bibinfo{person}{Masanari Kondo}, \bibinfo{person}{Yasutaka Kamei}, {and} \bibinfo{person}{Naoyasu Ubayashi}.} \bibinfo{year}{2024}\natexlab{}.
\newblock \showarticletitle{An empirical study on self-admitted technical debt in quantum software}. In \bibinfo{booktitle}{\emph{Proceedings of the 31st Asia-Pacific Software Engineering Conference (APSEC)}}. IEEE, \bibinfo{address}{Chongqing, China}, \bibinfo{pages}{41--50}.
\newblock


\bibitem[Jacob and Mani(2018)]%
        {jacob2018software}
\bibfield{author}{\bibinfo{person}{Pramod~Mathew Jacob} {and} \bibinfo{person}{Prasanna Mani}.} \bibinfo{year}{2018}\natexlab{}.
\newblock \showarticletitle{Software Architecture Pattern Selection Model for Internet of Things Based Systems}.
\newblock \bibinfo{journal}{\emph{IET Software}} \bibinfo{volume}{12}, \bibinfo{number}{5} (\bibinfo{year}{2018}), \bibinfo{pages}{390--396}.
\newblock


\bibitem[Jim{\'e}nez-Fern{\'a}ndez et~al\mbox{.}(2023)]%
        {jimenez2023systematic}
\bibfield{author}{\bibinfo{person}{Sergio Jim{\'e}nez-Fern{\'a}ndez}, \bibinfo{person}{Jos{\'e}~A Cruz-Lemus}, {and} \bibinfo{person}{Mario Piattini}.} \bibinfo{year}{2023}\natexlab{}.
\newblock \showarticletitle{A Systematic Mapping Study on Quantum Circuits Design Patterns.}
\newblock \bibinfo{journal}{\emph{ICEIS (2)}} (\bibinfo{year}{2023}), \bibinfo{pages}{109--116}.
\newblock


\bibitem[Kashif et~al\mbox{.}(2026)]%
        {kashif2025developers}
\bibfield{author}{\bibinfo{person}{Syed~Mohammad Kashif}, \bibinfo{person}{Peng Liang}, {and} \bibinfo{person}{Amjed Tahir}.} \bibinfo{year}{2026}\natexlab{}.
\newblock \showarticletitle{On developers’ self-declaration of ai-generated code: An analysis of practices}.
\newblock \bibinfo{journal}{\emph{ACM Transactions on Software Engineering and Methodology}} \bibinfo{volume}{35}, \bibinfo{number}{7} (\bibinfo{year}{2026}), \bibinfo{pages}{1--37}.
\newblock


\bibitem[Khan et~al\mbox{.}(2023)]%
        {khan2023software}
\bibfield{author}{\bibinfo{person}{Arif~Ali Khan}, \bibinfo{person}{Aakash Ahmad}, \bibinfo{person}{Muhammad Waseem}, \bibinfo{person}{Peng Liang}, \bibinfo{person}{Mahdi Fahmideh}, \bibinfo{person}{Tommi Mikkonen}, {and} \bibinfo{person}{Pekka Abrahamsson}.} \bibinfo{year}{2023}\natexlab{}.
\newblock \showarticletitle{Software Architecture for Quantum Computing Systems—A Systematic Review}.
\newblock \bibinfo{journal}{\emph{Journal of Systems and Software}}  \bibinfo{volume}{201} (\bibinfo{year}{2023}), \bibinfo{pages}{111682}.
\newblock
\showISSN{0164-1212}


\bibitem[Khan et~al\mbox{.}(2024)]%
        {khan2024advancing}
\bibfield{author}{\bibinfo{person}{Arif~Ali Khan}, \bibinfo{person}{Davide Taibi}, \bibinfo{person}{C{\'e}cile~M Perrault}, {and} \bibinfo{person}{Asif~Ali Khan}.} \bibinfo{year}{2024}\natexlab{}.
\newblock \showarticletitle{Advancing Quantum Software Engineering: A Vision of Hybrid Full-Stack Iterative Model}. In \bibinfo{booktitle}{\emph{Proceedings of the 40th ACM/SIGAPP Symposium on Applied Computing (SAC)}}. \bibinfo{publisher}{ACM}, \bibinfo{address}{Catania, Italy}, \bibinfo{pages}{1444--1448}.
\newblock


\bibitem[Khan et~al\mbox{.}(2025)]%
        {khan2025mining}
\bibfield{author}{\bibinfo{person}{Arif~Ali Khan}, \bibinfo{person}{Boshuai Ye}, \bibinfo{person}{Muhammad~Azeem Akbar}, \bibinfo{person}{Javed~Ali Khan}, \bibinfo{person}{Davoud Mougouei}, {and} \bibinfo{person}{Xinyuan Ma}.} \bibinfo{year}{2025}\natexlab{}.
\newblock \showarticletitle{Mining Q\&A Platforms for Empirical Evidence on Quantum Software Programming}.
\newblock \bibinfo{journal}{\emph{arXiv preprint arXiv:2503.05240}} (\bibinfo{year}{2025}).
\newblock


\bibitem[Kitchenham et~al\mbox{.}(2007)]%
        {keele2007guidelines}
\bibfield{author}{\bibinfo{person}{Barbara Kitchenham}, \bibinfo{person}{Stuart Charters}, {et~al\mbox{.}}} \bibinfo{year}{2007}\natexlab{}.
\newblock \bibinfo{booktitle}{\emph{Guidelines for Performing Systematic Literature Reviews in Software Engineering}}.
\newblock \bibinfo{type}{{T}echnical {R}eport}. \bibinfo{institution}{EBSE Technical Report EBSE-2007-01 ver. 2.3}.
\newblock


\bibitem[Kitchenham et~al\mbox{.}(2009)]%
        {KITCHENHAM20097}
\bibfield{author}{\bibinfo{person}{Barbara Kitchenham}, \bibinfo{person}{O. {Pearl Brereton}}, \bibinfo{person}{David Budgen}, \bibinfo{person}{Mark Turner}, \bibinfo{person}{John Bailey}, {and} \bibinfo{person}{Stephen Linkman}.} \bibinfo{year}{2009}\natexlab{}.
\newblock \showarticletitle{Systematic literature reviews in software engineering – A systematic literature review}.
\newblock \bibinfo{journal}{\emph{Information and Software Technology}} \bibinfo{volume}{51}, \bibinfo{number}{1} (\bibinfo{year}{2009}), \bibinfo{pages}{7--15}.
\newblock


\bibitem[Klymenko et~al\mbox{.}(2025)]%
        {klymenko2024architectural}
\bibfield{author}{\bibinfo{person}{Mykhailo Klymenko}, \bibinfo{person}{Thong Hoang}, \bibinfo{person}{Xiwei Xu}, \bibinfo{person}{Zhenchang Xing}, \bibinfo{person}{Muhammad Usman}, \bibinfo{person}{Qinghua Lu}, {and} \bibinfo{person}{Liming Zhu}.} \bibinfo{year}{2025}\natexlab{}.
\newblock \showarticletitle{Architectural Patterns for Designing Quantum Artificial Intelligence Systems}.
\newblock \bibinfo{journal}{\emph{Journal of Systems and Software}}  \bibinfo{volume}{227} (\bibinfo{year}{2025}), \bibinfo{pages}{112456}.
\newblock


\bibitem[Landis and Koch(1977)]%
        {landis1977measurement}
\bibfield{author}{\bibinfo{person}{J~Richard Landis} {and} \bibinfo{person}{Gary~G Koch}.} \bibinfo{year}{1977}\natexlab{}.
\newblock \showarticletitle{The measurement of observer agreement for categorical data}.
\newblock \bibinfo{journal}{\emph{Biometrics}} (\bibinfo{year}{1977}), \bibinfo{pages}{159--174}.
\newblock


\bibitem[Lewis et~al\mbox{.}(2016)]%
        {lewis2016decision}
\bibfield{author}{\bibinfo{person}{Grace~A Lewis}, \bibinfo{person}{Patricia Lago}, {and} \bibinfo{person}{Paris Avgeriou}.} \bibinfo{year}{2016}\natexlab{}.
\newblock \showarticletitle{A Decision Model for Cyber-Foraging Systems}. In \bibinfo{booktitle}{\emph{Proceedings of the 13th Working IEEE/IFIP Conference on Software Architecture (WICSA)}}. \bibinfo{publisher}{IEEE}, \bibinfo{address}{Venice, Italy}, \bibinfo{pages}{51--60}.
\newblock


\bibitem[Leymann(2019)]%
        {leymann2019towards}
\bibfield{author}{\bibinfo{person}{Frank Leymann}.} \bibinfo{year}{2019}\natexlab{}.
\newblock \showarticletitle{Towards a pattern language for quantum algorithms}. In \bibinfo{booktitle}{\emph{Proceedings of the 1st International Workshop on Quantum Technology and Optimization Problems (QTOP)}}. Springer, \bibinfo{address}{Munich, Germany}, \bibinfo{pages}{218--230}.
\newblock


\bibitem[Li et~al\mbox{.}(2021)]%
        {li2021understanding}
\bibfield{author}{\bibinfo{person}{Heng Li}, \bibinfo{person}{Foutse Khomh}, \bibinfo{person}{Moses Openja}, {et~al\mbox{.}}} \bibinfo{year}{2021}\natexlab{}.
\newblock \showarticletitle{Understanding Quantum Software Engineering Challenges An Empirical Study on Stack Exchange Forums and GitHub Issues}. In \bibinfo{booktitle}{\emph{Proceedings of the 37th International Conference on Software Maintenance and Evolution (ICSME)}}. \bibinfo{publisher}{IEEE}, \bibinfo{address}{Luxembourg}, \bibinfo{pages}{343--354}.
\newblock


\bibitem[Liu et~al\mbox{.}(2023)]%
        {liu2023decision}
\bibfield{author}{\bibinfo{person}{Yue Liu}, \bibinfo{person}{Qinghua Lu}, \bibinfo{person}{Hye-Young Paik}, \bibinfo{person}{Guangsheng Yu}, {and} \bibinfo{person}{Liming Zhu}.} \bibinfo{year}{2023}\natexlab{}.
\newblock \showarticletitle{Decision Models for Selecting Patterns in Governance-driven Blockchain Systems}. In \bibinfo{booktitle}{\emph{Proceedings of the 6th IEEE International Conference on Blockchain (ICBC)}}. \bibinfo{publisher}{IEEE}, \bibinfo{address}{Hainan, China}, \bibinfo{pages}{307--314}.
\newblock


\bibitem[M{\'a}rquez et~al\mbox{.}(2025)]%
        {marquez2025security}
\bibfield{author}{\bibinfo{person}{Gast{\'o}n M{\'a}rquez}, \bibinfo{person}{Muhammad Waseem}, {and} \bibinfo{person}{Tommi Mikkonen}.} \bibinfo{year}{2025}\natexlab{}.
\newblock \showarticletitle{Security discussions in quantum software projects on GitHub}.
\newblock \bibinfo{journal}{\emph{Journal of Systems and Software}} (\bibinfo{year}{2025}), \bibinfo{pages}{112585}.
\newblock


\bibitem[Mintz et~al\mbox{.}(2020)]%
        {mintz2020qcor}
\bibfield{author}{\bibinfo{person}{Tiffany~M Mintz}, \bibinfo{person}{Alexander~J Mccaskey}, \bibinfo{person}{Eugene~F Dumitrescu}, \bibinfo{person}{Shirley~V Moore}, \bibinfo{person}{Sarah Powers}, {and} \bibinfo{person}{Pavel Lougovski}.} \bibinfo{year}{2020}\natexlab{}.
\newblock \showarticletitle{Qcor: A language extension specification for the heterogeneous quantum-classical model of computation}.
\newblock \bibinfo{journal}{\emph{ACM Journal on Emerging Technologies in Computing Systems}} \bibinfo{volume}{16}, \bibinfo{number}{2} (\bibinfo{year}{2020}), \bibinfo{pages}{1--17}.
\newblock


\bibitem[Moguel et~al\mbox{.}(2022)]%
        {moguel2022quantum}
\bibfield{author}{\bibinfo{person}{Enrique Moguel}, \bibinfo{person}{Javier Rojo}, \bibinfo{person}{David Valencia}, \bibinfo{person}{Javier Berrocal}, \bibinfo{person}{Jose Garcia-Alonso}, {and} \bibinfo{person}{Juan~M Murillo}.} \bibinfo{year}{2022}\natexlab{}.
\newblock \showarticletitle{Quantum service-oriented computing: current landscape and challenges}.
\newblock \bibinfo{journal}{\emph{Software Quality Journal}} \bibinfo{volume}{30}, \bibinfo{number}{4} (\bibinfo{year}{2022}), \bibinfo{pages}{983--1002}.
\newblock


\bibitem[Murillo et~al\mbox{.}(2025)]%
        {murillo2024quantum}
\bibfield{author}{\bibinfo{person}{Juan~M Murillo}, \bibinfo{person}{Jose Garcia-Alonso}, \bibinfo{person}{Enrique Moguel}, \bibinfo{person}{Johanna Barzen}, \bibinfo{person}{Frank Leymann}, \bibinfo{person}{Shaukat Ali}, \bibinfo{person}{Tao Yue}, \bibinfo{person}{Paolo Arcaini}, \bibinfo{person}{Ricardo~P{\'e}rez Castillo}, \bibinfo{person}{Ignacio Garc{\'\i}a~Rodr{\'\i}guez de Guzm{\'a}n}, {et~al\mbox{.}}} \bibinfo{year}{2025}\natexlab{}.
\newblock \showarticletitle{Quantum Software Engineering: Roadmap and Challenges Ahead}.
\newblock \bibinfo{journal}{\emph{ACM Transactions on Software Engineering and Methodology}} \bibinfo{volume}{34}, \bibinfo{number}{5} (\bibinfo{year}{2025}), \bibinfo{pages}{1--48}.
\newblock


\bibitem[Nallamothula(2020)]%
        {nallamothula2020selection}
\bibfield{author}{\bibinfo{person}{Lalitha Nallamothula}.} \bibinfo{year}{2020}\natexlab{}.
\newblock \showarticletitle{Selection of Quantum Computing Architecture Using a Decision Tree Approach}. In \bibinfo{booktitle}{\emph{Proceedings of the 3rd International Conference on Intelligent Sustainable Systems (ICISS)}}. \bibinfo{publisher}{IEEE}, \bibinfo{address}{Coimbatore, India}, \bibinfo{pages}{644--649}.
\newblock


\bibitem[Nielsen and Chuang(2010)]%
        {nielsen2010quantum}
\bibfield{author}{\bibinfo{person}{Michael~A Nielsen} {and} \bibinfo{person}{Isaac~L Chuang}.} \bibinfo{year}{2010}\natexlab{}.
\newblock \bibinfo{booktitle}{\emph{Quantum Computation and Quantum Information}}.
\newblock \bibinfo{publisher}{Cambridge University Press}.
\newblock


\bibitem[Openja et~al\mbox{.}(2022)]%
        {openja2022technical}
\bibfield{author}{\bibinfo{person}{Moses Openja}, \bibinfo{person}{Mohammad~Mehdi Morovati}, \bibinfo{person}{Le An}, \bibinfo{person}{Foutse Khomh}, {and} \bibinfo{person}{Mouna Abidi}.} \bibinfo{year}{2022}\natexlab{}.
\newblock \showarticletitle{Technical Debts and Faults in Open-Source Quantum Software Systems: An Empirical Study}.
\newblock \bibinfo{journal}{\emph{Journal of Systems and Software}}  \bibinfo{volume}{193} (\bibinfo{year}{2022}), \bibinfo{pages}{111458}.
\newblock


\bibitem[Palesi et~al\mbox{.}(2024)]%
        {palesi2024assessing}
\bibfield{author}{\bibinfo{person}{Maurizio Palesi}, \bibinfo{person}{Enrico Russo}, \bibinfo{person}{Davide Patti}, \bibinfo{person}{Giuseppe Ascia}, {and} \bibinfo{person}{Vincenzo Catania}.} \bibinfo{year}{2024}\natexlab{}.
\newblock \showarticletitle{Assessing the role of communication in scalable multi-core quantum architectures}. In \bibinfo{booktitle}{\emph{Proceedings of the 17th IEEE International Symposium on Embedded Multicore/Many-core Systems-on-Chip (MCSoC)}}. \bibinfo{publisher}{IEEE}, \bibinfo{address}{Kuala Lumpur, Malaysia}, \bibinfo{pages}{482--489}.
\newblock


\bibitem[P{\'e}rez-Castillo et~al\mbox{.}(2024)]%
        {perez2024preliminary}
\bibfield{author}{\bibinfo{person}{Ricardo P{\'e}rez-Castillo}, \bibinfo{person}{Miriam Fern{\'a}ndez-Osuna}, \bibinfo{person}{Jos{\'e}~A Cruz-Lemus}, {and} \bibinfo{person}{Mario Piattini}.} \bibinfo{year}{2024}\natexlab{}.
\newblock \showarticletitle{A Preliminary Study of the Usage of Design Patterns in Quantum Software}. In \bibinfo{booktitle}{\emph{Proceedings of the 5th ACM/IEEE International Workshop on Quantum Software Engineering (Q-SE)}}. \bibinfo{publisher}{ACM}, \bibinfo{address}{Lisbon, Portugal}, \bibinfo{pages}{41--48}.
\newblock


\bibitem[P{\'e}rez-Castillo et~al\mbox{.}(2023)]%
        {perez2023generation}
\bibfield{author}{\bibinfo{person}{Ricardo P{\'e}rez-Castillo}, \bibinfo{person}{Luis Jim{\'e}nez-Navajas}, \bibinfo{person}{Iv{\'a}n Cantalejo}, {and} \bibinfo{person}{Mario Piattini}.} \bibinfo{year}{2023}\natexlab{}.
\newblock \showarticletitle{Generation of Classical-Quantum Code from UML models}. In \bibinfo{booktitle}{\emph{Proceedings of the 4th IEEE International Conference on Quantum Computing and Engineering (QCE)}}. \bibinfo{publisher}{IEEE}, \bibinfo{address}{Bellevue, WA, USA}, \bibinfo{pages}{165--168}.
\newblock


\bibitem[P{\'e}rez-Castillo et~al\mbox{.}(2021)]%
        {perez2021modelling}
\bibfield{author}{\bibinfo{person}{Ricardo P{\'e}rez-Castillo}, \bibinfo{person}{Luis Jim{\'e}nez-Navajas}, {and} \bibinfo{person}{Mario Piattini}.} \bibinfo{year}{2021}\natexlab{}.
\newblock \showarticletitle{Modelling Quantum Circuits with UML}. In \bibinfo{booktitle}{\emph{Proceedings of the IEEE/ACM 2nd International Workshop on Quantum Software Engineering (Q-SE)}}. \bibinfo{publisher}{IEEE}, \bibinfo{address}{Madrid, Spain}, \bibinfo{pages}{7--12}.
\newblock


\bibitem[Pezz{\`e} et~al\mbox{.}(2025)]%
        {pezze20252030}
\bibfield{author}{\bibinfo{person}{Mauro Pezz{\`e}}, \bibinfo{person}{Silvia Abrah{\~a}o}, \bibinfo{person}{Birgit Penzenstadler}, \bibinfo{person}{Denys Poshyvanyk}, \bibinfo{person}{Abhik Roychoudhury}, {and} \bibinfo{person}{Tao Yue}.} \bibinfo{year}{2025}\natexlab{}.
\newblock \showarticletitle{A 2030 roadmap for software engineering}.
\newblock \bibinfo{journal}{\emph{ACM Transactions on Software Engineering and Methodology}} \bibinfo{volume}{34}, \bibinfo{number}{5} (\bibinfo{year}{2025}), \bibinfo{pages}{1--55}.
\newblock


\bibitem[Piattini et~al\mbox{.}(2021)]%
        {piattini2021toward}
\bibfield{author}{\bibinfo{person}{Mario Piattini}, \bibinfo{person}{Manuel Serrano}, \bibinfo{person}{Ricardo Perez-Castillo}, \bibinfo{person}{Guido Petersen}, {and} \bibinfo{person}{Jose~Luis Hevia}.} \bibinfo{year}{2021}\natexlab{}.
\newblock \showarticletitle{Toward a Quantum Software Engineering}.
\newblock \bibinfo{journal}{\emph{IT Professional}} \bibinfo{volume}{23}, \bibinfo{number}{1} (\bibinfo{year}{2021}), \bibinfo{pages}{62--66}.
\newblock


\bibitem[Ray et~al\mbox{.}(2014)]%
        {ray2014large}
\bibfield{author}{\bibinfo{person}{Baishakhi Ray}, \bibinfo{person}{Daryl Posnett}, \bibinfo{person}{Vladimir Filkov}, {and} \bibinfo{person}{Premkumar Devanbu}.} \bibinfo{year}{2014}\natexlab{}.
\newblock \showarticletitle{A large-scale study of programming languages and code quality in GitHub}. In \bibinfo{booktitle}{\emph{Proceedings of the 22nd ACM SIGSOFT international symposium on foundations of software engineering (FSE)}}. \bibinfo{publisher}{ACM}, \bibinfo{address}{Hong Kong, China}, \bibinfo{pages}{155--165}.
\newblock


\bibitem[Richards(2015)]%
        {richards2015software}
\bibfield{author}{\bibinfo{person}{Mark Richards}.} \bibinfo{year}{2015}\natexlab{}.
\newblock \bibinfo{booktitle}{\emph{Software Architecture Patterns}}. Vol.~\bibinfo{volume}{4}.
\newblock \bibinfo{publisher}{O'Reilly Media}.
\newblock


\bibitem[Richardson(2018)]%
        {richardson2018microservices}
\bibfield{author}{\bibinfo{person}{Chris Richardson}.} \bibinfo{year}{2018}\natexlab{}.
\newblock \bibinfo{booktitle}{\emph{Microservices Patterns: With Examples in Java}}.
\newblock \bibinfo{publisher}{Simon and Schuster}.
\newblock


\bibitem[Rojo et~al\mbox{.}(2021)]%
        {rojo2021trials}
\bibfield{author}{\bibinfo{person}{Javier Rojo}, \bibinfo{person}{David Valencia}, \bibinfo{person}{Javier Berrocal}, \bibinfo{person}{Enrique Moguel}, \bibinfo{person}{Jose Garcia-Alonso}, {and} \bibinfo{person}{Juan Manuel~Murillo Rodriguez}.} \bibinfo{year}{2021}\natexlab{}.
\newblock \showarticletitle{Trials and Tribulations of Developing Hybrid Quantum-Classical Microservices Systems}.
\newblock \bibinfo{journal}{\emph{arXiv preprint arXiv:2105.04421}} (\bibinfo{year}{2021}).
\newblock


\bibitem[Salloum et~al\mbox{.}(2024)]%
        {salloum2024integration}
\bibfield{author}{\bibinfo{person}{Hadi Salloum}, \bibinfo{person}{Hamza~Shafee Aldaghstany}, \bibinfo{person}{Osama Orabi}, \bibinfo{person}{Ahmad Haidar}, \bibinfo{person}{Mohammad~Reza Bahrami}, {and} \bibinfo{person}{Manuel Mazzara}.} \bibinfo{year}{2024}\natexlab{}.
\newblock \showarticletitle{Integration of Machine Learning with Quantum Annealing}. In \bibinfo{booktitle}{\emph{Proceedings of the 38th International Conference on Advanced Information Networking and Applications (AINA)}}. \bibinfo{publisher}{Springer}, \bibinfo{address}{Kitakyushu, Japan}, \bibinfo{pages}{338--348}.
\newblock


\bibitem[Saurabh et~al\mbox{.}(2023)]%
        {saurabh2023conceptual}
\bibfield{author}{\bibinfo{person}{Nishant Saurabh}, \bibinfo{person}{Shantenu Jha}, {and} \bibinfo{person}{Andre Luckow}.} \bibinfo{year}{2023}\natexlab{}.
\newblock \showarticletitle{A conceptual architecture for a quantum-hpc middleware}. In \bibinfo{booktitle}{\emph{Proceedings of the 2nd IEEE International Conference on Quantum Software (QSW)}}. IEEE, \bibinfo{address}{Chicago, IL, USA}, \bibinfo{pages}{116--127}.
\newblock


\bibitem[Scheerer et~al\mbox{.}(2022)]%
        {scheerer2022fault}
\bibfield{author}{\bibinfo{person}{Max Scheerer}, \bibinfo{person}{Jonas Klamroth}, {and} \bibinfo{person}{Oliver Denninger}.} \bibinfo{year}{2022}\natexlab{}.
\newblock \showarticletitle{Fault-tolerant hybrid quantum software systems}. In \bibinfo{booktitle}{\emph{Proceedings of the 1st IEEE International Conference on Quantum Software (QSW)}}. IEEE, \bibinfo{address}{Barcelona, Spain}, \bibinfo{pages}{52--57}.
\newblock


\bibitem[Sch{\"o}nberger et~al\mbox{.}(2022)]%
        {schonberger2022peel}
\bibfield{author}{\bibinfo{person}{Manuel Sch{\"o}nberger}, \bibinfo{person}{Maja Franz}, \bibinfo{person}{Stefanie Scherzinger}, {and} \bibinfo{person}{Wolfgang Mauerer}.} \bibinfo{year}{2022}\natexlab{}.
\newblock \showarticletitle{Peel| pile? cross-framework portability of quantum software}. In \bibinfo{booktitle}{\emph{Proceedings of the IEEE 19th international conference on software architecture companion (ICSA-C)}}. IEEE, \bibinfo{address}{Honolulu, HI, USA}, \bibinfo{pages}{164--169}.
\newblock


\bibitem[Seaman(1999)]%
        {seaman1999qualitative}
\bibfield{author}{\bibinfo{person}{Carolyn~B. Seaman}.} \bibinfo{year}{1999}\natexlab{}.
\newblock \showarticletitle{Qualitative Methods in Empirical Studies of Software Engineering}.
\newblock \bibinfo{journal}{\emph{IEEE Transactions on Software Engineering}} \bibinfo{volume}{25}, \bibinfo{number}{4} (\bibinfo{year}{1999}), \bibinfo{pages}{557--572}.
\newblock


\bibitem[Serrano et~al\mbox{.}(2022)]%
        {serrano2022quantum}
\bibfield{author}{\bibinfo{person}{Manuel~A Serrano}, \bibinfo{person}{Jose~A Cruz-Lemus}, \bibinfo{person}{Ricardo Perez-Castillo}, {and} \bibinfo{person}{Mario Piattini}.} \bibinfo{year}{2022}\natexlab{}.
\newblock \showarticletitle{Quantum Software Components and Platforms: Overview and Quality Assessment}.
\newblock \bibinfo{journal}{\emph{Comput. Surveys}} \bibinfo{volume}{55}, \bibinfo{number}{8} (\bibinfo{year}{2022}), \bibinfo{pages}{1--31}.
\newblock


\bibitem[Shor(1994)]%
        {shor1994algorithms}
\bibfield{author}{\bibinfo{person}{Peter~W Shor}.} \bibinfo{year}{1994}\natexlab{}.
\newblock \showarticletitle{Algorithms for Quantum Computation: Discrete Logarithms and Factoring}. In \bibinfo{booktitle}{\emph{Proceedings of the 35th Annual Symposium on Foundations of Computer Science (FOCS)}}. \bibinfo{publisher}{IEEE}, \bibinfo{address}{Santa Fe, New Mexico, USA}, \bibinfo{pages}{124--134}.
\newblock


\bibitem[Song et~al\mbox{.}(2024)]%
        {song2024impact}
\bibfield{author}{\bibinfo{person}{Fangchen Song}, \bibinfo{person}{Ashish Agarwal}, {and} \bibinfo{person}{Wen Wen}.} \bibinfo{year}{2024}\natexlab{}.
\newblock \showarticletitle{The impact of generative AI on collaborative open-source software development: Evidence from GitHub Copilot}.
\newblock \bibinfo{journal}{\emph{arXiv preprint arXiv:2410.02091}} (\bibinfo{year}{2024}).
\newblock


\bibitem[Stol et~al\mbox{.}(2016)]%
        {stol2016grounded}
\bibfield{author}{\bibinfo{person}{Klaas-Jan Stol}, \bibinfo{person}{Paul Ralph}, {and} \bibinfo{person}{Brian Fitzgerald}.} \bibinfo{year}{2016}\natexlab{}.
\newblock \showarticletitle{Grounded Theory in Software Engineering Research: a Critical Review and Guidelines}. In \bibinfo{booktitle}{\emph{Proceedings of the 38th International Conference on Software Engineering (ICSE)}}. \bibinfo{address}{Austin, TX, USA}, \bibinfo{pages}{120--131}.
\newblock


\bibitem[Svore et~al\mbox{.}(2006)]%
        {svore2006layered}
\bibfield{author}{\bibinfo{person}{Krysta~M Svore}, \bibinfo{person}{Alfred~V Aho}, \bibinfo{person}{Andrew~W Cross}, \bibinfo{person}{Isaac Chuang}, {and} \bibinfo{person}{Igor~L Markov}.} \bibinfo{year}{2006}\natexlab{}.
\newblock \showarticletitle{A Layered Software Architecture for Quantum Computing Design Tools}.
\newblock \bibinfo{journal}{\emph{IEEE Computer}} \bibinfo{volume}{39}, \bibinfo{number}{1} (\bibinfo{year}{2006}), \bibinfo{pages}{74--83}.
\newblock


\bibitem[Tsymbalista and Katernyak(2025)]%
        {tsymbalista2025toward}
\bibfield{author}{\bibinfo{person}{Markiian Tsymbalista} {and} \bibinfo{person}{Ihor Katernyak}.} \bibinfo{year}{2025}\natexlab{}.
\newblock \showarticletitle{Toward an ecosystem-agnostic standard for quantum runtime architecture}.
\newblock \bibinfo{journal}{\emph{Academia Quantum}} \bibinfo{volume}{2}, \bibinfo{number}{2} (\bibinfo{year}{2025}).
\newblock


\bibitem[Upadhyay et~al\mbox{.}(2025)]%
        {upadhyay2025analyzing}
\bibfield{author}{\bibinfo{person}{Krishna Upadhyay}, \bibinfo{person}{Vinaik Chhetri}, \bibinfo{person}{AB Siddique}, {and} \bibinfo{person}{Umar Farooq}.} \bibinfo{year}{2025}\natexlab{}.
\newblock \showarticletitle{Analyzing the Evolution and Maintenance of Quantum Software Repositories}. In \bibinfo{booktitle}{\emph{Proceedings of the 4th IEEE International Conference on Quantum Software (QSW)}}. \bibinfo{publisher}{IEEE}, \bibinfo{address}{Helsinki, Finland}, \bibinfo{pages}{173--184}.
\newblock


\bibitem[Valle et~al\mbox{.}(2021)]%
        {valle2021architectural}
\bibfield{author}{\bibinfo{person}{Pedro Henrique~Dias Valle}, \bibinfo{person}{Lina Garc{\'e}s}, {and} \bibinfo{person}{Elisa~Yumi Nakagawa}.} \bibinfo{year}{2021}\natexlab{}.
\newblock \showarticletitle{Architectural strategies for interoperability of software-intensive systems: practitioners' perspective}. In \bibinfo{booktitle}{\emph{Proceedings of the 36th Annual ACM Symposium on Applied Computing (SAC)}}. \bibinfo{publisher}{ACM}, \bibinfo{address}{Virtual Event, Republic of Korea}, \bibinfo{pages}{1399--1408}.
\newblock


\bibitem[Van~Dinter et~al\mbox{.}(2021)]%
        {van2021automation}
\bibfield{author}{\bibinfo{person}{Raymon Van~Dinter}, \bibinfo{person}{Bedir Tekinerdogan}, {and} \bibinfo{person}{Cagatay Catal}.} \bibinfo{year}{2021}\natexlab{}.
\newblock \showarticletitle{Automation of Systematic Literature Reviews: A Systematic Literature Review}.
\newblock \bibinfo{journal}{\emph{Information and Software Technology}}  \bibinfo{volume}{136} (\bibinfo{year}{2021}), \bibinfo{pages}{106589}.
\newblock


\bibitem[Vietz et~al\mbox{.}(2021)]%
        {vietz2021decision}
\bibfield{author}{\bibinfo{person}{Daniel Vietz}, \bibinfo{person}{Johanna Barzen}, \bibinfo{person}{Frank Leymann}, {and} \bibinfo{person}{Karoline Wild}.} \bibinfo{year}{2021}\natexlab{}.
\newblock \showarticletitle{On Decision Support for Quantum Application Developers: Categorization, Comparison, and Analysis of Existing Technologies}. In \bibinfo{booktitle}{\emph{Proceedings of the 21st International Conference on Computational Science (ICCS)}}. \bibinfo{publisher}{Springer}, \bibinfo{address}{Krakow, Poland}, \bibinfo{pages}{127--141}.
\newblock


\bibitem[Wang et~al\mbox{.}(2025)]%
        {wang2025decision}
\bibfield{author}{\bibinfo{person}{Yanze Wang}, \bibinfo{person}{Yiling Huang}, \bibinfo{person}{Jingyue Li}, \bibinfo{person}{Shanshan Li}, \bibinfo{person}{He Zhang}, \bibinfo{person}{Chenxing Zhong}, \bibinfo{person}{Xiaodong Liu}, \bibinfo{person}{Bohan Liu}, \bibinfo{person}{Yue Liu}, \bibinfo{person}{Qinghua Lu}, {et~al\mbox{.}}} \bibinfo{year}{2025}\natexlab{}.
\newblock \showarticletitle{Decision Support for Selecting Blockchain-Based Application Design Patterns with Layered Taxonomy and Quality Attributes}.
\newblock \bibinfo{journal}{\emph{IEEE Transactions on Software Engineering}} \bibinfo{volume}{51}, \bibinfo{number}{4} (\bibinfo{year}{2025}), \bibinfo{pages}{1039--1066}.
\newblock


\bibitem[Waseem et~al\mbox{.}(2022)]%
        {waseem2022decision}
\bibfield{author}{\bibinfo{person}{Muhammad Waseem}, \bibinfo{person}{Peng Liang}, \bibinfo{person}{Aakash Ahmad}, \bibinfo{person}{Mojtaba Shahin}, \bibinfo{person}{Arif~Ali Khan}, {and} \bibinfo{person}{Gast{\'o}n M{\'a}rquez}.} \bibinfo{year}{2022}\natexlab{}.
\newblock \showarticletitle{Decision Models for Selecting Patterns and Strategies in Microservices Systems and Their Evaluation by Practitioners}. In \bibinfo{booktitle}{\emph{Proceedings of the 44th IEEE/ACM International Conference on Software Engineering: Software Engineering in Practice (ICSE-SEIP)}}. \bibinfo{publisher}{ACM}, \bibinfo{address}{Pittsburgh, PA, USA}, \bibinfo{pages}{135--144}.
\newblock


\bibitem[Waseem et~al\mbox{.}(2020)]%
        {waseem2020systematic}
\bibfield{author}{\bibinfo{person}{Muhammad Waseem}, \bibinfo{person}{Peng Liang}, {and} \bibinfo{person}{Mojtaba Shahin}.} \bibinfo{year}{2020}\natexlab{}.
\newblock \showarticletitle{A Systematic Mapping study on Microservices Architecture in DevOps}.
\newblock \bibinfo{journal}{\emph{Journal of Systems and Software}}  \bibinfo{volume}{170} (\bibinfo{year}{2020}), \bibinfo{pages}{110798}.
\newblock


\bibitem[Waseem et~al\mbox{.}(2021)]%
        {waseem2021nature}
\bibfield{author}{\bibinfo{person}{Muhammad Waseem}, \bibinfo{person}{Peng Liang}, \bibinfo{person}{Mojtaba Shahin}, \bibinfo{person}{Aakash Ahmad}, {and} \bibinfo{person}{Ali~Rezaei Nassab}.} \bibinfo{year}{2021}\natexlab{}.
\newblock \showarticletitle{On the Nature of Issues in Five Open Source Microservices Systems: An Empirical Study}. In \bibinfo{booktitle}{\emph{Proceedings of the 25th International Conference on Evaluation and Assessment in Software Engineering (EASE)}}. \bibinfo{publisher}{ACM}, \bibinfo{address}{Trondheim, Norway}, \bibinfo{pages}{201--210}.
\newblock


\bibitem[Waseem et~al\mbox{.}(2025)]%
        {waseem2025qadl}
\bibfield{author}{\bibinfo{person}{Muhammad Waseem}, \bibinfo{person}{Tommi Mikkonen}, \bibinfo{person}{Aakash Ahmad}, \bibinfo{person}{Muhammad~Taimoor Khan}, \bibinfo{person}{Majid Haghparast}, \bibinfo{person}{Vlad Stirbu}, {and} \bibinfo{person}{Peng Liang}.} \bibinfo{year}{2025}\natexlab{}.
\newblock \showarticletitle{QADL: Prototype of Quantum Architecture Description Language}. In \bibinfo{booktitle}{\emph{Proceedings of the 29th International Conference on Evaluation and Assessment in Software Engineering (EASE)}}. \bibinfo{publisher}{ACM}, \bibinfo{address}{Istanbul, Turkey}, \bibinfo{pages}{714--719}.
\newblock


\bibitem[Weder et~al\mbox{.}(2021)]%
        {weder2021hybrid}
\bibfield{author}{\bibinfo{person}{Benjamin Weder}, \bibinfo{person}{Johanna Barzen}, \bibinfo{person}{Frank Leymann}, {and} \bibinfo{person}{Michael Zimmermann}.} \bibinfo{year}{2021}\natexlab{}.
\newblock \showarticletitle{Hybrid Quantum Applications Need Two Orchestrations in Superposition: A Software Architecture Perspective}. In \bibinfo{booktitle}{\emph{Proceedings of the 28th International Conference on Web Services (ICWS)}}. \bibinfo{publisher}{IEEE}, \bibinfo{address}{Chicago, IL, USA}, \bibinfo{pages}{1--13}.
\newblock


\bibitem[Weder et~al\mbox{.}(2020)]%
        {weder2020integrating}
\bibfield{author}{\bibinfo{person}{Benjamin Weder}, \bibinfo{person}{Uwe Breitenb{\"u}cher}, \bibinfo{person}{Frank Leymann}, {and} \bibinfo{person}{Karoline Wild}.} \bibinfo{year}{2020}\natexlab{}.
\newblock \showarticletitle{Integrating Quantum Computing into Workflow Modeling and Execution}. In \bibinfo{booktitle}{\emph{Proceedings of the IEEE/ACM 13th International Conference on Utility and Cloud Computing (UCC)}}. \bibinfo{publisher}{IEEE}, \bibinfo{address}{Leicester, United Kingdom}, \bibinfo{pages}{279--291}.
\newblock


\bibitem[Weigold et~al\mbox{.}(2020)]%
        {weigold2020data}
\bibfield{author}{\bibinfo{person}{Manuela Weigold}, \bibinfo{person}{Johanna Barzen}, \bibinfo{person}{Frank Leymann}, {and} \bibinfo{person}{Marie Salm}.} \bibinfo{year}{2020}\natexlab{}.
\newblock \showarticletitle{Data Encoding Patterns for Quantum Computing}. In \bibinfo{booktitle}{\emph{Proceedings of the 27th Conference on Pattern Languages of Programs (PLOP)}}. \bibinfo{publisher}{ACM}, \bibinfo{address}{Irsee, Germany}, \bibinfo{pages}{1--11}.
\newblock


\bibitem[Weigold et~al\mbox{.}(2021a)]%
        {weigold2021encoding}
\bibfield{author}{\bibinfo{person}{Manuela Weigold}, \bibinfo{person}{Johanna Barzen}, \bibinfo{person}{Frank Leymann}, {and} \bibinfo{person}{Marie Salm}.} \bibinfo{year}{2021}\natexlab{a}.
\newblock \showarticletitle{Encoding patterns for quantum algorithms}.
\newblock \bibinfo{journal}{\emph{IET Quantum Communication}} \bibinfo{volume}{2}, \bibinfo{number}{4} (\bibinfo{year}{2021}), \bibinfo{pages}{141--152}.
\newblock


\bibitem[Weigold et~al\mbox{.}(2021b)]%
        {weigold2021expanding}
\bibfield{author}{\bibinfo{person}{Manuela Weigold}, \bibinfo{person}{Johanna Barzen}, \bibinfo{person}{Frank Leymann}, {and} \bibinfo{person}{Marie Salm}.} \bibinfo{year}{2021}\natexlab{b}.
\newblock \showarticletitle{Expanding data encoding patterns for quantum algorithms}. In \bibinfo{booktitle}{\emph{Proceedings of the IEEE 18th International Conference on Software Architecture Companion (ICSA-C)}}. IEEE, \bibinfo{address}{Stuttgart, Germany}, \bibinfo{pages}{95--101}.
\newblock


\bibitem[Wohlin(2014)]%
        {wohlin2014guidelines}
\bibfield{author}{\bibinfo{person}{Claes Wohlin}.} \bibinfo{year}{2014}\natexlab{}.
\newblock \showarticletitle{Guidelines for Snowballing in Systematic Literature Studies and a Replication in Software Engineering}. In \bibinfo{booktitle}{\emph{Proceedings of the 18th International Conference on Evaluation and Assessment in Software Engineering (EASE)}}. \bibinfo{publisher}{ACM}, \bibinfo{address}{London, England, UK}, \bibinfo{pages}{1--10}.
\newblock


\bibitem[Wu et~al\mbox{.}(2023)]%
        {wu2023qucomm}
\bibfield{author}{\bibinfo{person}{Anbang Wu}, \bibinfo{person}{Yufei Ding}, {and} \bibinfo{person}{Ang Li}.} \bibinfo{year}{2023}\natexlab{}.
\newblock \showarticletitle{Qucomm: Optimizing collective communication for distributed quantum computing}. In \bibinfo{booktitle}{\emph{Proceedings of the 56th Annual IEEE/ACM International Symposium on Microarchitecture}}. \bibinfo{pages}{479--493}.
\newblock


\bibitem[Xu et~al\mbox{.}(2021)]%
        {xu2021decision}
\bibfield{author}{\bibinfo{person}{Xiwei Xu}, \bibinfo{person}{HMN~Dilum Bandara}, \bibinfo{person}{Qinghua Lu}, \bibinfo{person}{Ingo Weber}, \bibinfo{person}{Len Bass}, {and} \bibinfo{person}{Liming Zhu}.} \bibinfo{year}{2021}\natexlab{}.
\newblock \showarticletitle{A Decision Model for Choosing Patterns in Blockchain-Based Applications}. In \bibinfo{booktitle}{\emph{Proceedings of the 18th IEEE International Conference on Software Architecture (ICSA)}}. \bibinfo{publisher}{IEEE}, \bibinfo{address}{Stuttgart, Germany}, \bibinfo{pages}{47--57}.
\newblock


\bibitem[Ye et~al\mbox{.}(2026)]%
        {ye2026c2}
\bibfield{author}{\bibinfo{person}{Boshuai Ye}, \bibinfo{person}{Arif~Ali Khan}, \bibinfo{person}{Teemu Pihkakoski}, \bibinfo{person}{Peng Liang}, \bibinfo{person}{Muhammad Azeem~Akbar}, \bibinfo{person}{Matti Silveri}, {and} \bibinfo{person}{Lauri Malmi}.} \bibinfo{year}{2026}\natexlab{}.
\newblock \showarticletitle{C2|Q>: A Robust Framework for Bridging Classical and Quantum Software Development}.
\newblock \bibinfo{journal}{\emph{ACM Transactions on Software Engineering and Methodology}} (\bibinfo{year}{2026}).
\newblock


\bibitem[Yousuf and Sofi(2026)]%
        {yousuf2026bug}
\bibfield{author}{\bibinfo{person}{Mir~Mohammad Yousuf} {and} \bibinfo{person}{Shabir~Ahmad Sofi}.} \bibinfo{year}{2026}\natexlab{}.
\newblock \showarticletitle{Bug Classification in quantum software: a rule-based framework and its evaluation}.
\newblock \bibinfo{journal}{\emph{Automated Software Engineering}} \bibinfo{volume}{33}, \bibinfo{number}{1} (\bibinfo{year}{2026}), \bibinfo{pages}{1--46}.
\newblock


\bibitem[Yue et~al\mbox{.}(2023)]%
        {yue2023challenges}
\bibfield{author}{\bibinfo{person}{Tao Yue}, \bibinfo{person}{Wolfgang Mauerer}, \bibinfo{person}{Shaukat Ali}, {and} \bibinfo{person}{Davide Taibi}.} \bibinfo{year}{2023}\natexlab{}.
\newblock \showarticletitle{Challenges and Opportunities in Quantum Software Architecture}.
\newblock In \bibinfo{booktitle}{\emph{Software Architecture: Recent Trends in Software Architecture}}. \bibinfo{publisher}{Springer}, \bibinfo{pages}{45--52}.
\newblock


\bibitem[Zhang et~al\mbox{.}(2011)]%
        {zhang2011empirical}
\bibfield{author}{\bibinfo{person}{He Zhang}, \bibinfo{person}{Muhammad~Ali Babar}, \bibinfo{person}{Xu Bai}, \bibinfo{person}{Juan Li}, {and} \bibinfo{person}{Liguo Huang}.} \bibinfo{year}{2011}\natexlab{}.
\newblock \showarticletitle{An Empirical Assessment of A Systematic Search Process for Systematic Reviews}. In \bibinfo{booktitle}{\emph{Proceedings of the 15th Annual Conference on Evaluation and Assessment in Software Engineering (EASE)}}. \bibinfo{publisher}{IEEE}, \bibinfo{address}{Durham, UK}, \bibinfo{pages}{56--65}.
\newblock


\bibitem[Zhao(2024)]%
        {zhao2024towards}
\bibfield{author}{\bibinfo{person}{Jianjun Zhao}.} \bibinfo{year}{2024}\natexlab{}.
\newblock \showarticletitle{Towards an Architecture Description Language for Hybrid Quantum-Classical Systems}. In \bibinfo{booktitle}{\emph{Proceedings of the 3rd IEEE International Conference on Quantum Software (QSW)}}. \bibinfo{publisher}{IEEE}, \bibinfo{address}{Shenzhen, China}, \bibinfo{pages}{19--23}.
\newblock


\bibitem[Zimmermann et~al\mbox{.}(2008)]%
        {zimmermann2008combining}
\bibfield{author}{\bibinfo{person}{Olaf Zimmermann}, \bibinfo{person}{Uwe Zdun}, \bibinfo{person}{Thomas Gschwind}, {et~al\mbox{.}}} \bibinfo{year}{2008}\natexlab{}.
\newblock \showarticletitle{Combining Pattern Languages and Reusable Architectural Decision Models into a Comprehensive and Comprehensible Design Method}. In \bibinfo{booktitle}{\emph{Proceedings of the 7th Working IEEE/IFIP Conference on Software Architecture (WICSA)}}. \bibinfo{publisher}{IEEE}, \bibinfo{address}{Vancouver, BC, Canada}, \bibinfo{pages}{157--166}.
\newblock


\end{thebibliography}










\end{document}